\documentclass[prd,nofootinbib,twocolumn]{revtex4}
\usepackage{graphicx}
\usepackage{bm}
\usepackage{float}
\usepackage{subfigure}
\usepackage{amsmath}
\usepackage{amsfonts}
\usepackage{color}
\usepackage{slashed}
\usepackage{fancyhdr}

\newcommand{\mpl}{M_{\text{pl}}}

\newcommand{\Dd}[1]{\frac{{\rm d}}{{\rm d} #1}}

\newcommand{\intk}[2]{\int \frac{{\rm d}^{#1}#2}{(2\pi)^{#1}}}
\newcommand{\intx}[2]{\int {\rm d}^{#1}#2}
\newcommand{\intd}[1]{\int {\rm d}#1}

\newcommand{\dx}[1]{\text{d}#1}
\newcommand{\dd}{\text{d}}

\newcommand{\intdt}{i\int {\rm d}t}

\newcommand{\mma}{{\sf Mathematica}}

\begin{document}

\title{The $n$-body problem in General Relativity up to the second post-Newtonian order from perturbative field
theory}

\author{Yi-Zen Chu}
\affiliation{CERCA, Department of Physics, Case Western Reserve University,
10900 Euclid Avenue, Cleveland, OH 44106-7079, USA}

\begin{abstract}
\noindent Motivated by experimental probes of general relativity, we adopt
methods from perturbative (quantum) field theory to compute, up to certain
integrals, the effective lagrangian for its $n$-body problem. Perturbation
theory is performed about a background Minkowski spacetime to
$\mathcal{O}[(v/c)^4]$ beyond Newtonian gravity, where $v$ is the typical speed
of these $n$ particles in their center of energy frame. For the specific case
of the 2 body problem, the major efforts underway to measure gravitational
waves produced by in-spiraling compact astrophysical binaries require their
gravitational interactions to be computed beyond the currently known
$\mathcal{O}[(v/c)^7]$. We argue that such higher order post-Newtonian
calculations must be automated for these field theoretic methods to be applied
successfully to achieve this goal. In view of this, we outline an algorithm
that would in principle generate the relevant Feynman diagrams to an arbitrary
order in $v/c$ and take steps to develop the necessary software. The Feynman
diagrams contributing to the $n$-body effective action at
$\mathcal{O}[(v/c)^6]$ beyond Newton are derived.
\end{abstract}

\maketitle

\section{Introduction and motivation}

In this paper we are concerned with the problem of describing the gravitational
dynamics of arbitrary $n \geq 2$ compact non-rotating bodies moving in a
background Minkowski spacetime. By assuming non-relativistic motion, this
problem can be approached in a perturbative manner, by approximating these
compact objects as point masses and calculating the effective lagrangian
$L_{\text{eff}}[\{ \vec{x}_a, \vec{v}_a, \dot{\vec{v}}_a, \dots \}]$ for their
coordinates $\{\vec{x}_a| a = 1,2,\dots,n\}$ and their time derivatives
$\{\vec{v}_a, \dot{\vec{v}}_a, \dots\}$, up to some given order in the typical
speed $v$ of these $n$ objects:\footnote{We use units where all speeds or
velocities are measured in multiples of the speed of light, i.e. $c = 1$.}
Newtonian gravity starts at $\mathcal{O}[v^0]$ and the Einstein-Infeld-Hoffman
lagrangian \cite{EinsteinInfeldHoffman}, that describes the precession of the
perihelion of elliptical orbits, is of $\mathcal{O}[v^2]$ (1 PN).\footnote{The
nomenclature is: $\mathcal{O}[v^{2Q}] \leftrightarrow Q$ PN.} The $n$ body
problem at $\mathcal{O}[v^4]$ was first tackled by Ohta et al. \cite{Ohta}.
Some computational and coordinate issues encountered there were clarified by
Damour and Sch\"{a}fer \cite{DamourSchaeferNBody}. In the latter, some
integrals could not be evaluated. A portion of these were later performed by
Sch\"{a}fer \cite{3BodySchaefer}, so that currently, up to the $n=3$ case is
known. But to know the effective lagrangian for arbitrary $n$ at this order,
one needs to further calculate the integrals for the $n=4$ case. As we will see
later, once $L_{\text{eff}}$ is known up to $n=4$, the arbitrary $n$-body
lagrangian will follow from a limited form of superposition.

We will examine this problem using perturbative field theory techniques
introduced in \cite{nrgr}. The motivations are two-fold, both of them stemming
from experimental probes of gravitational physics: one requires the 2-body
effective lagrangian to higher than $\mathcal{O}[v^7]$, and the other may need
the $n$-body counterpart at $\mathcal{O}[v^4]$.

\textsl{Gravitational Waves Detection} \quad The recent years have seen an
array of gravitational wave detectors such as GEO, LIGO, TAMA, and VIRGO coming
online. These experiments seek to detect gravitational waves produced by binary
black holes and/or neutron stars as they spiral towards each other. Within
their frequency bandwidth, these detectors are able to track the frequency
evolution of the gravitational waves from these binaries over
$\mathcal{O}[10^4]$ orbital cycles and hence make very accurate measurements.
To be able to do so, however, theoretical templates need to be constructed so
that the raw data can be integrated against them to determine if there is a
significant correlation. Via a generalized Kepler's third law relating orbital
frequency to the binary separation distance, these templates are based on
energy balance: the rate of energy loss of these binaries is equal to the power
in the gravitational radiation emitted. Both the notion of energy and
expressions for the flux of gravitational radiation require the knowledge of
the dynamics of these binaries, which in turn is encapsulated in their
effective lagrangian. Due to the high accuracy to be attained, this effective
lagrangian needs to be computed up to 3 PN and higher.\footnote{Blanchet
\cite{Blanchet2BodyLADM} offers a review of the post-Newtonian framework and
its relation to gravitational wave experimental observables.}

Currently, the dynamics of compact astrophysical binaries is known up to 3.5
PN.\footnote{The half integer PN order lagrangians, scaling as odd powers of
$v$ relative to Newtonian gravity, describe dissipation -- gravitational waves
produced by and interacting with the $n$ compact objects. In this present
paper, we shall focus only on the conservative part of their dynamics up to 2
PN.} (See \S1.3 of Blanchet \cite{Blanchet2BodyLADM} and the references
therein.) To obtain the dynamics at 4 PN and beyond is a challenging task.
Because of the need to regularize the divergences that arise from approximating
compact objects as point particles, one may wish to engage field theoretic
methods to handle them. Such a pursuit was initiated in \cite{nrgr}, where it
was shown how to carry out the field theory effective lagrangian calculation in
a systematic manner by first doing some dimensional analysis. One of the main
thrusts of this present work is to attempt to make as methodical as possible
such a route in post-Newtonian calculations. In particular, we advocate using
the computer to automate the process, so that at the end only those Feynman
diagrams that truly require human intervention are left for manual evaluation.
Given the computational effort required at 4 PN and beyond, we believe this is
necessary not only to save time and energy, but also to reduce human errors.
For example, at such a high PN order, even the derivation of the necessary
diagrams will itself be non-trivial -- the reader not convinced of this fact is
encouraged to look at appendix \ref{3PNDiagramsSection} containing the 3 PN
diagrams -- but an efficient implementation of the algorithm that we will
sketch in the main body of this work will allow automatic generation of Feynman
diagrams to arbitrary PN order, modulo computing power.

\textsl{Solar System Gravity} \quad Closer to Earth, the
Einstein-Infeld-Hoffman lagrangian, at $\mathcal{O}[v^2]$ beyond Newton, is
routinely used to compute the solar system ephemerides, and to analyze
spacecraft trajectories and space based gravitational experiments. A range of
experiments, such as the new lunar ranging observatory APOLLO, proposals to
land laser ranging missions on Mars and/or Mercury, and spacecraft laboratories
-- GTDM, LATOR, BEACON, etc. -- will begin to probe the non-Euclidean nature of
the solar system's spacetime geometry beyond 1 PN by measuring the timing and
deflection of light propagation more precisely than before. (See, for instance,
Turyshev \cite{TuryshevTestGR} for a recent review.)

Within the point particle approximation, both the solar system dynamics and its
geometry can be gotten at simultaneously by computing from general relativity
the effective $n$-body lagrangian. Because general relativity is a non-linear
field theory, knowledge of the 2 body lagrangian is not sufficient to deduce
its $n$-body counterpart, as superposition is not obeyed. That the $n$-body
$L_{\text{eff}}$ encodes not only dynamics $\{\vec{x}_a[t]\}$ but also the
geometry $g_{\mu\nu}[t,\vec{x}]$ can be seen by adding a test particle to the
$n$-body system.\footnote{This observation can be found, for example, in Damour
and Esposito-Farese \cite{Damour_Esposito-Farese_ReadgOffLeff}.} Denoting the
latter's mass and coordinate vector as $M_\epsilon$ and $y^\mu \equiv (t,
\vec{y})^\mu$ respectively, in the limit as $M_\epsilon$ tends to zero relative
to the rest of the other masses in the system, we know its exact action has to
be\footnote{We work in cartesian coordinates and employ the $\eta_{\mu\nu} =
\text{diag}[1,-1,\dots,-1]$ sign convention. The Einstein summation convention
is adopted. Greek letters run from 0 to $d-1$ while English alphabets run from
$1$ to $d-1$.}

{\allowdisplaybreaks
\begin{align*}
- M_\epsilon & \int \dx{t} \sqrt{ \bar{g}_{\mu\nu} \frac{\text{d} y^\mu}{\dx{t}} \frac{\text{d} y^\nu}{\dx{t}} } \\
&= - M_\epsilon \int \dx{t} \bigg( 1 - \frac{1}{2} \left( \frac{\dd \vec{y}}{\dx{t}} \right)^2 + \frac{1}{2} \delta g_{00}[z] \nonumber \\
&\qquad + \delta g_{0i}[z] \frac{\text{d} y^i}{\dx{t}} + \frac{1}{2} \delta g_{ij}[z] \frac{\text{d} y^i}{\dx{t}} \frac{\text{d} y^j}{\dx{t}} + \dots \bigg), \\
\bar{g}_{\mu\nu} &\equiv \eta_{\mu\nu} + \delta g_{\mu\nu} \\
z &\equiv \{\vec{x}_a,\vec{v}_a,\dot{\vec{v}}_a,\dots\},t,\vec{y}; \quad a =
1,2,\dots,n,
\end{align*}}

since it now moves along a geodesic on the spacetime metric generated by the
rest of the $n$ masses. Therefore, if $L_{n+1}$ is the $(n+1)$-body lagrangian
less the $M_\epsilon(-1 + (1/2) \left( \dd \vec{y}/\dx{t} \right)^2)$, the
deviation of the spacetime metric from Minkowski $\delta g_{\mu\nu}$ can be
read off the action of the test particle using the prescription:

{\allowdisplaybreaks
\begin{align*}
&\delta g_{00}[t, \vec{x}] = -\left. 2 \frac{\partial}{\partial M_\epsilon} L_{n+1}[\vec{y} = \vec{x}] \right\vert_{\dot{\vec{y}} = \ddot{\vec{y}} = \dots = M_\epsilon = 0} \\
&\delta g_{0i}[t, \vec{x}] = -\left. \frac{\partial}{\partial M_\epsilon} \frac{\partial L_{n+1}}{\partial (\dd y^i/\dx{t})} [\vec{y} = \vec{x}] \right\vert_{\dot{\vec{y}} = \ddot{\vec{y}} = \dots = M_\epsilon = 0} \\
&\delta g_{ij}[t, \vec{x}] \\
&= -\left. \frac{\partial}{\partial M_\epsilon} \frac{\partial^2
L_{n+1}}{\partial (\dd y^i/\dx{t}) \partial (\dd y^j/\dx{t})}[\vec{y} =
\vec{x}]\right\vert_{\dot{\vec{y}} = \ddot{\vec{y}} = \dots = M_\epsilon = 0}
\end{align*}}

We see that understanding and testing the dynamics -- the equations that govern
the time evolution of the $\{\vec{x}_a\}$ -- is intimately tied to
understanding and testing the spacetime geometry $g_{\mu\nu}$ of the solar
system.

The outline of the paper is as follows. In section \ref{setup}, we set up a
lagrangian description of the system of $n$ compact astrophysical objects by
approximating them as $n$ point particles. Einstein's equations can then be
solved perturbatively as a Born series, whose graphical representation are the
Feynman diagrams containing no graviton loops; the result of summing the
diagrams yield the $n$-body effective action we seek. (Our description will be
brief because it will merely be an overview of the methods developed in
\cite{nrgr}.) We then sketch the algorithm that could be used to generate the
necessary Feynman diagrams contributing to the effective action up to an
arbitrary PN order. In section \ref{diagramresults}, we calculate the
individual diagrams that occur at the Newtonian thru 2 PN order and present the
effective action of the $n$-body system up to certain integrals (\ref{0PNLeff},
\ref{1PNLeff}, \ref{2PNnBody}). As a by-product, we reproduce the known 2 PN 2
body lagrangian. In the appendixes, we discuss the integrals encountered in the
diagrams; the algorithm for generating the $N \geq 2$ graviton Feynman rules on
a computer; and also display the Feynman diagrams occurring at the 3 PN order.

\section{The $n$-body system}
\label{setup}

The assumption that we have a system of $n$ compact objects, with their typical
size $r_{\rm s}$ much smaller than their typical separation distance $r$, i.e.
$r_s \ll r$, suggests that the detailed structure of these objects ought not
affect their gravitational dynamics, at least to leading order. These $n$
objects could then be viewed as point particles. Because the most general
action for a point particle must be some scalar functional of its $d$-velocity
$u_a^\mu \equiv \dx{x}_a^\mu/\dx{s}_a$ and geometric tensors\footnote{The
conventions for the Christoffel symbols $\Gamma^\mu_{\phantom{\mu}
\alpha\beta}$, Riemann tensor $R_{\mu\nu \alpha\beta} \equiv g_{\mu\lambda}
R^\lambda_{\phantom{\lambda}\nu \alpha\beta}$, Ricci tensor $R_{\mu\nu}$ and
Ricci scalar $\mathcal{R}$ can be inferred from the formulae in appendix
\ref{ngravitonformulas}.} (and possibly the electromagnetic tensor
$F_{\mu\nu}$, if large scale magnetic fields are present) integrated over the
world line of the said particle; part of the action is already fixed to be of
the form

{\allowdisplaybreaks
\begin{align}
\label{pointparticleactionform} S_{\text{pp}} &= -\sum_{a = 1}^n M_a \int
\dx{s}_a \bigg( 1 + c^{(a)}_4 R_{\mu\nu
\alpha\beta} R^{\mu\nu \alpha\beta} \nonumber \\
&\qquad + c^{(a)}_6 R_{\mu\nu\alpha\beta}
R^{\mu\phantom{\sigma}\alpha}_{\phantom{\mu}\sigma\phantom{\alpha}\rho} u_a^\nu
u_a^\beta u_a^\rho u_a^\sigma \nonumber \\
&\qquad + c^{(a)}_{\rm FR} F_{\mu\alpha} F_{\nu\beta} R^{\mu\alpha \nu\beta} +
\dots \bigg),
\end{align}}

where $\dx{s}_a$ is the infinitesimal proper time of the $a$th point particle
and the ``$\dots$" means one really has an infinite number of terms to
consider, since the only constraints at this point are that each of them is a
coordinate scalar and that none of them can be removed by a re-definition of
either the metric $g_{\mu\nu}$ or the photon field $A_\mu$.

However, as argued in \cite{nrgr}, unless the $n$ objects are very large or
have very large dipole and higher mass moments, it is expected that the minimal
terms $\{-M_a \int \dx{s}_a\}$ would suffice up to 4 PN order (see also \S1.2
of Blanchet \cite{Blanchet2BodyLADM} for a discussion), and in what follows we
will compute with them only. (We will also ignore electromagnetic
interactions.) Here, the $\{M_a\}$ lend themselves to a natural interpretation
as the masses of the astrophysical objects and $-M_a \int \dx{s}_a$ describes a
structure-less, mathematical point particle. At the 5 PN order and beyond, one
would be compelled to include as many of the non-minimal terms as is required
to maintain the consistency of the field theory up to a given level of
accuracy. Physically, this means one has to begin accounting for the fact that,
even if one neglects their rotation, astrophysical objects are not really point
particles and their individual mass distributions and sub-structures do produce
gravitational effects. To give the coefficients $\{c^{(a)}_{\text{X}}\}$ of
these non-minimal terms physical meaning, one would have to compute
(in-principle) measure-able quantities both in the actual physical setup and
with the point particle terms in (\ref{pointparticleactionform}). The
$\{c^{(a)}_{\text{X}}\}$ are then fixed by requiring the results of the latter
match the former: for instance, if we have multiple non-rotating black holes
bound by their mutual gravity, then one could calculate the partial wave
amplitudes of gravitational waves scattering off the Schwarzschild metric and
match the point particle computation onto it by tuning the coefficients $\{M_a,
c^{(a)}_{\text{X}}\}$ appropriately.

Up to 2 PN, the gravitational dynamics of the $n$-body system, with $x_a^\mu$
denoting the $\mu$th component of the coordinate vector of the $a$th point
particle, is therefore encoded in the action $\mathcal{S}$, where

{\allowdisplaybreaks
\begin{align}
\label{fullaction} \mathcal{S} &= S_{\text{GR}} + S_{\text{pp}} \\
\label{vtx} S_{\text{GR}} &= -2 \mpl^{d-2} \int_{\mathbb{R}^d} \text{d}^d x \sqrt{|g|} \mathcal{R} \\
\label{wl} S_{\text{pp}} &= - \sum_{1 \leq a \leq n} M_a \int \dx{t}_a \sqrt{
g_{\mu \nu} v_a^{\mu} v_a^{\nu} } \\
v_a^\mu &\equiv \frac{\text{d} x_a^\mu}{\dx{t}_a}[t_a] \nonumber \\
\mpl &\equiv (32 \pi G_{\rm N})^{-1/2} \nonumber
\end{align}}

Moreover, we expect the metric of spacetime to depart markedly from Minkowski
only close to one of these $n$ compact objects, where it is irrelevant for the
problem at hand, and thus we can expand the metric about
$\eta_{\mu\nu}$:\footnote{The factor of $\mpl^{1-(d/2)}$ ensures that the
graviton kinetic term does not contain $\mpl$. Also, for the rest of this
paper, we will raise and lower indices with $\eta_{\mu\nu}$.}

\begin{align*}
g_{\mu\nu} &= \eta_{\mu\nu} + \frac{h_{\mu\nu}}{\mpl^{(d/2)-1}}.
\end{align*}

The general relativistic effective lagrangian $L_{\text{eff}}[\{\vec{x}_a,
\vec{v}_a, \dot{\vec{v}}_a, \dots\}]$ for $n$ objects can now be computed via
the prescription usually associated with perturbative quantum field theory,
namely, as the sum of fully connected diagrams:

{\allowdisplaybreaks
\begin{align}
\label{effectiveaction} \exp\left[ i \int \dx{t} \ L_{\text{eff}} \right]
&= \left(\prod_{\mu \leq \nu = 0}^{d-1} \int \mathcal{D}h_{\mu\nu} \exp\left[ i \mathcal{S} + i S_{\text{gf}} \right]\right)_{\text{cl}} \\
\label{fullyconnected} &= \exp\left[ \sum \binom{\text{Fully connected}}{\text{diagrams}} \right] \\
S_{\text{gf}} &= \int \text{d}^d x \ \eta^{\alpha \beta} \left(
\partial^{\lambda} h_{\lambda \alpha} - \frac{1}{2} \partial_\alpha
h \right) \nonumber \\
\label{gaugefixing} & \quad \times \left( \partial^{\lambda} h_{\lambda \beta}
- \frac{1}{2} \partial_\beta h \right), \\
h &\equiv h^{\lambda}_{\phantom{\alpha} \lambda} \nonumber
\end{align}}

or alternatively,

{\allowdisplaybreaks
\begin{align}
&\exp\left[ i \int \dx{t} \ L_{\text{eff}} \right] \nonumber \\
\label{pathintegral} &= \mathcal{N} \left. \exp \left[ i S_{\text{I}} \left[\frac{1}{i} \frac{\delta}{\delta J^{\mu \nu}} \right] \right] \right\vert_{J=0} \nonumber \\
&\quad \times \exp \bigg[ -\frac{1}{2} \intx{d}{x} \intx{d}{y} \ J^{\alpha
\beta}_x D^{\text{F}}_{\alpha \beta ; \lambda \tau}[x-y] J^{\lambda \tau}_y
\bigg],
\end{align}}

with $D^{\text{F}}_{\alpha \beta ; \lambda \tau}[x-y]$ being the Feynman
graviton Green's function and $S_{\text{I}}[(1/i) \delta/\delta J^{\mu\nu}]$
indicates we are replacing every graviton field in (\ref{fullaction}), less the
graviton kinetic term, with the corresponding functional derivative with
respect to $J_{\mu \nu}$ with the same indices. In particular, because the
graviton field is symmetric in its indices, we have

\begin{align*}
\frac{\delta J_{\mu \nu}[y]}{\delta J_{\alpha \beta}[x]} = \frac{1}{2} \left(
\delta^{\alpha}_{\phantom{\alpha}\mu} \delta^{\beta}_{\phantom{\beta}\nu} +
\delta^{\alpha}_{\phantom{\alpha}\nu} \delta^{\beta}_{\phantom{\beta}\mu}
\right) \delta^d[x-y].
\end{align*}

The expression in (\ref{pathintegral}) is the functional integral version of
the statement that, up to an (for current purposes) irrelevant factor
$\mathcal{N}$, to compute the effective action, one needs to expand $\exp[i
S_{\text{I}}]$ and, for each term in the series, consider all possible Wick
contractions between the graviton fields. In the next subsection, we will use
it as a guide to devise an algorithm for generating the necessary Feynman
diagrams at a given PN order.

A gauge fixing term $S_{\text{gf}}$ (\ref{gaugefixing}) has been added to make
invertible the graviton kinetic term in the Einstein-Hilbert action
(\ref{vtx}), whose explicit form then reads

{\allowdisplaybreaks
\begin{align}
\label{gravitonkinetic_time} &S_{\text{GR}}[h^2] + S_{\text{gf}}[h^2] \nonumber \\
&= \int \text{d}^dx \bigg( \frac{1}{2} \partial_0 h^{\beta\nu} \partial_0
h_{\beta\nu} - \frac{1}{4}
\partial_0 h \partial_0 h \\
\label{gravitonkinetic_space} &\qquad + \frac{1}{2} \partial^i h^{\beta\nu}
\partial_i h_{\beta\nu} - \frac{1}{4} \partial^i h \partial_i h \bigg).
\end{align}}

This choice corresponds to the linearized de Donder gauge $\eta^{\mu \nu}
\Gamma^{(1)}_{\alpha \mu \nu} = \partial^\mu h_{\mu\alpha} - \frac{1}{2}
\partial_\alpha h^\mu_{\phantom{a}\mu} = 0$.

The subscript ``cl" (short for ``classical") in (\ref{effectiveaction})
indicates the Feynman diagrams with graviton loops are excluded. As already
remarked, evaluating these classical Feynman graphs amounts to solving
Einstein's equations for $h_{\mu\nu}$ via an iterative Born series expansion.

\subsection{Physical Scales In The $n$-Body Problem}

It is possible to begin computing the diagrams in (\ref{fullyconnected}) after
only expanding (\ref{fullaction}) in powers of graviton fields, performing a
non-relativistic expansion afterwards and keeping the terms needed up to a
given PN order. However, we will now show that it is more efficient if one also
expands the action (\ref{fullaction}) in terms of the number of time
derivatives and powers of velocities $\{v_a\}$ they contain, before any
diagrams are drawn and calculated, as this will allow one to keep only the
necessary terms in (\ref{fullaction}) such that every Feynman diagram generated
from them scales exactly as $v^{2Q}$, for a given PN order $Q \in
\mathbb{Z}^+$.

To this end, we note that, because we are assuming that the $n$ objects are
moving non-relativistically, with their typical speed $v \ll 1$, we already
know that the lowest order effective action must give us Newtonian gravity:

{\allowdisplaybreaks
\begin{align*}
S_{\text{eff}} &= \int \dx{t} \bigg( \sum_{1 \leq a \leq n} \frac{1}{2} M_a \vec{v}_a^2 \nonumber \\
& \quad + \frac{1}{2} \sum_{1 \leq a, b \leq n} \kappa \frac{G_{\rm
N}^{(d/2)-1} M_a M_b}{R_{ab}^{d-3}} + \dots \bigg), \\
R_{ab} &\equiv |\vec{x}_a - \vec{x}_b|
\end{align*}}

with $\kappa$ being some (presently unimportant) dimensionless number. This
prompts us to associate with this lowest order action $\mathcal{S}_c$ whenever
this particular product of masses $M$, separation distances $r$ and time occur
in the action:

\begin{align}
\label{scale1} \mathcal{S}_c \sim \int \dx{t} \ M v^2 \sim \int \dx{t} \
\mpl^{2-d} M^2 r^{3-d}
\end{align}

We may then obtain from (\ref{scale1})

\begin{align}
\label{scale2} \mathcal{S}_c \sim M v r, \quad
\frac{G_\text{N}^{(d/2)-1}M}{r^{d-3}} \sim v^2,
\end{align}

where the first relation holds because the only physical length and time scales
in the problem for a fixed coordinate frame are the typical separation distance
$r$ and orbital period $r/v$. Similarly, we relate all time and space
derivatives and integrals to appropriate powers of $r$ and frequency $v/r$,

{\allowdisplaybreaks
\begin{align}
\label{scale3} \int \text{d}^d x &\sim r^d v^{-1} \\
\label{scale4} \frac{\text{d}}{\dx{x}^0} &\sim \delta[x^0 - x'^0] \sim
\frac{v}{r},
\end{align}}

where the $\delta[x^0 - x'^0]$ relation will be needed shortly.

Next, we observe that the real part of the Feynman graviton Green's function
obtained from inverting (\ref{gravitonkinetic_time}) and
(\ref{gravitonkinetic_space}) can be expressed as an infinite series in time
derivatives:\footnote{\label{momentumspacefootnote}The momentum space
representation in the following expressions -- which is related to its position
space counterpart via (\ref{integral_PropI}) -- will be useful when contracting
graviton vertices coming from the cubic and higher in $h_{\mu\nu}$ terms in the
Einstein-Hilbert action (\ref{vtx}), because manipulation of spatial
derivatives on $h_{\mu\nu}$ become algebraic manipulation of momentum dot
products in the numerator.}

{\allowdisplaybreaks
\begin{align}
\label{gravitonpropagator} \text{Re} & \left\langle 0 \left\vert \text{T}
\left\{ h_{\mu \nu}[x^0, \vec{x}] h_{\alpha \beta}[x'^0,\vec{x}'] \right\}
\right\vert 0 \right\rangle \\ &= - \frac{i P_{\mu \nu ; \alpha \beta}}{2}
\sum_{m = 0}^\infty \frac{\Gamma[\frac{d-3-2m}{2}]}{4^{1+m} \pi^{\frac{d-1}{2}}
\Gamma[1+m]} \nonumber \\ &\quad \times \frac{1}{|\vec{x} - \vec{x}'|^{d-3-2m}}
\left( \frac{\text{d}}{\dx{x}'^0} \frac{\text{d}}{\dx{x}^0} \right)^m
\delta[x^0 - x'^0] \nonumber \\ &= - \frac{i P_{\mu \nu ; \alpha \beta}}{2}
\sum_{m = 0}^\infty \intk{d-1}{p} \frac{e^{i \vec{p} \cdot (\vec{x} -
\vec{x}')}}{[\vec{p}^2]^{1+m}} \nonumber \\ &\quad \times \left(
\frac{\text{d}}{\dx{x}'^0} \frac{\text{d}}{\dx{x}^0} \right)^m \delta[x^0 -
x'^0] \nonumber \\ P_{\mu \nu; \alpha \beta} &\equiv \eta_{\alpha \mu}
\eta_{\beta \nu} + \eta_{\alpha \nu} \eta_{\beta \mu} - \frac{2}{d-2} \eta_{\mu
\nu} \eta_{\alpha \beta}, \nonumber
\end{align}}

where our notation alludes to the fact that the classical Feynman graviton
Green's function is the non-interacting-vacuum expectation value of the time
ordered product of two graviton fields.

That there are only even number of time derivatives reflects the relationship
that the real part of the Feynman Green's function for a massless graviton is
equal to half its retarded plus half its advanced counterpart; see Poisson
\cite{PoissonGreensFunction} for a discussion. The introduction of an
additional background field $\overline{h}_{\mu\nu}$ in \cite{nrgr} takes into
account the imaginary part of $D^{\text{F}}_{\mu\nu; \alpha\beta} [x-x']$,
which describes the dissipative part of the dynamics -- the interaction of
gravitational waves produced by and interacting with the $n$ point masses. In
this paper, we are focusing only on the conservative part of the dynamics, and
hence will ignore Im $D^{\text{F}}_{\mu\nu; \alpha\beta} [x-x']$.

The zeroth order term in Re $D^{\text{F}}_{\mu\nu; \alpha\beta} [x-x']$
(\ref{gravitonpropagator}), with no time derivatives, can be obtained by
inverting (\ref{gravitonkinetic_space}), i.e. the graviton kinetic term with
only spatial derivatives. Diagrams involving the higher order terms in
(\ref{gravitonpropagator}), with time derivatives, can be gotten by treating
(\ref{gravitonkinetic_time}), the graviton kinetic term with only time
derivatives, as a perturbation. For instance, the first correction to the
Newtonian gravitational potential due to the finite speed of graviton
propagation is proportional to $\int \dx{t} \dx{t}_a \dx{t}_b \delta[t - t_a]
\delta[t - t_b] (\text{d}/\dx{t}_a) (\text{d}/\dx{t}_b) |\vec{x}_a[t_a] -
\vec{x}_b[t_b]|^{5-d}$, which could also be viewed as a contraction between two
distinct world line operators of the form $-\mpl^{1-(d/2)} (M_a/2) \int
\dx{t}_a h_{00}[x_a]$, from (\ref{wl}), with one insertion of
(\ref{gravitonkinetic_time}).

Keeping in mind that each diagram is built out of contracting graviton fields
$\langle h_{\mu\nu}[x] h_{\alpha\beta}[x'] \rangle = D^{\text{F}}_{\mu\nu;
\alpha \beta}[x-x']$ from distinct terms in (\ref{fullaction}), this implies
every graviton field in (\ref{fullaction}) should be assigned a scale that is
square root that of the lowest order non-relativistic Green's function
containing no time derivatives. The $|\vec{x} - \vec{x}'|^{3-d} \sim r^{3-d}$
dependence implies that spatial derivatives on $h_{\mu\nu}$ ought to scale with
one less one power of $r$ of the same. By recalling (\ref{scale4}), we then
have

{\allowdisplaybreaks
\begin{align}
\label{scale5} h_{\mu\nu}[x^0, \vec{x}] &\sim r^{1-d/2} v^{1/2} \\
\label{scale6} \partial_i h_{\mu\nu}[x^0, \vec{x}] &\sim r^{-d/2} v^{1/2} \\
\label{scale7} \partial_0 h_{\mu\nu}[x^0, \vec{x}] &\sim r^{-d/2} v^{3/2}
\end{align}}

Putting the scaling relations from (\ref{scale1}), (\ref{scale2}),
(\ref{scale3}), (\ref{scale4}), (\ref{scale5}), (\ref{scale6}) and
(\ref{scale7}) into the action (\ref{fullaction}), we then see that, upon
expanding (\ref{fullaction}) in powers of graviton fields, velocities
$\vec{v}_a$, and the number of time derivatives on $h_{\mu\nu}$, each term in
the action now scales homogeneously with $\mathcal{S}_c$ and $v$:

{\allowdisplaybreaks
\begin{align}
\int \dx{t}_a \mathcal{O}^{(a)}_w[n,\sigma, \epsilon_\Sigma] \vec{v}_a^{2 \sigma} &\sim \mathcal{S}_c^{1-\frac{n}{2}} v^{2 n - 2 + 2 \sigma + \epsilon_\Sigma} \nonumber \\
\label{scalingrelations} \int \text{d}^dx \ \mathcal{O}_v[ m, \psi] &\sim
\mathcal{S}_c^{1-\frac{m}{2}} v^{2 m - 4 + \psi}
\end{align}}

where $\mathcal{O}^{(a)}_w[n,\sigma, \epsilon_\Sigma]$ denotes the world line
term in (\ref{wl}) associated with the $a$th particle containing exactly $n$
graviton fields, less the $\vec{v}_a^{2 \sigma}$ term from $\eta_{\mu\nu}
v_a^\mu v_a^\nu$, with a total of $\epsilon_\Sigma$ spatial indices (for
example, a term with $h_{ij} v^i v^j h_{0k} v^k$ has $\epsilon_\Sigma = 3$).
Note that there is usually more than one term for a given $\epsilon_\Sigma$, so
one has to sum over all possible terms. $\mathcal{O}_v[ m, \psi]$ denotes the
term in (\ref{vtx}) containing exactly $m$ graviton fields ($m \geq 2$), with
precisely $\psi$ time derivatives ($\psi = 0, 1$, or $2$).

Given these results in (\ref{scalingrelations}), and given $n_{(v)}$ number of
graviton vertices from (\ref{vtx}), $n_{(w)}$ number of world line operators
from (\ref{wl}), and $N$ total number of graviton fields (so that $N/2$ is
really the number of Green's functions in the diagram) one can work out that
every Feynman diagram in the theory arising from products of these operators
must scale as

\begin{align}
\label{diagramscaling} \mathcal{S}_c^{n_{(v)} + n_{(w)} - \frac{N}{2}} v^{2(
n_{(w)} - 2 + \lambda_{\Sigma}/2)},
\end{align}

so that such a diagram contributes to the $(n_{(w)} - 2 + \lambda_{\Sigma}/2)$
PN effective action; and $n_{(v)} + n_{(w)} - \frac{N}{2} = 1$, as the
non-relativistic expansion, for the conservative part of the dynamics, is a
series of the schematic form $S_{\text{eff}} = S_0 + S_2 v^2 + S_4 v^4 +
\dots$, where each term of the effective action has to contain the appropriate
products of masses, velocities and time integrals such that $S_n \sim
\mathcal{S}_c$, with the $v^n$ factored out. Here, $\lambda_{\Sigma}$ is a
positive integer that is the result of summing powers of speeds coming from
time derivatives, velocities contracted with graviton fields (such as $h_{ij}
v^i v^j$), number of graviton kinetic terms with time derivatives
(\ref{gravitonkinetic_time}) inserted, and the factors of $\vec{v}_a^2$ arising
from the term  $\eta_{\mu \nu} v_a^{\mu} v_a^{\nu}$ inside the proper time
$\dx{s}_a$. We have used, in deriving the exponent of $v$, the constraint that
$n_{(v)} + n_{(w)} - \frac{N}{2} = 1$. The fact that no diagram can scale
greater than the first power of $\mathcal{S}_c$ has been proven in \cite{nrgr}.
Observe that, with only the minimal $-\sum_a M_a \int\dx{s}_a$ terms included,
these scaling relations are independent of the number of spacetime dimensions.

\subsection{Algorithm}

We are now in a position to describe an algorithm that could, with an efficient
implementation and sufficient computing resources, generate the necessary
Feynman diagrams, for a given subset of world line terms in
(\ref{pointparticleactionform}), up to an arbitrary order in the
non-relativistic PN expansion.

For a desired scaling (\ref{diagramscaling}), corresponding to a specific PN
order, one can insert in (\ref{fullaction}) explicit factors of $\mathcal{S}_c$
and $v$ according to the results in (\ref{scalingrelations}), so that one may
employ \mma\footnote{We frame this discussion around \mma, but this algorithm
can most likely be implemented with any software with similar symbolic
differentiation and combinatorial capabilities.} \cite{mathematica} to extract
the relevant products of operators, i.e. pre-contraction, in the taylor series
expansion of the exponential in the path integral (\ref{pathintegral}) using
either its {\sf Series} or {\sf Coefficient} command. Observe that every fully
connected diagram $\mathcal{D}_\text{F}$ constructed out of each term in the
taylor series expansion of $\exp[i S_{\text{I}}]$ in (\ref{pathintegral}) is
the first non-trivial term of the series expansion of
$\exp[\mathcal{D}_\text{F}]$. Hence the gravitational effective action is the
sum of all fully connected diagrams constructed from the terms in the series
expansion of $\exp[iS_{\text{I}}]$. Moreover, one does not need to include in
the code the explicit form of the graviton fields, velocities, etc., but it
suffices to have placeholders, such as the ones used in
(\ref{scalingrelations}), containing enough information to re-construct at the
end the relevant types of graviton fields considered ($h_{00}, h_{0i}$ or
$h_{ij}$), which point mass the field(s) belongs to, factors of velocities,
number of time derivatives in the $m$-graviton term(s), numerical constants
from taylor expanding the exponential and the square root in the infinitesimal
proper time, and so on.

Next, the combinatorics of contraction can be handled by \mma, by assigning to
every graviton field in a given product a distinct number, so that such a
product corresponds to some list, say $\{1,2,\dots,s\}$ for a product of $s$
graviton fields. A permutation of $\{1,\dots,s\}$ is equivalent to a Feynman
diagram if and only if it leaves no numbers fixed and the resulting permutation
operation $\widehat{\pi}$ can be factored into products of disjoint 2-cycles,
i.e. $\widehat{\pi} = (a_1 \ a_2)(a_3 \ a_4)\dots(a_{s-1} \ a_s)$ with
$\{a_1,a_2,\dots,a_{s-1},a_s\}$ being a re-arrangement of the original set
$\{1,\dots,s\}$. For instance, the set $\{3,4,1,2\}$ means one would have to
contract graviton field ``1" with graviton field ``3"; graviton field ``2" with
graviton field ``4". (The requirement that each fully connected diagram scales
as $\mathcal{S}_c^1$ ensures there will be no quantum corrections.) What
remains is removing those diagrams that are not fully connected. One method of
achieving this is to check if the permuted set $\{a_1, a_2, \dots, a_s\}$ can
be factorized into two or more disjoint sets, where each of these individual
sets contain only terms that are contracted amongst themselves. The term
$\int\dx{t}_a \int\dx{t}_b \langle h_{00}[x_a] h_{00}[x_b] \rangle \int\dx{t}_c
\int\dx{t}_e \langle h_{00}[x_c] h_{00}[x_e] \rangle$, for example, can be
represented as $\{2,1,4,3\} = \{2,1\} \{4,3\}$, i.e. factorize-able; whereas
$\int\dx{t}_a \int\dx{t}_b \int\dx{t}_c \langle h_{00}[x_a] h_{00}[x_b] \rangle
\langle h_{00}[x_a] h_{00}[x_c] \rangle$ is equivalent to $\{3,4,1,2\}$ and not
factorize-able. Such a prescription can be implemented with a suitable
adaptation of the command {\sf Permutations}.

Once the contractions are determined for a given product of terms from
expanding $\exp[i S_{\text{I}}]$, if the particular diagram does not involve
terms from the Einstein-Hilbert action, it can be computed automatically
because it would be built out of products of the graviton Green's function. The
scalar portion is $|\vec{x}_a - \vec{x}_b|^{2s + 3 - d}$, where $s$ is the
number of (\ref{gravitonkinetic_time}) inserted; while there will also be
factors of velocities proportional to $i P_{\mu\nu;\alpha\beta} v_a^\mu v_a^\nu
v_b^\alpha v_b^\beta$ from (\ref{gravitonpropagator}). When insertions of
(\ref{gravitonkinetic_time}) are present, one would have to take the
appropriate time derivatives afterwards. (Some care needs to be exercised in
keeping track of the time $\delta$-functions when doing so -- see
(\ref{2PN_Term_96_eqn}) for an example.) For diagrams with graviton vertices,
although they may not be calculated automatically, the required Wick
contractions and permutations of particle labels can be displayed so that the
user does not have to figure out the combinatorics manually, but rather focus
only on the tensor contractions of the graviton Feynman rules, manipulation of
the momentum dot products and the ensuing Feynman integrals. Furthermore, some
of the higher PN diagrams involving graviton vertices will be products of lower
PN graviton vertex diagrams with other expressions that can also be
automatically calculated -- such as factors of $\vec{v}_a^2$, the graviton
Green's function with or without insertions of (\ref{gravitonkinetic_time})
contracted into velocities, $\langle h_{\mu\nu}[x_a] h_{\alpha\beta}[x_b]
\rangle v_a^\mu v_a^\nu v_b^\alpha v_b^\beta$, etc. (See, for instance,
Fig.(\ref{2PN_Term_65}$\vert$a) and Fig.(\ref{2PN_Term_126}$\vert$c).) This
implies the evaluation of such higher PN diagrams with vertices may most likely
be automated if a repository of these lower PN graviton vertex diagrams is
kept.

As an illustration of the utility of such an algorithmic approach, we have
generated in appendix \ref{3PNDiagramsSection} the 3 PN Feynman diagrams for
the minimal point particle action $-\sum_a M_a \int \dx{s}_a$. We also maintain
a web page at \cite{PNCode}, where the \mma \ code used in this paper can be
found.

\subsection{$n$-body diagrams and superposition}

Now suppose one wants to calculate the lagrangian for $n$ point particles up to
the $Q$th post-Newtonian order. Then the exponent of $v$ in
(\ref{diagramscaling}) tells us that the maximum number of distinct particles
that can appear in a given Feynman diagram is

\begin{align*}
\text{max}[n_{(w)}] &= Q + 2
\end{align*}

and so at a fixed post-Newtonian order $Q$, obtaining the Feynman diagrams for
the $(Q+2)$-body problem is sufficient for obtaining the lagrangian for the
arbitrary $n$-body problem. In particular, at the 2 PN order, we see that the
$n$-body problem is equivalent to the 4-body problem. For a general $Q$ PN
order, the diagrams for $n > Q+2$ point particles can be obtained by summing
the diagrams for the $n = Q+2$ case over all the particles in the system, since
no additional distinct diagrams are needed. For $n < Q+2$ point particles, the
relevant diagrams can be gotten from the $n = Q+2$ diagrams by setting the
masses $M_{Q+2}, M_{Q+1}, \dots, M_{n+1}$ to zero. Even with the non-minimal
terms beyond the $-\sum_a M_a\int\dx{s}_a$ included, it is apparent that
superposition will continue to hold at any given PN order once we have computed
the effective lagrangian for a sufficient number of distinct point particles,
since each Feynman diagram can only contain a finite number of world line
operators.

\section{Results}
\label{diagramresults}

We now present the diagram-by-diagram results for the computation of the
effective action up to 2 PN. Because the calculation is long and saturated with
technicalities, the reader only interested in the final results may simply
refer to (\ref{0PNLeff}) for $\mathcal{O}[v^0]$, (\ref{1PNLeff}) for
$\mathcal{O}[v^2]$, and (\ref{2PNnBody}, \ref{2PN2Body}, \ref{2PN3Body},
\ref{2PN4Body}) for the $\mathcal{O}[v^4]$ effective lagrangians.

All Feynman diagram integrals are evaluated within the framework of dimensional
regularization, where the number of space-time dimensions, $d = m - 2
\varepsilon$, is some infinitesimal deviation from a positive integer: i.e.,
$m\in\mathbb{Z}^+$ and $\varepsilon = 0^+$. For some of the more difficult
integrals encountered at the 2 PN level, we will restrict our interest to that
of the physically relevant case when $d = 4 - 2 \varepsilon$. Within
dimensional regularization, integrals such as $\int \text{d}^{d-1}p \
(\vec{p}^{2})^{-\sigma}$, $\int \text{d}^{d-1}p \ p^i (\vec{p}^{2})^{-\sigma}$
and $\int \text{d}^{d-1}p \ p^i p^j (\vec{p}^{2})^{-\sigma}$ are set to zero.
This can be justified formally by setting to zero the appropriate $\sigma$,
$\rho$, or $\tau$ exponent of (\ref{integral_OneLoopRd}),
(\ref{integral_OneLoopRvmu}), (\ref{integral_OneLoopRvmunu}) or
(\ref{integral_OneLoopRvmunuAB}), since $\Gamma[z]$ diverges as $z \to 0$.

Because it is easier to manipulate momentum dot products than derivatives, both
the Feynman rules for the graviton vertices are derived and the tensor
contractions of graviton vertices are performed in fourier space. (See appendix
\ref{ngravitonformulas} for an algorithm that could generate the $N$-graviton
Feynman rule for $N \geq 2$.) We will thus present the master integrals for
each diagram first in momentum space.

\textsl{Notation} \quad A few words about the notation used: the time argument
of the $a$th particle is $t_a$, so that $\vec{x}_a = \vec{x}_a[t_a]$. However,
if the spatial coordinate vectors $\{ \vec{x}_a | a=1,2,\dots,n \}$ and their
time derivatives occur within a single time integral $\int \dx{t}$, then it is
implied that they all share the same time argument $t$. $\vec{R}_{ab} \equiv
\vec{x}_a - \vec{x}_b$ and its Euclidean length is $R_{ab} \equiv |\vec{x}_a -
\vec{x}_b| = (-\eta_{ij} (x_a^i - x_b^i) (x_a^j - x_b^j))^{1/2}$. The partial
derivative $\partial_i^a \equiv \partial/\partial x^i_a$ refers to the
derivative with respect to the $i$th component of the spatial coordinate vector
of the $a$th particle. The spatial velocity of the $a$th particle is $\vec{v}_a
= \vec{v}_a[t_a] \equiv \dx{\vec{x}}_a/\dx{t}_a \equiv \dot{\vec{x}}_a$, and
its acceleration is $\dot{\vec{v}}_a = \dot{\vec{v}}_a[t_a] \equiv \text{d}^2
\vec{x}_a/\dx{t}_a^2 \equiv \ddot{\vec{x}}_a$. Whenever we compute in fourier
space, the relevant sign and $\pi$ conventions are encoded in the following
definition: $f[x] \equiv (2\pi)^{-d} \int \text{d}^d p \ \tilde{f}[p] \exp[ip_0
x^0] \exp[-i \vec{p} \cdot \vec{x}]$, where $f$ is some arbitrary function, and
$x$ and $p$ are its coordinate and momentum space arguments respectively.

\textsl{Feynman diagrams} \quad A blob with some letter ``$a$" at its center
represent a world line operator from (\ref{wl}) belonging to the $a$th
particle, with the indices of its various graviton fields $h_{\mu\nu} \in
\{h_{00}, h_{0i}, h_{ij}\}$ indicated on the side. $\{a,b,c,e\}$ are distinct
labels. A line represents the lowest order graviton Green's function with no
time derivatives. The $\times$ on a line represent an insertion of
(\ref{gravitonkinetic_time}). A black dot with $k$ lines attached to it is the
$k$-graviton piece of (\ref{vtx}) with zero time derivatives. The $k$-graviton
piece of (\ref{vtx}) with 1 or 2 time derivatives will be indicated with a ``1"
or ``2" respectively; see for example Fig.(\ref{2PN_Term_70}$\vert$a) and
Fig.(\ref{2PN_Term_106}$\vert$a). The $v^k$ appearing alongside the graviton
indices of the $a$th world line operator indicates which power of $\vec{v}_a^2$
from expanding $-M_a \int \dx{x}^0_a ( 1 - \vec{v}_a^2 + \dots)^{1/2}$ needs to
be included. Every Feynman diagram displayed serves dual purposes: it
represents the class of diagrams that can be obtained from it by permuting
particle labels; but the result of the diagram shown in the body of the text is
always for the specific choice of labels in the figure. (The exceptions are the
diagrams where 2 3-graviton vertices are contracted:
Fig.(\ref{2PN_Term_146}$\vert$c), Fig.(\ref{2PN_Term_148}$\vert$d),
Fig.(\ref{2PN_Term_149}$\vert$d) and Fig.(\ref{2PN_Term_152}$\vert$c). We will
discuss the notations there.)

\subsection{0 PN}

\begin{figure}
\begin{center}
\includegraphics[width=1.5in]{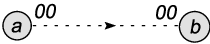}
\end{center}
\caption{Newtonian gravity.} \label{NewtonianGravity}
\end{figure}

At the lowest order, we have Newtonian gravity coming from the single diagram
in Fig.(\ref{NewtonianGravity}) and the usual kinetic energy.

\begin{align}
\label{0PNLeff} L^{(0 \text{ PN})}_{\text{eff}} &= \sum_{1 \leq a \leq n} \frac{1}{2} M_a \vec{v}_a^2 \nonumber \\
&\ + \frac{1}{2} \sum_{\substack{1 \leq a,b \leq n \\ a \neq b}}
\frac{2^{\frac{5d}{2}-8} \Gamma[\frac{d-1}{2}] }{ \pi^{1/2} (d-2) } \frac{
G_{\rm N}^{\frac{d-2}{2} } M_a M_b }{ R_{ab}^{d-3} }
\end{align}

\subsection{1 PN}

At 1 PN order, we have 2- and 3-body diagrams. Since the lagrangian at this
order has been computed numerous times in the literature, we will merely
present the results and not discuss any of the calculation in detail.

\subsubsection{2 body diagrams}

\begin{figure}
\begin{center}
\includegraphics[width=1.5in]{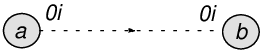} \\
\text{(a) No permutations necessary.} \\
\includegraphics[width=1.5in]{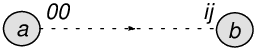} \\
\text{(b) 2 permutations from swapping $a \leftrightarrow b$.} \\
\includegraphics[width=1.5in]{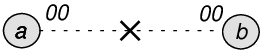} \\
\text{(c) No permutations necessary.} \\
\includegraphics[width=1.5in]{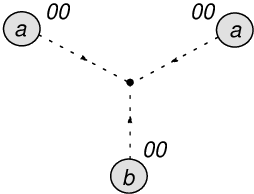} \\
\text{(d) 2 permutations from swapping $a \leftrightarrow b$.} \\
\includegraphics[width=2.5in]{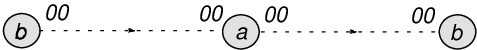} \\
\text{(e) 2 permutations from swapping $a \leftrightarrow b$.} \\
\includegraphics[width=1.5in]{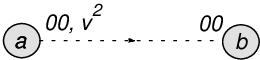} \\
\text{(f) 2 permutations from swapping $a \leftrightarrow b$.}
\end{center}
\caption{1 PN two body diagrams} \label{1PN2Body}
\end{figure}

The 2 body diagrams are displayed in Fig.(\ref{1PN2Body}).

{\allowdisplaybreaks
\begin{align*}
\text{Fig.} (\ref{1PN2Body}\vert\text{a}) &= -\intdt \frac{\Gamma\left[
\frac{d-3}{2} \right]}{8 \pi^{\frac{d-1}{2}}} \frac{M_a M_b}{\mpl^{d-2}
R_{ab}^{d-3}} \vec{v}_a \cdot \vec{v}_b \\
\text{Fig.} (\ref{1PN2Body}\vert\text{b}) &= \intdt \frac{\Gamma\left[
\frac{d-3}{2} \right]}{16 (d-2) \pi^{\frac{d-1}{2}}} \frac{M_a M_b}{\mpl^{d-2}
R_{ab}^{d-3}} \vec{v}_b^2 \\
\text{Fig.} (\ref{1PN2Body}\vert\text{c}) &= \intdt \frac{(5-d)(d-3)
\Gamma\left[ \frac{d-5}{2} \right]}{64 (d-2) \pi^{\frac{d-1}{2}}} \frac{M_a
M_b}{\mpl^{d-2}
R_{ab}^{d-1}} \\
&\quad \times \left( (3-d) \vec{v}_a \cdot \vec{R}_{ab} \vec{v}_b \cdot
\vec{R}_{ba} - \vec{v}_a \cdot \vec{v}_b R_{ab}^2 \right) \\
\text{Fig.} (\ref{1PN2Body}\vert\text{d}) &= -\intdt \frac{(d-3)^2 \Gamma^2
\left[ \frac{d-3}{2} \right]}{256 (d-2)^2 \pi^{d-1}} \frac{M_a^2
M_b}{\mpl^{2(d-2)}
R_{ab}^{2(d-3)}} \\
\text{Fig.} (\ref{1PN2Body}\vert\text{e}) &= \intdt \frac{\Gamma^2 \left[
\frac{d-1}{2} \right]}{ 128 (d-2)^2 \pi^{d-1} } \frac{M_a M_b^2}{\mpl^{2(d-2)}
R_{ab}^{2(d-3)}} \\
\text{Fig.} (\ref{1PN2Body}\vert\text{f}) &= \intdt \frac{\Gamma\left[
\frac{d-1}{2} \right]}{16 (d-2) \pi^{\frac{d-1}{2}}} \frac{M_a M_b}{\mpl^{d-2}
R_{ab}^{d-3}} \vec{v}_a^2
\end{align*}}

\subsubsection{3 body diagrams}

\begin{figure}
\begin{center}
\includegraphics[width=1.5in]{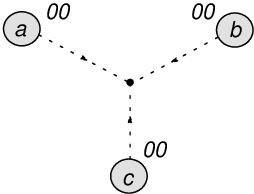} \\
\text{(a) No permutations necessary.} \\
\includegraphics[width=2.5in]{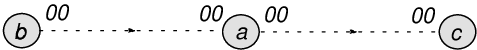} \\
\text{(b) 3 permutations: $a$, $b$, or $c$ for the middle label.}
\end{center}
\caption{1 PN three body diagrams} \label{1PN3Body}
\end{figure}

The 3 body diagrams are found in Fig.(\ref{1PN3Body}).

{\allowdisplaybreaks
\begin{align*}
\text{Fig.} (\ref{1PN3Body}\vert\text{a}) &= -\intdt \frac{(d-3)^2
\Gamma^2[\frac{d-3}{2}]}{128 (d-2)^2 \pi^{d-1}} \frac{M_a M_b M_c}{
\mpl^{2(d-2)} } \\
&\quad \times \left( R_{ab}^{3-d} R_{ac}^{3-d} + R_{bc}^{3-d} R_{ac}^{3-d} +
R_{ab}^{3-d} R_{bc}^{3-d} \right) \\
\text{Fig.} (\ref{1PN3Body}\vert\text{b}) &= \intdt
\frac{\Gamma^2[\frac{d-1}{2}]}{64 (d-2)^2 \pi^{d-1}} \frac{M_a M_b M_c}{
\mpl^{2(d-2)} R_{ab}^{d-3} R_{ac}^{d-3}}
\end{align*}}

For later use, we note that the master integral for the 3-graviton diagram is

{\allowdisplaybreaks
\begin{align}
\label{masterintegralI000000} &I_{000000}[q,r,s] \nonumber \\
&\equiv -\frac{i}{\mpl^{(d-2)/2}} \left(\frac{d-3}{d-2}\right)^2 \nonumber \\
&\times \left( \prod_{\ell = 1}^3 \int \frac{\dd^{d-1}p_\ell}{(2\pi)^\ell} \right) \frac{\vec{p}_1^2 + \vec{p}_2^2 + \vec{p}_3^2}{\vec{p}_1^2 \vec{p}_2^2 \vec{p}_3^2} \nonumber \\
&\times \exp\left[ i \vec{p}_1 \cdot \vec{x}_q[t_q] + i \vec{p}_2 \cdot \vec{x}_r[t_r] + i \vec{p}_3 \cdot \vec{x}_s[t_s] \right] \nonumber \\
&\times (2\pi)^{d-1} \delta^{(d-1)}\left[ \vec{p}_1 + \vec{p}_2 + \vec{p}_3
\right]
\end{align}}

\subsubsection{$\mathcal{O}[v^2]$ Effective Lagrangian}

Summing the first order relativistic correction to kinetic energy from the
$\eta_{\mu\nu}v^\mu v^\nu$ in the infinitesimal proper time $\dx{s}$ and the
diagrams from Fig.(\ref{1PN2Body}) and Fig.(\ref{1PN3Body}) hands us the 1 PN
order, $d \geq 4$ dimensional $n$-body effective lagrangian:

{\allowdisplaybreaks
\begin{align}
\label{1PNLeff} &L^{(1 \text{ PN})}_{\text{eff}} \nonumber \\
&= \sum_{a=1}^n \frac{1}{8} M_a \vec{v}_a^4 \nonumber \\
&+ \frac{1}{2} \sum_{\substack{1 \leq a, b \leq n \\ a \neq b}}
\frac{2^{\frac{5(d-4)}{2}} \Gamma[\frac{d-3}{2}] }{ (d-2) \pi^{1/2} }
\frac{G_{\rm N}^{\frac{d}{2}-1} M_a M_b}{R_{ab}^{d-3}} \nonumber \\
&\quad \times \bigg( (d-3)^2 \frac{\vec{R}_{ab} \cdot \vec{v}_a \vec{R}_{ba}
\cdot \vec{v}_b}{R_{ab}^2} \nonumber \\
&\qquad + (d-1) \left( \vec{v}_a^2 + \vec{v}_b^2 \right) -
(3d-5) \vec{v}_a \cdot \vec{v}_b \bigg) \nonumber \\
&- \frac{1}{2} \sum_{\substack{1 \leq a, b \leq n \\ a \neq b}} \frac{2^{5d -
17} \Gamma^2 \left[ \frac{d-1}{2} \right]}{(d-2)^2 \pi}
\frac{G_{\text{N}}^{d-2} M_a M_b(M_a + M_b)}{R_{ab}^{2(d-3)}} \nonumber \\
&- \frac{1}{3!} \sum_{\substack{1 \leq a, b, c \leq n \\ a, b, c \text{
distinct}}} \frac{2^{5d - 16} \Gamma^2 \left[ \frac{d-1}{2} \right]}{(d-2)^2
\pi} G_{\text{N}}^{d-2} M_a M_b M_c \nonumber \\
&\quad \times \left( R_{ab}^{3-d} R_{ac}^{3-d} + R_{ba}^{3-d} R_{bc}^{3-d} +
R_{ca}^{3-d} R_{cb}^{3-d} \right) \nonumber \\
&\vec{R}_{ab} \equiv \vec{x}_a - \vec{x}_b, \ R_{ab} \equiv |\vec{R}_{ab}|
\end{align}}

Setting $d=4$ recovers the known result in the literature; for instance,
equation (38c) of Damour and Sch\"{a}fer \cite{DamourSchaeferNBody}. The $d
\geq 4$, 2 body version of (\ref{1PNLeff}) has been computed by Cardoso et. al.
\cite{Cardoso_ddimensions}.

\subsection{2 PN}

At 2 PN, we have 2-, 3- and 4-body diagrams. We shall classify the diagrams
according to whether they involve terms from the Einstein-Hilbert action, i.e.
diagrams with or without graviton vertices. Whenever there are time derivatives
acting on $\delta$-functions, for example $(\dd/\dx{t}_a) \delta[t_a - t]$, it
is implied that integration by parts is to be carried out. To save space, we
will not display the explicit result of differentiation.

\subsubsection{2 body diagrams}


\textsl{No graviton vertices} \quad The diagrams that do not involve graviton
vertices are:

{\allowdisplaybreaks
\begin{figure}
\begin{center}
\includegraphics[width=1.5in]{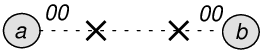} \\
\text{(a) No permutations necessary.} \\
\includegraphics[width=1.5in]{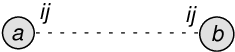} \\
\text{(b) No permutations necessary.} \\
\includegraphics[width=1.5in]{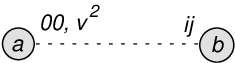} \\
\text{(c) 2 permutations from swapping $a \leftrightarrow b$.} \\
\includegraphics[width=1.5in]{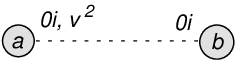} \\
\text{(d) 2 permutations from swapping $a \leftrightarrow b$.} \\
\includegraphics[width=1.5in]{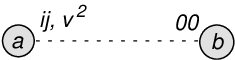} \\
\text{(e) 2 permutations from swapping $a \leftrightarrow b$.} \\
\includegraphics[width=1.5in]{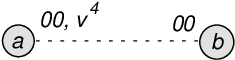} \\
\text{(f) 2 permutations from swapping $a \leftrightarrow b$.} \\
\includegraphics[width=1.5in]{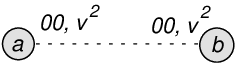} \\
\text{(g) No permutations necessary.} \\
\includegraphics[width=1.5in]{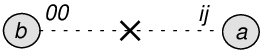} \\
\text{(h) 2 permutations from swapping $a \leftrightarrow b$.} \\
\includegraphics[width=1.5in]{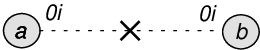} \\
\text{(i) No permutations necessary.} \\
\includegraphics[width=1.5in]{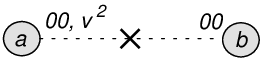} \\
\text{(j) 2 permutations from swapping $a \leftrightarrow b$.} \\
\caption{2 PN two body diagrams with no graviton vertices: 1 of 2}
\label{2PN_Term_95} \label{2PN_Term_1} \label{2PN_Term_2} \label{2PN_Term_4}
\label{2PN_Term_6} \label{2PN_Term_7} \label{2PN_Term_10} \label{2PN_Term_11}
\label{2PN_Term_13} \label{2PN_Term_14}
\end{center}
\end{figure}}

{\allowdisplaybreaks
\begin{figure}
\begin{center}
\includegraphics[width=2.5in]{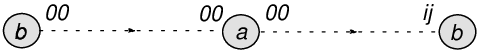} \\
\text{(a) 2 permutations from swapping $a \leftrightarrow b$.} \\
\includegraphics[width=2.5in]{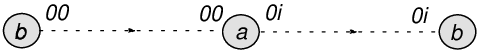} \\
\text{(b) 2 permutations from swapping $a \leftrightarrow b$.} \\
\includegraphics[width=2.5in]{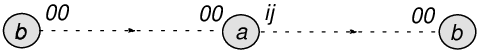} \\
\text{(c) 2 permutations from swapping $a \leftrightarrow b$.} \\
\includegraphics[width=2.5in]{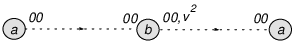} \\
\text{(d) 2 permutations from swapping $a \leftrightarrow b$.} \\
\includegraphics[width=2.5in]{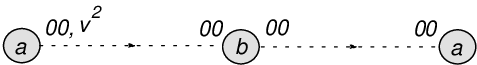} \\
\text{(e) 2 permutations from swapping $a \leftrightarrow b$.} \\
\includegraphics[width=2.5in]{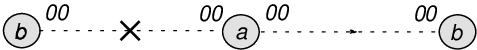} \\
\text{(f) 2 permutations from swapping $a \leftrightarrow b$.} \\
\includegraphics[width=1.5in]{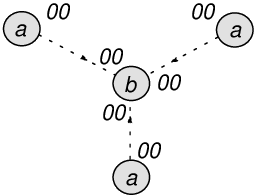} \\
\text{(g) 2 permutations from swapping $a \leftrightarrow b$.} \\
\includegraphics[width=2.5in]{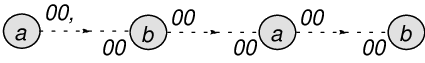} \\
\text{(h) No permutations necessary.} \\
\caption{2 PN two body diagrams with no graviton vertices: 2 of 2}
\label{2PN_Term_16} \label{2PN_Term_24} \label{2PN_Term_46} \label{2PN_Term_47}
\label{2PN_Term_36} \label{2PN_Term_96} \label{2PN_Term_93} \label{2PN_Term_78}
\end{center}
\end{figure}}

{\allowdisplaybreaks
\begin{align*}
\text{Fig.(\ref{2PN_Term_95}$\vert$a)} &= \intdt \int \dx{t}_a \int \dx{t}_b \frac{(d-3) \Gamma\left[ \frac{d-7}{2} \right]}{512 \pi^{\frac{d-1}{2}} (d-2)} \nonumber \\
&\times \frac{M_a M_b}{ \mpl^{d-2} R_{ab}^{d-7} } \left(
\frac{\text{d}}{\dx{t}_a} \frac{\text{d}}{\dx{t}_b} \right)^2 \delta[t-t_a]
\delta[t-t_b] \nonumber \\
\text{Fig.(\ref{2PN_Term_1}$\vert$b)} &= \intdt \frac{\Gamma\left[
\frac{d-3}{2} \right]}{ 16 \pi^{\frac{d-1}{2}} } \frac{M_a
M_b}{\mpl^{d-2} R_{ab}^{d-3}} \nonumber \\
&\qquad \times \left( (\vec{v}_a \cdot \vec{v}_b)^2-\frac{\vec{v}_a^2
\vec{v}_b^2}{d-2} \right) \nonumber \\
\text{Fig.(\ref{2PN_Term_2}$\vert$c)} &= \intdt
\frac{\Gamma\left[\frac{d-3}{2}\right]}{ 32 \pi^{\frac{d-1}{2}} (d-2) }
\frac{M_a M_b}{\mpl^{d-2} R_{ab}^{d-3}} \vec{v}_a^2 \vec{v}_b^2 \nonumber \\
\text{Fig.(\ref{2PN_Term_4}$\vert$d)} &= -\intdt
\frac{\Gamma\left[\frac{d-3}{2}\right]}{16 \pi^{\frac{d-1}{2}}} \frac{M_a M_b
}{\mpl^{d-2} R_{ab}^{d-3}} \vec{v}_a \cdot \vec{v}_b \ \vec{v}_a^2 \nonumber \\
\text{Fig.(\ref{2PN_Term_6}$\vert$e)} &= \intdt
\frac{\Gamma\left[\frac{d-3}{2}\right]}{32 (d-2) \pi^{\frac{d-1}{2}}} \frac{M_a
M_b}{\mpl^{d-2} R_{ab}^{d-3}} \vec{v}_a^4 \nonumber \\
\text{Fig.(\ref{2PN_Term_7}$\vert$f)} &= \intdt \frac{3
\Gamma\left[\frac{d-1}{2}\right]}{64 (d-2)\pi^{\frac{d-1}{2}}} \frac{M_a
M_b}{\mpl^{d-2} R_{ab}^{d-3}} \vec{v}_a^4 \nonumber \\
\text{Fig.(\ref{2PN_Term_10}$\vert$g)} &= \intdt
\frac{\Gamma\left[\frac{d-1}{2}\right]}{32 (d-2) \pi^{\frac{d-1}{2}}} \frac{M_a
M_b}{\mpl^{d-2} R_{ab}^{d-3}} \vec{v}_a^2 \vec{v}_b^2 \nonumber \\
\text{Fig.(\ref{2PN_Term_11}$\vert$h)} &= \intdt \int \dx{t}_a \int \dx{t}_b
\frac{\Gamma\left[ \frac{d-5}{2} \right]}{ 64 (d-2)
\pi^{\frac{d-1}{2}} }\nonumber \\
&\qquad \times \frac{M_a M_b}{\mpl^{d-2} R_{ab}^{d-5}} \vec{v}_a^2  \Dd{t_a}
\Dd{t_b} \delta[t - t_a] \delta[t - t_b] \nonumber \\
\text{Fig.(\ref{2PN_Term_13}$\vert$i)} &= -\intdt \int \dx{t}_a \int \dx{t}_b
\frac{\Gamma\left[ \frac{d-5}{2} \right]}{32
\pi^{\frac{d-1}{2}}} \nonumber \\
&\times \frac{M_a M_b}{\mpl^{d-2} R_{ab}^{d-5}} \vec{v}_a \cdot \vec{v}_b
\Dd{t_a} \Dd{t_b} \delta[t - t_a] \delta[t - t_b] \nonumber \\
\text{Fig.(\ref{2PN_Term_14}$\vert$j)} &= \intdt \int \dx{t}_a \int \dx{t}_b \frac{(d-3) \Gamma\left[ \frac{d-5}{2} \right]}{ 128 (d-2) \pi^{\frac{d-1}{2}} } \nonumber \\
&\qquad \times \frac{M_a M_b}{ \mpl^{d-2} R_{ab}^{d-5} } \vec{v}_a^2 \Dd{t_a}
\Dd{t_b}\delta[t-t_a] \delta[t-t_b] \nonumber \\
\text{Fig.(\ref{2PN_Term_16}$\vert$a)} &= \intdt \frac{(d-3)
\Gamma\left[\frac{d-3}{2}\right]^2}{256 (d-2)^2 \pi^{d-1}}
\frac{M_a M_b^2}{\mpl^{2 (d-2)} R_{ab}^{2(d-3)} } \vec{v}_b^2 \nonumber \\
\text{Fig.(\ref{2PN_Term_24}$\vert$b)} &= -\intdt \frac{(d-3)
\Gamma\left[\frac{d-3}{2}\right]^2}{128 (d-2) \pi^{d-1}} \nonumber \\
&\quad \times \frac{M_a M_b^2}{\mpl^{2(d-2)} R_{ab}^{2(d-3)}} \vec{v}_a \cdot \vec{v}_b \nonumber \\
\text{Fig.(\ref{2PN_Term_46}$\vert$c)} &= \intdt \frac{(d-3)
\Gamma\left[\frac{d-3}{2}\right]^2}{256 (d-2)^2 \pi^{d-1}}
\frac{M_a M_b^2}{\mpl^{2(d-2)} R_{ab}^{2(d-3)} } \vec{v}_a^2 \nonumber \\
\text{Fig.(\ref{2PN_Term_47}$\vert$d)} &= \intdt \frac{3
\Gamma\left[\frac{d-1}{2}\right]^2}{256 (d-2)^2 \pi^{d-1}} \frac{M_a^2
M_b}{ \mpl^{2(d-2)} R_{ab}^{2(d-3)} } \vec{v}_b^2 \nonumber \\
\text{Fig.(\ref{2PN_Term_36}$\vert$e)} &= \intdt
\frac{\Gamma\left[\frac{d-1}{2}\right]^2}{128 (d-2)^2 \pi^{d-1}} \frac{M_a^2
M_b}{ \mpl^{2(d-2)} R_{ab}^{2(d-3)} } \vec{v}_a^2
\end{align*}}

Fig.(\ref{2PN_Term_96}$\vert$f) contains a first order relativistic correction
to the graviton Green's function. Integrating over the time $\delta$-function
with no time derivatives acting on it, before integrating by parts, the
resulting integral becomes

{\allowdisplaybreaks
\begin{align}
\label{2PN_Term_96_eqn} &\text{Fig.(\ref{2PN_Term_96}$\vert$f)} \nonumber \\
&= \frac{i (d-3) \Gamma\left[\frac{d-5}{2}\right] \Gamma\left[\frac{d-1}{2}\right] M_a M_b^2}{512 (d-2)^2 \mpl^{2(d-2)} \pi^{d-1}} \int\dx{t}_a \int\dx{t}_b \delta[t_a-t_b] \nonumber \\
&\times \bigg( \frac{\dd |\vec{x}_b[t_a] - \vec{x}_a[t_a]|^{3-d}}{\dx{t}_a} \frac{\dd |\vec{x}_b[t_b] - \vec{x}_a[t_a]|^{5-d}}{\dx{t}_b} \nonumber \\
&\qquad + |\vec{x}_b[t_a] - \vec{x}_a[t_a]|^{3-d} \frac{\dd^2 |\vec{x}_b[t_b] -
\vec{x}_a[t_a]|^{5-d}}{\dx{t}_a \dx{t}_b} \bigg),
\end{align}}

with a common time argument $t_a$ for both $\vec{x}_a$ and $\vec{x}_b$ in the
factor $|\dots|^{3-d}$.

{\allowdisplaybreaks
\begin{align*}
\text{Fig.(\ref{2PN_Term_93}$\vert$g)} &= \intdt
\frac{\Gamma\left[\frac{d-1}{2}\right]^3}{ 1024 (d-2)^3 \pi^{\frac{3}{2} (d-1)}
} \frac{M_a^3 M_b}{ \mpl^{3(d-2)} R_{ab}^{3(d-3)} } \nonumber \\
\text{Fig.(\ref{2PN_Term_78}$\vert$h)} &= \intdt
\frac{\Gamma\left[\frac{d-1}{2}\right]^3}{512 (d-2)^3 \pi^{ \frac{3}{2}(d-1) }}
\frac{M_a^2 M_b^2}{ \mpl^{3(d-2)} R_{ab}^{3(d-3)} }
\end{align*}}

\begin{figure}
\begin{center}
\includegraphics[width=1.5in]{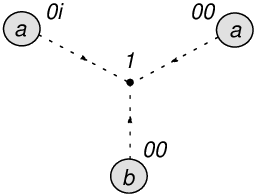} \\
\text{(a) 2 permutations from swapping $a \leftrightarrow b$.} \\
\includegraphics[width=1.5in]{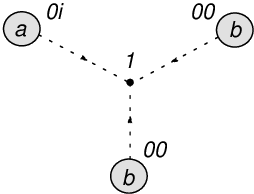} \\
\text{(b) 2 permutations from swapping $a \leftrightarrow b$.} \\
\includegraphics[width=1.5in]{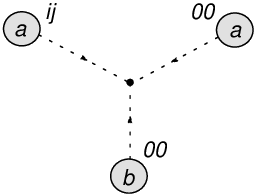} \\
\text{(c) 2 permutations from swapping $a \leftrightarrow b$.} \\
\includegraphics[width=1.5in]{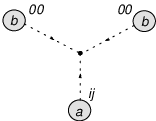} \\
\text{(d) 2 permutations from swapping $a \leftrightarrow b$.} \\
\caption{2 PN 2 body diagrams with graviton vertices: 1 of 4}
\label{2PN_Term_70} \label{2PN_Term_104} \label{2PN_Term_55}
\label{2PN_Term_98}
\end{center}
\end{figure}

\begin{figure}
\begin{center}
\includegraphics[width=1.5in]{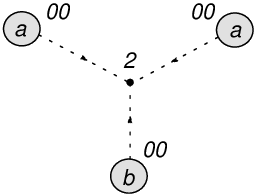} \\
\text{(a) 2 permutations from swapping $a \leftrightarrow b$.} \\
\includegraphics[width=3in]{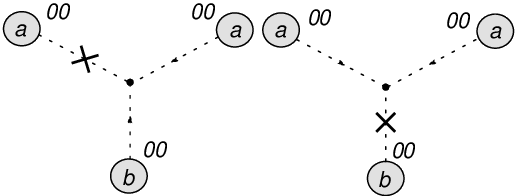} \\
\text{(b) 2 permutations from swapping $a \leftrightarrow b$.} \\
\includegraphics[width=1.5in]{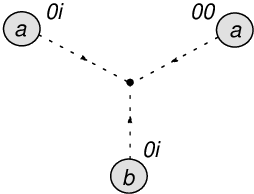} \\
\text{(c) 2 permutations from swapping $a \leftrightarrow b$.} \\
\includegraphics[width=1.5in]{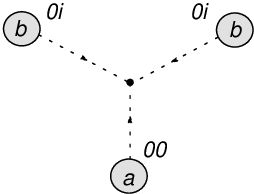} \\
\text{(d) 2 permutations from swapping $a \leftrightarrow b$.} \\
\caption{2 PN 2 body diagrams with graviton vertices: 2 of 4}
\label{2PN_Term_106} \label{2PN_Term_131} \label{2PN_Term_60}
\label{2PN_Term_100}
\end{center}
\end{figure}

\begin{figure}
\begin{center}
\includegraphics[width=1.5in]{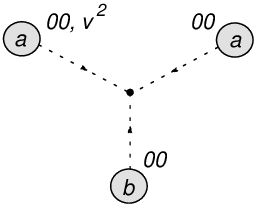} \\
\text{(a) 2 permutations from swapping $a \leftrightarrow b$.} \\
\includegraphics[width=1.5in]{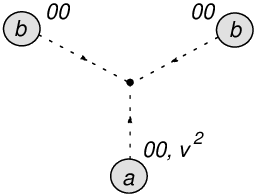} \\
\text{(b) 2 permutations from swapping $a \leftrightarrow b$.} \\
\includegraphics[width=1.5in]{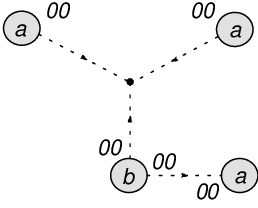} \\
\text{(c) 2 permutations from swapping $a \leftrightarrow b$.} \\
\includegraphics[width=1.5in]{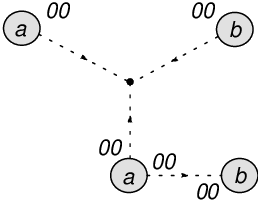} \\
\text{(d) 2 permutations from swapping $a \leftrightarrow b$.} \\
\caption{2 PN 2 body diagrams with graviton vertices: 3 of 4}
\label{2PN_Term_65} \label{2PN_Term_102} \label{2PN_Term_126}
\label{2PN_Term_134}
\end{center}
\end{figure}

\begin{figure}
\begin{center}
\includegraphics[width=1.0in]{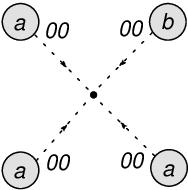} \\
\text{(a) 2 permutations from swapping $a \leftrightarrow b$.} \\
\includegraphics[width=1.0in]{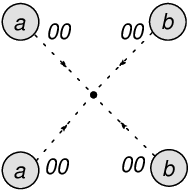} \\
\text{(b) No permutations necessary.} \\
\includegraphics[width=1.8in]{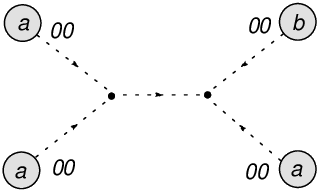} \\
\text{(c) 6 permutations. See text for discussion.} \\
\includegraphics[width=1.8in]{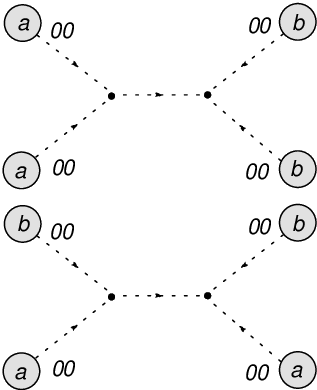} \\
\text{(d) 6 permutations. See text for discussion.} \\
\caption{2 PN 2 body diagrams with graviton vertices: 4 of 4}
\label{2PN_Term_128} \label{2PN_Term_130} \label{2PN_Term_146}
\label{2PN_Term_148}
\end{center}
\end{figure}


\textsl{Graviton vertices} \quad The rest of the 2 body diagrams contain terms
from the Einstein-Hilbert action.

Note that the form of the fourier space master integrals associated with each
class of diagrams usually comes about after some manipulation of momentum dot
products, application of the identity $2 \vec{p}_a \cdot \vec{p}_b = \mp
(\vec{p}_a \mp \vec{p}_b)^2 \pm \vec{p}_a^2 \pm \vec{p}_b^2$ and its analogs,
and the use of momentum conservation $\sum_{r=1}^\ell \vec{p}_r = 0$, $\ell =
3$ or $4$.

{\bf Fig.(\ref{2PN_Term_70}$\vert$a, b)} \quad The 2- and 3-body version of
Fig.(\ref{2PN_Term_70}$\vert$a, b) requires the following master integral:

{\allowdisplaybreaks
\begin{align}
\label{masterintegralI0i0000} &I_{0i0000}[q,r,s] \nonumber \\
&\equiv \intdt \int \dx{t}_q \int \dx{t}_r \int \dx{t}_s \frac{\delta[t-t_q]}{ 2(d-2) \mpl^{(d-2)/2} } \nonumber \\
&\times \left( \prod_{\ell=1}^3 \int \frac{\dd^{d-1}p_\ell}{(2\pi)^{d-1}} \right) \nonumber \\
&\times \vec{v}_a \cdot [(2(d-4) \vec{p}_2 + (d-5) \vec{p}_3) \nonumber \\
&\qquad \times \delta[t-t_s](i(\dd/\dx{t}_r)\delta[t-t_r]) \nonumber \\
&\qquad + ((d-5) \vec{p}_2 + 2(d-4) \vec{p}_3) \nonumber \\
&\qquad \times \delta[t-t_r](i(\dd/\dx{t}_s)\delta[t-t_s])] \nonumber \\
&\times \frac{\exp\left[ i \left( \vec{p}_1 \cdot \vec{x}_q[t_q] + \vec{p}_2
\cdot \vec{x}_r[t_r] + \vec{p}_3 \cdot \vec{x}_s[t_s] \right)
\right]}{\vec{p}_1^2 \vec{p}_2^2 \vec{p}_3^2} \nonumber \\
&\times (2\pi)^{d-1} \delta^{d-1}\left[ \vec{p}_1 + \vec{p}_2 + \vec{p}_3
\right]
\end{align}}

Notice that a $p_s^j$ (with $s=1,2$ or $3$) in the numerator can be obtained by
differentiating the appropriate exponential, i.e. $p_s^j \exp[i \vec{p}_s \cdot
\vec{x}_r] = -i (\partial/\partial x_r^j) \exp[i \vec{p}_s \cdot \vec{x}_r]$.
Our approach to the fourier integrals arising from this and the rest of the
diagrams with graviton vertices is to first substitute the momentum
$\delta$-function(s) with its (their) integral representation(s),

\begin{align}
\label{deltafunctionintegralrep} (2\pi)^{d-1} \delta^{d-1}\left[ \sum_{r=1}^s
\vec{p}_r \right] &= \int \dd^{d-1}z \exp\left[ - i \vec{z} \cdot \sum_{r=1}^s
\vec{p}_r \right],
\end{align}

and next use (\ref{integral_PropI}) to re-express the original momentum
integrals as position space ones, with the momentum dot products in the
numerator converted into derivatives on the resulting integrand.

In this regard, the 2-distinct particles case usually requires more care than
the 3- and 4-distinct particles cases. We shall illustrate this with
Fig.(\ref{2PN_Term_70}$\vert$a), where $\vec{x}_q = \vec{x}_r \equiv \vec{x}_a$
and $\vec{x}_s \equiv \vec{x}_b$. The time derivatives occurring in
Fig.(\ref{2PN_Term_70}$\vert$a) are

{\allowdisplaybreaks
\begin{align*}
\frac{\dd}{\dx{t}_a} &= v_a^j[t_a] \frac{\partial}{\partial x_a^j[t_a]}, \nonumber \\
\frac{\dd}{\dx{t}_b} &= v_a^j[t_b] \frac{\partial}{\partial x_a^j[t_b]}, \nonumber \\
\frac{\dd}{\dx{t}_c} &= v_b^j[t_c] \frac{\partial}{\partial x_b^j[t_c]},
\end{align*}}

with each partial derivative acting only on the appropriate $|\vec{x}_a -
\vec{z}|$ or $|\vec{x}_b - \vec{z}|$ with the same time argument as the
velocity vector contracted into it.

When $d=4 - 2 \varepsilon$, we therefore have

{\allowdisplaybreaks
\begin{align*}
&\text{Fig.(\ref{2PN_Term_70}$\vert$a)} \nonumber \\
&= \frac{M_a^2 M_b}{4 \mpl^{(3/2)(d-2)}} I_{0i0000}[a,a,b] \nonumber \\
&\stackrel{d=4}{=} \lim_{\varepsilon \to 0} \intdt \int \dx{t}_a \int \dx{t}_b \int \dx{t}_c \frac{M_a^2 M_b}{ 16 \mpl^4 } \nonumber \\
&\times \delta[t-t_a]\delta[t-t_b]\delta[t-t_c] \left( \frac{\Gamma[(1/2)-\varepsilon]}{4\pi^{(3-2\varepsilon)/2}} \right)^2 \int \dd^{3-2\varepsilon} z \nonumber \\
&\times v_a^i[t_a] \big( v_a^j[t_b] |\vec{x}_a[t_a] - \vec{z}|^{-1+2\varepsilon} \nonumber \\
&\qquad \times (\partial_j^a |\vec{x}_a[t_b] - \vec{z}|^{-1+2\varepsilon})(\partial_i^b |\vec{x}_b[t_c] - \vec{z}|^{-1+2\varepsilon}) \nonumber \\
&\ + v_b^j[t_c] |\vec{x}_a[t_a] - \vec{z}|^{-1+2\varepsilon} \nonumber \\
&\qquad \times (\partial_j^b |\vec{x}_b[t_c] -
\vec{z}|^{-1+2\varepsilon})(\partial_i^a |\vec{x}_a[t_b] -
\vec{z}|^{-1+2\varepsilon}) \big),
\end{align*}}

After differentiation, these integrals can then be evaluated using
(\ref{integral_OneLoopRvmunuAB}). This leads us to

{\allowdisplaybreaks
\begin{align*}
\text{Fig.(\ref{2PN_Term_70}$\vert$a)} &\stackrel{d=4}{=} \intdt \frac{M_a^2 M_b}{512 \mpl^4 \pi^2 R_{ab}^4} \nonumber \\
&\times ( -2 (\vec{R}_{ab} \cdot \vec{v}_a)^2 - 2 \vec{R}_{ab} \cdot \vec{v}_b \ \vec{R}_{ab} \cdot \vec{v}_a \nonumber \\
&\qquad + (\vec{v}_a^2 + \vec{v}_a \cdot \vec{v}_b) R_{ab}^2 )
\end{align*}}

A similar analysis for Fig.(\ref{2PN_Term_104}$\vert$b), making use of
(\ref{integral_OneLoopRd}) and (\ref{integral_OneLoopRvmunu}), gives

{\allowdisplaybreaks
\begin{align*}
\text{Fig.(\ref{2PN_Term_104}$\vert$b)} &= \frac{M_a M_b^2}{8 \mpl^{(3/2)(d-2)}} I_{0i0000}[a,b,b] \nonumber \\
&\stackrel{d=4}{=} \intdt \frac{M_a M_b^2}{1024 \mpl^4 \pi^2 R_{ab}^4} \nonumber \\
&\times \left(\vec{R}_{ba} \cdot \vec{v}_a \ \vec{R}_{ba} \cdot \vec{v}_b -
\vec{v}_a \cdot \vec{v}_b \ R_{ab}^2\right)
\end{align*}}

Before proceeding further, it is useful to introduce the following master
integral that would occur in several 2-, 3- and 4-body Feynman integrals:

{\allowdisplaybreaks
\begin{align}
\label{masterintegralI3} &I_3[a,b,c] \nonumber \\
&= \left( \prod_{s=1}^3 \intk{d-1}{p_s}
\right) (2\pi)^{d-1} \delta^{(d-1)}\left[ \vec{p}_1 + \vec{p}_2 + \vec{p}_3 \right] \nonumber \\
&\quad \times \frac{ \exp \left[ i \left( \vec{p}_1 \cdot \vec{x}_a + \vec{p}_2
\cdot \vec{x}_b + \vec{p}_3 \cdot \vec{x}_c \right) \right] }{ \vec{p}_1^2
\vec{p}_2^2 \vec{p}_3^2 } \nonumber \\
&= \left( \frac{\Gamma\left[ \frac{d-3}{2} \right]}{4 \pi^{\frac{d-1}{2}}}
\right)^3 \int \frac{{\rm d}^{d-1} z}{|\vec{z}-\vec{x}_a|^{d-3}
|\vec{z}-\vec{x}_b|^{d-3} |\vec{z}-\vec{x}_c|^{d-3}},
\end{align}}

where we have provided both its fourier and position space representations. In
appendix \ref{integralssection}, we obtain $I_3[a,b,c]$ in closed form when
$d=4-2\varepsilon$, up to $\mathcal{O}[\varepsilon^0]$
(\ref{3PointMasslessVertexIntegralLog}).

{\bf Fig.(\ref{2PN_Term_55}$\vert$c,d)} \quad We now turn to
Fig.(\ref{2PN_Term_55}$\vert$c) and Fig.(\ref{2PN_Term_55}$\vert$d). Its
associated master integral is

{\allowdisplaybreaks
\begin{align}
&\label{masterintegralIij0000} I_{ij0000}[q,r,s] \nonumber \\
&\equiv \intdt \int \dx{t}_q \int \dx{t}_r \int \dx{t}_s \frac{\delta[t-t_q] \delta[t-t_r] \delta[t-t_s]}{\mpl^{(d-2)/2}} \nonumber \\
&\times \left( \prod_{\ell=1}^3 \int \frac{\dd^{d-1}p_\ell}{(2\pi)^{d-1}} \right) \nonumber \\
&\times \bigg\{ \frac{\vec{v}_q^2}{(d-2)^2} \frac{\vec{p}_1^2 - (d-3)(\vec{p}_2^2 + \vec{p}_3^2)}{\vec{p}_1^2 \vec{p}_2^2 \vec{p}_3^2} \nonumber \\
&\qquad + \frac{d-3}{2(d-2)} \frac{(\vec{p}_1 \cdot \vec{v}_q)^2 + (\vec{p}_2 \cdot \vec{v}_q)^2 + (\vec{p}_3 \cdot \vec{v}_q)^2}{\vec{p}_1^2 \vec{p}_2^2 \vec{p}_3^2} \bigg\} \nonumber \\
&\times \exp\left[ i \left( \vec{p}_1 \cdot \vec{x}_q[t_q] + \vec{p}_2 \cdot
\vec{x}_r[t_r] + \vec{p}_3 \cdot \vec{x}_s[t_s] \right)
\right] \nonumber \\
&\times (2\pi)^{d-1} \delta^{d-1}\left[ \vec{p}_1 + \vec{p}_2 + \vec{p}_3
\right]
\end{align}}

The term proportional to $\vec{v}_q^2$ may be integrated in arbitrary
$d$-dimensions by integrating over the momentum that is absent in the
denominator (after cancelation), followed by an application of
(\ref{integral_PropI}), because it reduces to a product of the form $\prod_s
(2\pi)^{1-d} \int \dd^{d-1}p_s \exp[i\vec{p}_s \cdot
\vec{R}_{su}]/\vec{p}_s^2$. This type of fourier integral will occur
frequently.

The second term containing momenta dotted into velocities has the position
space representation

{\allowdisplaybreaks
\begin{align}
&\label{masterintegralIij0000_PartII}-i \frac{d-3}{2(d-2)\mpl^{(d-2)/2}} \int\dx{t} \left( \frac{\Gamma\left[ \frac{d-3}{2} \right]}{4\pi^{\frac{d-1}{2}}} \right)^3 \nonumber \\
&\times \int \dd^{d-1}z \ v_q^i v_q^j \big( (\partial_i^p \partial_j^p R_{pz}^{3-d}) R_{qz}^{3-d} R_{rz}^{3-d} \nonumber \\
&\qquad + R_{pz}^{3-d} (\partial_i^q \partial_j^q R_{qz}^{3-d}) R_{rz}^{3-d} \nonumber \\
&\qquad + R_{pz}^{3-d} R_{qz}^{3-d} (\partial_i^r \partial_j^r R_{rz}^{3-d})
\big)
\end{align}}

There is a subtlety when taking double spatial derivatives on a single factor
of the Euclidean distance raised to the $3-d$ power occurring within the
Feynman integrals, such as $\partial_i^a \partial_j^a R_{az}^{3-d}$. Strictly
speaking, because $R_{az}^{3-d}$ is the Green's function of the spatial
laplacian operator $\delta^{ij}\partial_i^a \partial_j^a$, one needs to employ
the formula

{\allowdisplaybreaks
\begin{align*}
&\frac{\partial}{\partial x^i} \frac{\partial}{\partial x^j} \frac{\Gamma\left[
\frac{d-3}{2} \right]}{4 \pi^{(d-1)/2} |\vec{x}|^{d-3}}
\nonumber \\
&= \frac{\Gamma\left[ \frac{d-1}{2} \right]}{2 \pi^{(d-1)/2}} \left( (d-1)
\frac{x^i x^j}{|\vec{x}|^{d+1}} - \frac{\delta^{ij}}{|\vec{x}|^{d-1}} \right) \nonumber \\
&\qquad \qquad - \frac{\delta^{ij}}{d-1} \delta^{d-1}\left[ \vec{x} \right],
\end{align*}}

where there is a $\delta$-function term in addition to those following from
straightforward differentiation so that one would obtain the correct result
upon taking the trace of both sides. Insofar as the Feynman integrals are
concerned, however, it appears the $\delta$-function term may be dropped as
long as proper regularization is used. For instance, if we try to compute the
integral

\begin{align*}
\int\dd^{D-1} z R_{bz}^{-\rho} \partial_i^a \partial_j^a R_{az}^{3-d}
\end{align*}

by first carrying out the differentiation (without including the
$\delta$-function term), we would obtain two terms, one with a $R_{az}^i
R_{az}^j$ in the integrand and the other with $\delta^{ij}$. If we simply set
$D = d$ and employ the formulae (\ref{integral_OneLoopRd}) and
(\ref{integral_OneLoopRvmunu}), each of the two terms will be ill-defined.
However, displacing $D = d + \kappa$ in (\ref{integral_OneLoopRd}) and
(\ref{integral_OneLoopRvmunu}), and performing a laurent expansion in $\kappa$
afterwards would yield a finite result that is not traceless and is furthermore
consistent with first doing the scalar integral (\ref{integral_OneLoopRd}) with
$\sigma_1 = \rho$ and $\sigma_2 = d-3$ and then carrying out the double
derivatives at the end.

If one is interested only in the higher PN 2-body problem, the existence of
such subtleties may be reason to stay within fourier space as far as possible
for the evaluation of Feynman integrals. For the 2 body case of the integrals
(\ref{masterintegralIij0000_PartII}) and analogous ones below, to avoid this
subtlety for the terms where there are 2 factors of $R_{az}^{3-d}$ (or 2
factors of $R_{bz}^{3-d}$) and the derivatives are acting on the $R_{bz}^{3-d}$
(or $R_{az}^{3-d}$), we shall first do the integral using $I_3$, before
differentiation.

Along this line, we further remark that

\begin{align*}
\int \dd^{d-1}z R_{az}^{2(3-d)} \partial_i^b \partial_j^b R_{bz}^{3-d}
\end{align*}

can be integrated and then differentiated, whereas

\begin{align*}
\int \dd^{d-1}z R_{az}^{3-d} R_{bz}^{3-d} \partial_i^b \partial_j^b
R_{bz}^{3-d}
\end{align*}

has to be differentiated first. One cannot begin with 3 distinct coordinate
vectors $\{\vec{x}_a,\vec{x}_b,\vec{x}_c\}$ in this integral, engage $I_3$, and
then set $\vec{x}_c = \vec{x}_b$: the last step involves terms like $\lim_{c
\to b} R_{bc}^i/R_{bc}$ and $\lim_{c \to b} R_{bc}^{-2}$ and hence is ill
defined.

We now employ (\ref{integral_PropI}), derivatives on $I_3$,
(\ref{integral_OneLoopRd}) and (\ref{integral_OneLoopRvmunu}) to deduce

{\allowdisplaybreaks
\begin{align*}
&\text{Fig.(\ref{2PN_Term_55}$\vert$c)} \nonumber \\
&= \frac{M_a^2 M_b}{8 \mpl^{(3/2)(d-2)}} I_{ij0000}[a,a,b] \nonumber \\
&\stackrel{d=4}{=} \intdt \frac{M_a^2 M_b}{1024 \mpl^4 \pi^2 R_{ab}^4} \nonumber \\
&\times \left(-3 (\vec{R}_{ab} \cdot \vec{v}_a)^2 - 4 (\vec{R}_{ba} \cdot
\vec{v}_a)^2 + \vec{v}_a^2 \ R_{ab}^2 \right),
\end{align*}}

and

{\allowdisplaybreaks
\begin{align*}
&\text{Fig.(\ref{2PN_Term_98}$\vert$d)} \nonumber \\
&= \frac{M_a M_b^2}{16 \mpl^{(3/2)(d-2)}} I_{ij0000}[a,b,b] \nonumber \\
&\stackrel{d=4}{=} \intdt \frac{M_a M_b^2}{2048 \mpl^4 \pi^2 R_{ab}^4} \nonumber \\
&\times \left(-4 (\vec{R}_{ab} \cdot \vec{v}_a)^2 - 3 (\vec{R}_{ba} \cdot
\vec{v}_a)^2 + 5 \vec{v}_a^2 \ R_{ab}^2 \right)
\end{align*}}

{\bf Fig.(\ref{2PN_Term_106}$\vert$a)} \quad It turns out
Fig.(\ref{2PN_Term_106}$\vert$a) and its 3-body counterpart are zero in $d = 4$
spacetime dimensions. Because it is not apparent, we display the results here
for the 3 body case. In terms of the master integral in
(\ref{masterintegralI3}), we have

{\allowdisplaybreaks
\begin{align}
\label{2PN_Term_106_eqn} &\text{Fig.(\ref{2PN_Term_75}$\vert$c)} \nonumber \\
&= -\intdt \intd{t_a} \intd{t_b} \intd{t_c} \nonumber \\
&\quad \times \frac{(d-4)(d-1) M_a M_b M_c}{16
(d-2)^2 \mpl^{2(d-2)} } I_3[a,b,c] \nonumber \\
&\quad \times \left( - \left( \Dd{t_a} \right)^2 - \left( \Dd{t_b} \right)^2 -
\left( \Dd{t_c} \right)^2 \right) \nonumber \\
&\quad \times \delta[t_a-t] \delta[t_b-t] \delta[t_c-t]
\end{align}}

To be sure, when the number of distinct particles changes from 3 to 2, the
integrals occurring in Fig.(\ref{2PN_Term_106}$\vert$a) would be different from
that in (\ref{2PN_Term_106_eqn}) -- which really is the result for
Fig.(\ref{2PN_Term_75}$\vert$c) -- but because it remains finite,
Fig.(\ref{2PN_Term_106}$\vert$a) still vanishes due to the coefficient $(d-4)$.

{\bf Fig.(\ref{2PN_Term_131}$\vert$b)} \quad Fig.(\ref{2PN_Term_131}$\vert$b)
corresponds to the first relativistic correction to each of the lowest order
graviton Green's functions in the 1 PN 3-graviton diagram
Fig.(\ref{1PN2Body}$\vert$d). Its associated master integral is

{\allowdisplaybreaks
\begin{align}
&\label{masterintegralI000000x} I_{000000\times}[q,r,s] \nonumber \\
&\equiv \intdt \int \dx{t}_q \int \dx{t}_r \int \dx{t}_s \frac{(d-3)^2}{(d-2)^2\mpl^{(d-2)/2}} \nonumber \\
&\times \left( \prod_{\ell=1}^3 \int \frac{\dd^{d-1}p_\ell}{(2\pi)^{d-1}} \right) \nonumber \\
&\times \bigg\{ \frac{\vec{p}_1^2 + \vec{p}_2^2 + \vec{p}_3^2}{\vec{p}_1^4 \vec{p}_2^2 \vec{p}_3^2} \delta[t-t_r] \delta[t-t_s] \left( \frac{\dd^2}{\dx{t}_q^2} \delta[t-t_q] \right) \nonumber \\
&\quad + \frac{\vec{p}_1^2 + \vec{p}_2^2 + \vec{p}_3^2}{\vec{p}_1^2 \vec{p}_2^4 \vec{p}_3^2} \delta[t-t_q] \delta[t-t_s] \left( \frac{\dd^2}{\dx{t}_r^2} \delta[t-t_r] \right) \nonumber \\
&\quad + \frac{\vec{p}_1^2 + \vec{p}_2^2 + \vec{p}_3^2}{\vec{p}_1^2 \vec{p}_2^2 \vec{p}_3^4} \delta[t-t_q] \delta[t-t_r] \left( \frac{\dd^2}{\dx{t}_s^2} \delta[t-t_s] \right) \bigg\} \nonumber \\
&\times \exp\left[ i \left( \vec{p}_1 \cdot \vec{x}_q[t_q] + \vec{p}_2 \cdot
\vec{x}_r[t_r] + \vec{p}_3 \cdot \vec{x}_s[t_s] \right)
\right] \nonumber \\
&\times (2\pi)^{d-1} \delta^{d-1}\left[ \vec{p}_1 + \vec{p}_2 + \vec{p}_3
\right]
\end{align}}

For the case of 2 distinct particles, $I_{000000\times}[a,a,b]$ becomes

{\allowdisplaybreaks
\begin{align*}
&I_{000000\times}[a,a,b] \nonumber \\
&\stackrel{d=4}{=} \frac{i}{128 \pi^3 \mpl} \int \dx{t} \int
\dd^{3-2\varepsilon}z
R_{az}^{-1+2\varepsilon} R_{bz}^{-1+2\varepsilon} \nonumber \\
&\qquad \times \left( \dot{v}_a^i \partial_i^a + v_a^i v_a^j \partial_i^a \partial_j^a \right) R_{az}^{-1+2\varepsilon} \nonumber \\
&+ \frac{i}{2\mpl} \int \dx{t} \int \dx{t}_a \int \dx{t}_b \int \dx{t}_c \nonumber \\
&\times \delta[t-t_a] \delta[t-t_b] \delta[t-t_c] \nonumber\\
&\times \int\int \frac{\dd^{3-2\varepsilon}p_1 \ \dd^{3-2\varepsilon}p_2}{(2\pi)^{2(3-2\varepsilon)}} \bigg( \frac{e^{i\vec{p}_2 \cdot (\vec{x}_a[t_b] - \vec{x}_b[t_c])}}{\vec{p}_2^2} \nonumber \\
&\times \frac{\dot{i \vec{v}}_a[t_a] \cdot \vec{p}_1 - \vec{v}_a[t_a] \cdot
\vec{p}_1 \
\vec{v}_a[t_a] \cdot \vec{p}_1}{\vec{p}_1^4} e^{i\vec{p}_1 \cdot (\vec{x}_a[t_a] - \vec{x}_a[t_b])} \bigg) \nonumber \\
&- \int \dx{t} \frac{i}{64 \pi^2 \mpl R_{ab}} \left( \dot{v}_a^i \partial_i^a + v_a^i v_a^j \partial_i^a \partial_j^a \right) R_{ab} \nonumber \\
&+ \int \dx{t} \frac{i}{4\mpl} \left( \dot{v}_b^i \partial_i^b + v_b^i v_b^j
\partial_i^b
\partial_j^b \right) I_3[a,a,b].
\end{align*}}

The first position space integral can be evaluated using
(\ref{integral_OneLoopRd}), (\ref{integral_OneLoopRvmu}) and
(\ref{integral_OneLoopRvmunu}). The fourier integral after it vanishes upon
integrating over the time $\delta$-functions because the exponential becomes
unity. We thus have

{\allowdisplaybreaks
\begin{align*}
&\text{Fig.(\ref{2PN_Term_131}$\vert$b)} \nonumber \\
&= \frac{M_a^2 M_b}{16 \mpl^{\frac{3}{2}(d-2)}} I_{000000\times}[a,a,b] \nonumber \\
&\stackrel{d=4}{=} \intdt \frac{M_a^2 M_b}{2048 \mpl^4 \pi^2 R_{ab}^4} \nonumber \\
&\times \bigg( 5 (\vec{R}_{ab} \cdot \vec{v}_a)^2 + 4 (\vec{R}_{ba} \cdot
\vec{v}_b)^2 \nonumber \\
&\quad - (4 \vec{R}_{ab} \cdot \dot{\vec{v}}_a + 2 \vec{R}_{ba} \cdot
\dot{\vec{v}}_b + 3 \vec{v}_a^2 + 2 \vec{v}_b^2) R_{ab}^2 \bigg)
\end{align*}}

{\bf Fig.(\ref{2PN_Term_60}$\vert$c,d)} \quad The master integral associated
with Fig.(\ref{2PN_Term_60}$\vert$c) and Fig.(\ref{2PN_Term_60}$\vert$d) is

{\allowdisplaybreaks
\begin{align}
&\label{masterintegralI0i0j00} I_{0i0j00}[q,r,s] \nonumber \\
&\equiv \intdt \int \dx{t}_q \int \dx{t}_r \int \dx{t}_s \frac{\delta[t-t_q] \delta[t-t_r] \delta[t-t_s]}{4(d-2)\mpl^{(d-2)/2}} \nonumber \\
&\times \left( \prod_{\ell=1}^3 \int \frac{\dd^{d-1}p_\ell}{(2\pi)^{d-1}} \right) \nonumber \\
&\times \bigg\{ \vec{v}_q \cdot \vec{v}_r \frac{(d-4) (\vec{p}_1^2 + \vec{p}_2^2) + (d-2) \vec{p}_3^2}{\vec{p}_1^2 \vec{p}_2^2 \vec{p}_3^2} \nonumber \\
&\qquad + 2 (d-4) \frac{\vec{p}_2 \cdot \vec{v}_q \ \vec{p}_1 \cdot \vec{v}_r}{\vec{p}_1^2 \vec{p}_2^2 \vec{p}_3^2} \nonumber \\
&\qquad + 2 (d-3) \frac{\vec{p}_3 \cdot \vec{v}_q \ \vec{p}_1 \cdot \vec{v}_r + \vec{p}_2 \cdot \vec{v}_q \ \vec{p}_3 \cdot \vec{v}_r}{\vec{p}_1^2 \vec{p}_2^2 \vec{p}_3^2} \bigg\} \nonumber \\
&\times \exp\left[ i \left( \vec{p}_1 \cdot \vec{x}_q[t_q] + \vec{p}_2 \cdot
\vec{x}_r[t_r] + \vec{p}_3 \cdot \vec{x}_s[t_s] \right)
\right] \nonumber \\
&\times (2\pi)^{d-1} \delta^{d-1}\left[ \vec{p}_1 + \vec{p}_2 + \vec{p}_3
\right]
\end{align}}

A similar approach to the one taken for (\ref{masterintegralI0i0000}) and
(\ref{masterintegralIij0000}), together with the integrals
(\ref{integral_OneLoopRvmunu}) and (\ref{integral_OneLoopRvmunuAB}), then
yields

{\allowdisplaybreaks
\begin{align*}
&\text{Fig.(\ref{2PN_Term_60}$\vert$c)} \nonumber \\
&= \frac{M_a^2 M_b}{2 \mpl^{(3/2)(d-2)}} I_{0i0j00}[a,b,a] \nonumber \\
&\stackrel{d=4}{=} -\intdt \frac{M_a^2 M_b}{512 \mpl^4 \pi^2 R_{ab}^4} \nonumber \\
&\times \left( \vec{v}_a \cdot \vec{v}_b \ R_{ab}^2 + \vec{R}_{ab} \cdot
\vec{v}_a \ \vec{R}_{ab} \cdot \vec{v}_b - 4 \vec{R}_{ba} \cdot \vec{v}_a \
\vec{R}_{ba} \cdot \vec{v}_b \right),
\end{align*}}

as well as

{\allowdisplaybreaks
\begin{align*}
\text{Fig.(\ref{2PN_Term_100}$\vert$d)} &= \frac{M_a^2 M_b}{4 \mpl^{(3/2)(d-2)}} I_{0i0j00}[a,a,b] \nonumber \\
&\stackrel{d=4}{=} \intdt \frac{M_a^2 M_b}{128 \mpl^4 \pi^2 R_{ab}^4}
(\vec{R}_{ab} \cdot \vec{v}_a)^2
\end{align*}}

{\bf Fig.(\ref{2PN_Term_65}$\vert$a,b)} \quad Fig.(\ref{2PN_Term_65}$\vert$a)
and Fig.(\ref{2PN_Term_102}$\vert$b) are, up to numerical constants, products
of $\vec{v}_a^2$ with the 1 PN 3-graviton diagram Fig.(\ref{1PN2Body}$\vert$d),
namely,

{\allowdisplaybreaks
\begin{align*}
&\text{Fig.(\ref{2PN_Term_65}$\vert$a)} \nonumber \\
&= \frac{M_a^2 M_b}{16 \mpl^{\frac{3}{2}(d-2)}} \int\dx{t} \ \vec{v}_a^2 I_{000000}[a,b,a] \nonumber \\
&= -\intdt \frac{(d-3)^2 \Gamma \left[\frac{d-3}{2}\right]^2 M_a^2 M_b}{256
(d-2)^2 \mpl^{2d-4} \pi^{d-1} R_{ab}^{2(d-3)}} \vec{v}_a^2,
\end{align*}}

and

{\allowdisplaybreaks
\begin{align*}
\text{Fig.(\ref{2PN_Term_102}$\vert$b)} &= \frac{M_a^2 M_b}{32 \mpl^{\frac{3}{2}(d-2)}} \int\dx{t} \ \vec{v}_b^2 I_{000000}[a,a,b] \nonumber \\
&= -\intdt \frac{(d-3)^2 \Gamma\left[ \frac{d-3}{2} \right]^2 M_a^2 M_b}{512
(d-2)^2 \mpl^{2(d-2)} \pi^{d-1} R_{ab}^{2(d-3)}} \vec{v}_b^2
\end{align*}}

{\bf Fig.(\ref{2PN_Term_126}$\vert$c,d)} \quad Fig.(\ref{2PN_Term_126}$\vert$c)
and Fig.(\ref{2PN_Term_134}$\vert$d) are, up to constant factors, the product
of the lowest order graviton Green's function $\langle h_{00} h_{00} \rangle$
with the 1 PN 3-graviton diagram Fig.(\ref{1PN2Body}$\vert$d), namely,

{\allowdisplaybreaks
\begin{align*}
&\text{Fig.(\ref{2PN_Term_126}$\vert$c)} \nonumber \\
&= \int\dx{t} \frac{\Gamma\left[ \frac{d-1}{2} \right] M_a^3 M_b}{128 (d-2)\mpl^{\frac{5}{2}(d-2)} \pi^{\frac{d-1}{2}} R_{ab}^{d-3}} I_{000000}[a,a,b] \nonumber \\
&= - \intdt \frac{\Gamma\left[\frac{d-1}{2}\right]^3 M_a^3 M_b}{512 (d-2)^3
\mpl^{3(d-2)} \pi^{\frac{3}{2} (d-1)} R_{ab}^{3(d-3)}} \nonumber \\
&\text{Fig.(\ref{2PN_Term_134}$\vert$d)} \nonumber \\
&= \int\dx{t} \frac{\Gamma\left[ \frac{d-1}{2} \right] M_a^2 M_b^2}{64 (d-2)\mpl^{\frac{5}{2}(d-2)} \pi^{\frac{d-1}{2}} R_{ab}^{d-3}} I_{000000}[a,b,b] \nonumber \\
&= -\intdt \frac{\Gamma\left[\frac{d-1}{2}\right]^3 M_a^2 M_b^2}{256 (d-2)^3
\mpl^{3(d-2)} \pi^{\frac{3}{2} (d-1)} R_{ab}^{3(d-3)}}
\end{align*}}

{\bf Fig.(\ref{2PN_Term_128}$\vert$a,b)} \quad The associated master integral
for Fig.(\ref{2PN_Term_128}$\vert$a) and Fig.(\ref{2PN_Term_130}$\vert$b) is

{\allowdisplaybreaks
\begin{align}
\label{masterintegralI00000000}&I_{00000000}[q,r,s,u] \nonumber \\
&= -i\int \dx{t} \int \dx{t}_q \int \dx{t}_r \int \dx{t}_s \int \dx{t}_u \nonumber \\
&\times \delta[t-t_q] \delta[t-t_r] \delta[t-t_s] \delta[t-t_u] \nonumber \\
&\times \frac{(d-3)(d(7d-51)+86)}{4(d-2)^3\mpl^{d-2}} \left( \prod_{\ell=1}^4 \int \frac{\dd^{d-1}p_\ell}{(2\pi)^{d-1}} \right) \nonumber \\
&\times \frac{\vec{p}_1^2 + \vec{p}_2^2 + \vec{p}_3^2 + \vec{p}_4^2}{\vec{p}_1^2 \vec{p}_2^2 \vec{p}_3^2 \vec{p}_4^2} \nonumber \\
&\times \exp\left[ i \left(\vec{p}_1 \cdot \vec{x}_q + \vec{p}_2 \cdot \vec{x}_r + \vec{p}_3 \cdot \vec{x}_s + \vec{p}_4 \cdot \vec{x}_u \right)\right] \nonumber \\
&\times (2\pi)^{d-1} \delta^{d-1}\left[ \vec{p}_1 + \vec{p}_2 + \vec{p}_3 +
\vec{p}_4 \right]
\end{align}}

Employing (\ref{integral_PropI}) hands us

{\allowdisplaybreaks
\begin{align*}
&\text{Fig.(\ref{2PN_Term_128}$\vert$a)} \nonumber \\
&= \frac{M_a^3 M_b}{96 \mpl^{2(d-2)}} I_{00000000}[a,a,a,b] \nonumber \\
&= -\intdt \frac{ (d-3) (d (7 d-51)+86) \Gamma^3\left[\frac{d-3}{2}\right]
M_a^3 M_b}{24576 (d-2)^3 \mpl^{3(d-2)} \pi^{\frac{3}{2}(d-1)} R_{ab}^{3(d-3)}}
\end{align*}}

and

{\allowdisplaybreaks
\begin{align*}
\text{Fig.(\ref{2PN_Term_130}$\vert$b)} &= \frac{M_a^2 M_b^2}{64 \mpl^{2(d-2)}} I_{00000000}[a,a,b,b] \nonumber \\
&= 0
\end{align*}}

{\bf Fig.(\ref{2PN_Term_146}$\vert$c,d)} \quad The evaluation of
Fig.(\ref{2PN_Term_146}$\vert$c) and Fig.(\ref{2PN_Term_148}$\vert$d),
involving the contraction of 2 3-graviton vertices, requires the most effort.
Its associated master integrals in momentum space, after substantial algebraic
manipulation reads:

{\allowdisplaybreaks
\begin{widetext}
\begin{align}
\label{masterintegralI0000-0000}&I_{0000-0000}[q,r,s,u] = I_{0000-0000}^{\text{I}}[q,r,s,u] + I_{0000-0000}^{\text{II}}[q,r,s,u] + I_{0000-0000}^{\text{III}}[q,r,s,u] \\
&\label{masterintegralI0000-0000_I} I_{0000-0000}^{\text{I}}[q,r,s,u] \\
&\equiv \intdt \int \dx{t}_q \int \dx{t}_r \int \dx{t}_s \int \dx{t}_u \frac{\delta[t-t_q] \delta[t-t_r] \delta[t-t_s] \delta[t-t_u]}{8(d-2)^3 \mpl^{d-2}} \left( \prod_{\ell=1}^4 \int \frac{\dd^{d-1}p_\ell}{(2\pi)^{d-1}} \right) \nonumber \\
&\quad \times \frac{1}{\vec{p}_1^2 \vec{p}_2^2 \vec{p}_3^2 \vec{p}_4^2} \bigg\{ (d-3)^2 (5d-18) \left( \vec{p}_1^2 + \vec{p}_2^2 + \vec{p}_3^2 + \vec{p}_4^2 \right) + (3 d^3 - 40 d^2 + 151 d - 174) (\vec{p}_1 + \vec{p}_2)^2 \nonumber \\
&\qquad \qquad -\frac{(d-3)^2}{(\vec{p}_1 + \vec{p}_2)^2} \left( 4(d-2)(\vec{p}_1^2 \vec{p}_2^2 + \vec{p}_3^2 \vec{p}_4^2) - (7d-22) (\vec{p}_1^2 + \vec{p}_2^2)(\vec{p}_3^2 + \vec{p}_4^2) \right) \bigg\} \nonumber \\
&\quad \times \exp\big[ i \vec{p}_1 \cdot \vec{x}_q[t_q] + i \vec{p}_2 \cdot
\vec{x}_r[t_r] + i \vec{p}_3 \cdot \vec{x}_s[t_s] + i \vec{p}_4 \cdot
\vec{x}_u[t_u] \big] (2\pi)^{d-1}
\delta^{d-1}\left[ \sum_{w=1}^4 \vec{p}_w \right] \nonumber \\
&+ \text{5 other permutations of } \{\vec{x}_q,\vec{x}_r,\vec{x}_s,\vec{x}_u\} \nonumber \\
&\label{masterintegralI0000-0000_II_orig} I_{0000-0000}^{\text{II}}[q,r,s,u] \\
&\equiv \intdt \int \dx{t}_q \int \dx{t}_r \int \dx{t}_s \int \dx{t}_u \frac{(d-3)^2 \delta[t-t_q] \delta[t-t_r] \delta[t-t_s] \delta[t-t_u]}{4(d-2)^2 \mpl^{d-2}} \left( \prod_{\ell=1}^4 \int \frac{\dd^{d-1}p_\ell}{(2\pi)^{d-1}} \right) \nonumber \\
&\quad \times \frac{\vec{p}_1^4 + \vec{p}_2^4 + \vec{p}_3^4 +
\vec{p}_4^4}{\vec{p}_1^2 \vec{p}_2^2 \vec{p}_3^2 \vec{p}_4^2 (\vec{p}_1 +
\vec{p}_2)^2} \exp\left[ i \vec{p}_1 \cdot \vec{x}_q[t_q] + i \vec{p}_2 \cdot
\vec{x}_r[t_r] + i \vec{p}_3 \cdot \vec{x}_s[t_s] + i \vec{p}_4 \cdot
\vec{x}_u[t_u] \right] (2\pi)^{d-1} \delta^{d-1}\left[ \sum_{w=1}^4 \vec{p}_w \right] \nonumber \\
&+ \text{5 other permutations of } \{\vec{x}_q,\vec{x}_r,\vec{x}_s,\vec{x}_u\} \nonumber \\
&\label{masterintegralI0000-0000_III} I_{0000-0000}^{\text{III}}[q,r,s,u] \\
&\equiv \intdt \int \dx{t}_q \int \dx{t}_r \int \dx{t}_s \int \dx{t}_u \frac{(d-3)^2 \delta[t-t_q] \delta[t-t_r] \delta[t-t_s] \delta[t-t_u]}{2(d-2)^2 \mpl^{d-2}} \left( \prod_{\ell=1}^4 \int \frac{\dd^{d-1}p_\ell}{(2\pi)^{d-1}} \right) \nonumber \\
&\times \frac{\vec{p}_1 \cdot \vec{p}_3 \ \vec{p}_2 \cdot \vec{p}_4 + \vec{p}_1
\cdot \vec{p}_4 \ \vec{p}_2 \cdot \vec{p}_3}{\vec{p}_1^2 \vec{p}_2^2
\vec{p}_3^2 \vec{p}_4^2 (\vec{p}_1 + \vec{p}_2)^2} \exp\left[ i \vec{p}_1 \cdot
\vec{x}_q[t_q] + i \vec{p}_2 \cdot \vec{x}_r[t_r] + i \vec{p}_3 \cdot
\vec{x}_s[t_s] + i \vec{p}_4 \cdot \vec{x}_u[t_u] \right] (2\pi)^{d-1}
\delta^{d-1}\left[ \sum_{w=1}^4 \vec{p}_w \right] \nonumber \\
&+ \text{5 other permutations of } \{\vec{x}_q,\vec{x}_r,\vec{x}_s,\vec{x}_u\}
\nonumber
\end{align}
\end{widetext}}

When contracting the 2 3-graviton Feynman rules in momentum space leading to
(\ref{masterintegralI0000-0000},\ref{masterintegralI0000-0000_I},\ref{masterintegralI0000-0000_II_orig},\ref{masterintegralI0000-0000_III}),
one makes a particular choice for the labels on the point particles' coordinate
vectors. If the two particle labels on the 3-graviton vertex on the left hand
side are $\{a,b\}$ and the two on the right hand side are $\{c,e\}$, so that
such a choice can be denoted as either $(ab|ce)$, $(ba|ce)$, $(ab|ec)$ or
$(ba|ec)$ -- there is a symmetry obeyed by the labels on either side -- then
the 6 permutations to be summed over in the definitions of
$I_{0000-0000}^{\text{I}}, I_{0000-0000}^{\text{II}}$ and
$I_{0000-0000}^{\text{III}}$ are: $2 \times (ab|ce)$, $2 \times (ac|be)$ and $2
\times (ae|bc)$. The particular permutation displayed in
Fig.(\ref{2PN_Term_146}$\vert$c) is $(aa|ba)$. It also represents the sum of
the 6 permutations, with each of the 6 terms giving the same result. In our
notation, the sum is $6 \times (aa|ab)$. As for the two diagrams in
Fig.(\ref{2PN_Term_148}$\vert$d), they represent the sum of the 6 permutations:
$4 \times (ab|ab)$ and $2 \times (aa|bb)$.

Turning to the integrals themselves, $I_{0000-0000}^{\text{I}}[q,r,s,u]$ can be
calculated in $d$-dimensions with (\ref{integral_PropI}). For the term
containing $(\vec{p}_1 + \vec{p}_2)^2$ in the numerator, one would first sum
over the 6 permutations of the coordinate vectors $\{\vec{x}_q\}$; this is
equivalent to holding fixed the $\{\vec{x}_q, \vec{x}_r, \vec{x}_s,
\vec{x}_u\}$ and permuting the names of the integration variables $\{\vec{p}_1,
\dots, \vec{p}_4\}$. Upon doing so, one would find the relevant part of the
integrand can be replaced as follows: $(\vec{p}_1 + \vec{p}_2)^2/(\vec{p}_1^2
\vec{p}_2^2 \vec{p}_3^2 \vec{p}_4^2) \to 2 (\vec{p}_1^2 + \vec{p}_2^2 +
\vec{p}_3^2 + \vec{p}_4^2)/(\vec{p}_1^2 \vec{p}_2^2 \vec{p}_3^2 \vec{p}_4^2)$.
After suitable re-definitions of the momentum variables of the form $\vec{p}'_a
\equiv \vec{p}_a + \vec{p}_b$, the $(\vec{p}_1^2 \vec{p}_2^2 + \vec{p}_3^2
\vec{p}_4^2)/(\vec{p}_1 + \vec{p}_2)^2$ portion of
$I_{0000-0000}^{\text{I}}[q,r,s,u]$ is, after integration, proportional to
$\delta^{d-1}[\vec{x}_q - \vec{x}_r]$ or $\delta^{d-1}[\vec{0}]$, i.e. zero
within dimension regularization; while the $(\vec{p}_1^2 +
\vec{p}_2^2)(\vec{p}_3^2 + \vec{p}_4^2)/(\vec{p}_1 + \vec{p}_2)^2$ integrand
again takes the (\ref{integral_PropI})-form $\exp[i\sum_r^3 \vec{p}_r \cdot
\vec{R}_{ru}]/[\prod_s^3 \vec{p}_s^2]$.

$I_{0000-0000}^{\text{II}}[q,r,s,u]$ can be dealt with by first integrating,
for each of its four terms, over the momentum appearing in the numerator. We
demonstrate this with the first term, dropping all numerical constants and
suppressing the time arguments.

{\allowdisplaybreaks
\begin{align}
\label{masterintegralI0000-0000_II} &\left( \prod_{\ell=1}^4 \int \frac{\dd^{d-1}p_\ell}{(2\pi)^{d-1}} \right) \frac{\vec{p}_1^4}{\vec{p}_1^2 \vec{p}_2^2 \vec{p}_3^2 \vec{p}_4^2 (\vec{p}_1 + \vec{p}_2)^2} \nonumber \\
&\times \exp\big[ i \big( \vec{p}_1 \cdot \vec{x}_q + \vec{p}_2 \cdot \vec{x}_r + \vec{p}_3 \cdot \vec{x}_s + \vec{p}_4 \cdot \vec{x}_u \big) \big] \nonumber \\
&\times (2\pi)^{d-1} \delta^{d-1}\left[ \vec{p}_1 + \vec{p}_2 + \vec{p}_3 + \vec{p}_4 \right] \nonumber \\
&= \left( \prod_{\ell=2}^4 \int \frac{\dd^{d-1}p_\ell}{(2\pi)^{d-1}} \right) \nonumber \\
&\times \bigg\{ \frac{1}{\vec{p}_2^2 \vec{p}_3^2 \vec{p}_4^2} e^{i \vec{p}_2 \cdot (\vec{x}_r - \vec{x}_q) + i \vec{p}_3 \cdot (\vec{x}_s - \vec{x}_q) + i \vec{p}_4 \cdot (\vec{x}_u - \vec{x}_q)} \nonumber \\
&+ 2 \delta_{ij} \frac{p_2^i}{\vec{p}_2^2} e^{ i \vec{p}_2 \cdot (\vec{x}_r - \vec{x}_q) } \frac{(p_3^j + p_4^j)}{\vec{p}_3^2 \vec{p}_4^2 (\vec{p}_3 + \vec{p}_4)^2} \nonumber \\
&\qquad \times e^{ i \vec{p}_3 \cdot (\vec{x}_s - \vec{x}_q) + i \vec{p}_4
\cdot (\vec{x}_u - \vec{x}_q) } \bigg\},
\end{align}}

where we have dropped the term proportional to $\int \dd^{d-1}p_2
\exp[i\vec{p}_2\cdot(\vec{x}_r[t_r] - \vec{x}_q[t_q])]$, which is zero even if
$\vec{x}_r = \vec{x}_q$, within dimensional regularization. The first piece in
the second equality containing only squares of momenta in the denominator can
be done with (\ref{integral_PropI}). For the second piece, the $\vec{p}_2$
integral is a $(1/i)\partial_i^r$ on $(\ref{integral_PropI})$; it is zero if
$\vec{x}_r = \vec{x}_q$. The $\vec{p}_3$ and $\vec{p}_4$ integrals translate to
an appropriate derivative on $I_3[a,a,b]$ or $I_3[a,b,b]$ if $\vec{x}_s =
\vec{x}_u$ and $\vec{x}_s \neq \vec{x}_q$.\footnote{The $\vec{p}_3$ and
$\vec{p}_4$ integrals with the numerator $p_3^i + p_4^i$ removed are
$I_3[q,s,u]$s because one can recover the form (\ref{masterintegralI3}) if one
introduces an additional variable $\vec{q} \equiv \vec{p}_3 + \vec{p}_4$ and a
corresponding integral and momentum conserving $\delta$-function.} If
$\vec{x}_s = \vec{x}_q$ and $\vec{x}_u \neq \vec{x}_q$ (or vice versa), the
$\vec{p}_3$ ($\vec{p}_4$) integrals take the form of
(\ref{integral_OneLoopRvmu}), and the remaining $\vec{p}_4$ ($\vec{p}_3$)
integral is then a derivative on (\ref{integral_PropI}). The $\vec{p}_3$ and
$\vec{p}_4$ integrals return zero if $\vec{x}_s = \vec{x}_u = \vec{x}_q$.

What remains is $I_{0000-0000}^{\text{III}}[q,r,s,u]$. First replace the
$(\vec{p}_1 + \vec{p}_2)^2$ in the denominator with $\vec{q}$ and introduce a
$(2\pi)^{1-d} \int \dd^{d-1}q \delta^{(d-1)}[\vec{q} - \vec{p}_1 - \vec{p}_2]$.
Applying (\ref{deltafunctionintegralrep}) on both the $\delta$-functions then
tells us that, in its position space representation -- again ignoring the
constant factors -- $I_{0000-0000}^{\text{III}}[q,r,s,u]$ becomes

{\allowdisplaybreaks
\begin{align*}
&\int \dd^{d-1}y \int \dd^{d-1}z \ \delta^{ij} \delta^{mn} \nonumber \\
&\times \big( (\partial_i^q R_{qy}^{3-d})(\partial_m^r R_{ry}^{3-d})(R_{yz}^{3-d})(\partial_j^s R_{sz}^{3-d})(\partial_n^u R_{uz}^{3-d}) \nonumber \\
&\quad + (\partial_i^q R_{qy}^{3-d})(\partial_m^r
R_{ry}^{3-d})(R_{yz}^{3-d})(\partial_n^s R_{sz}^{3-d})(\partial_j^u
R_{uz}^{3-d}) \big)
\end{align*}}

For the 2 body problem, we would either have one of the $\vec{y}$ or $\vec{z}$
integration involve only one of the coordinate vectors $\vec{x}_a$ or
$\vec{x}_b$, i.e. $(aa|ab)$ and $(aa|bb) = (bb|aa)$, or have 2 distinct
coordinate vectors occurring in each of them, i.e. $(ab|ab)$. For the former,
one may simply use (\ref{integral_OneLoopRd}), (\ref{integral_OneLoopRvmunu}),
(\ref{integral_OneLoopRvmunuAB}), together with the cosine rule $\vec{R}_{az}
\cdot \vec{R}_{bz} = -(1/2)R_{ab}^2 + (1/2) R_{az}^2 + (1/2) R_{bz}^2$. For the
latter $(ab|ab)$ case when $d=4-2\varepsilon$, one can first integrate over
$\vec{y}$ using $I_3$, carry out the differentiation, and by applying the
cosine rule the $(R_{az} + R_{bz} + R_{ab})$ appearing in the denominator right
after differentiation of $I_3$ will be removed. The integrals that remain are
tractable via (\ref{integral_OneLoopRd}).

Summing up the contributions from the 6 permutations for each of the
Fig.(\ref{2PN_Term_146}$\vert$c) and Fig.(\ref{2PN_Term_148}$\vert$d) now gives

{\allowdisplaybreaks
\begin{align*}
\text{Fig.(\ref{2PN_Term_146}$\vert$c)} &= \frac{M_a^3 M_b}{192 \mpl^{2(d-2)}} I_{0000-0000}[a,a,a,b] \nonumber \\
&\stackrel{d=4}{=} -\intdt \frac{M_a^3 M_b}{16384 \mpl^6 \pi^3 R_{ab}^3},
\end{align*}}

and

{\allowdisplaybreaks
\begin{align*}
\text{Fig.(\ref{2PN_Term_148}$\vert$d)} &= \frac{M_a^2 M_b^2}{128 \mpl^{2(d-2)}} I_{0000-0000}[a,a,b,b] \nonumber \\
&\stackrel{d=4}{=} \intdt \frac{M_a^2 M_b^2}{16384 \mpl^6 \pi^3 R_{ab}^3}
\end{align*}}

\subsubsection{3 body diagrams}

\begin{figure}
\begin{center}
\includegraphics[width=2.5in]{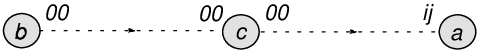} \\
\text{(a) 6 permutations: $a$, $b$, or $c$ in the middle.} \\
\text{Two choices for remaining two labels.} \\
\includegraphics[width=2.5in]{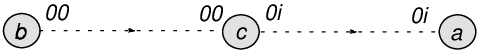} \\
\text{(b) 6 permutations: $a$, $b$, or $c$ in the middle.} \\
\text{Two choices for remaining two labels.} \\
\includegraphics[width=3in]{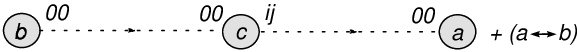} \\
\text{(c) 3 permutations: $a$, $b$, or $c$ in the middle.} \\
\includegraphics[width=2.5in]{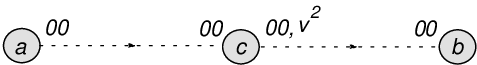} \\
\text{(d) 3 permutations: $a$, $b$, or $c$ in the middle.} \\
\includegraphics[width=2.5in]{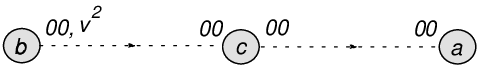} \\
\text{(e) 6 permutations: $a$, $b$, or $c$ in the middle.} \\
\text{Two choices for remaining two labels.} \\
\includegraphics[width=3in]{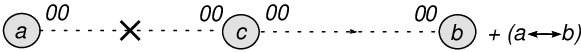} \\
\text{(f) 3 permutations: $a$, $b$, or $c$ in the middle.} \\
\includegraphics[width=1.5in]{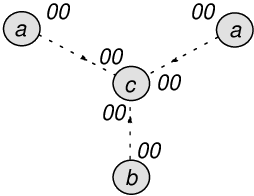} \\
\text{(g) 6 permutations: $a$, $b$, or $c$ in the center.}
\text{Two choices for remaining two labels.} \\
\includegraphics[width=2.5in]{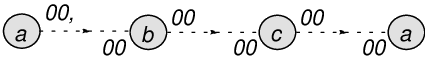} \\
\text{(h) 3 permutations: $a$, $b$, or $c$ at the two ends.} \\
\includegraphics[width=2.5in]{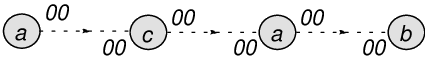} \\
\text{(i) 6 permutations: $a$, $b$, or $c$ for the repeated label.}
\text{Two choices for remaining two labels.} \\
\caption{2 PN three body diagrams with no graviton vertices.}
\label{2PN_Term_17} \label{2PN_Term_25} \label{2PN_Term_32} \label{2PN_Term_33}
\label{2PN_Term_37} \label{2PN_Term_52} \label{2PN_Term_108}
\label{2PN_Term_110} \label{2PN_Term_79}
\end{center}
\end{figure}


\textsl{No graviton vertices} \quad The 3 body diagrams that do not have
graviton vertices are:

{\allowdisplaybreaks
\begin{align*}
\text{Fig.(\ref{2PN_Term_17}$\vert$a)} &= \intdt \frac{(d-3)
\Gamma\left[\frac{d-3}{2}\right]^2}{256 (d-2)^2 \pi^{d-1}} \nonumber \\
&\qquad \times \frac{M_a M_b M_c}{\mpl^{2(d-2)} R_{ac}^{d-3} R_{bc}^{d-3}}
\vec{v}_a^2 \nonumber \\
\text{Fig.(\ref{2PN_Term_25}$\vert$b)} &= -\intdt \frac{(d-3)
\Gamma\left[\frac{d-3}{2}\right]^2}{128 (d-2) \pi^{d-1}} \nonumber \\
&\qquad \times \frac{M_a M_b M_c}{\mpl^{2(d-2)} R_{ac}^{d-3} R_{bc}^{d-3}}
\vec{v}_a \cdot \vec{v}_c \nonumber \\
\text{Fig.(\ref{2PN_Term_32}$\vert$c)} &= \intdt \frac{(d-3)
\Gamma\left[\frac{d-3}{2}\right]^2}{128 (d-2)^2 \pi^{d-1}} \nonumber \\
&\qquad \times \frac{M_a M_b M_c}{\mpl^{2(d-2)} R_{ac}^{d-3} R_{bc}^{d-3}}
\vec{v}_c^2 \nonumber \\
\text{Fig.(\ref{2PN_Term_33}$\vert$d)} &= \intdt \frac{3
\Gamma \left[\frac{d-1}{2}\right]^2}{128 (d-2)^2 \pi^{d-1}} \nonumber \\
&\qquad \times \frac{M_a M_b M_c}{\mpl^{2 (d-2)} R_{ac}^{d-3} R_{bc}^{d-3}}
\vec{v}_c^2 \nonumber \\
\text{Fig.(\ref{2PN_Term_37}$\vert$e)} &= \intdt
\frac{\Gamma\left[\frac{d-1}{2}\right]^2}{128 (d-2)^2 \pi^{d-1}} \nonumber \\
&\qquad \times \frac{M_a M_b M_c}{\mpl^{2 (d-2)} R_{ac}^{d-3} R_{bc}^{d-3}}
\vec{v}_b^2
\end{align*}}

As with its 2 body counterpart, Fig.(\ref{2PN_Term_52}$\vert$f) requires some
caution when taking the time derivatives. As an example, the integral in the
$a-\times-c-b$ diagram, without the constant factors, is

\begin{align*}
&\int\dx{t}_a\int\dx{t}_c \delta[t_a-t_c] \nonumber \\
&\times \bigg( \frac{\dd |\vec{x}_c[t_c] - \vec{x}_b[t_c]|^{3-d}}{\dx{t}_c} \frac{\dd |\vec{x}_a[t_a] - \vec{x}_c[t_c]|^{5-d}}{\dx{t}_a} \nonumber \\
&+ |\vec{x}_c[t_c] - \vec{x}_b[t_c]|^{3-d} \frac{\dd^2 |\vec{x}_a[t_a] -
\vec{x}_c[t_c]|^{5-d}}{\dx{t}_a \dx{t}_c} \bigg)
\end{align*}

Fig.(\ref{2PN_Term_52}$\vert$f) is the sum of $a-\times-c-b$ and
$b-\times-c-a$, but since differentiation and piecing together the relevant
constants are straightforward, we will not display the result.

{\allowdisplaybreaks

{\allowdisplaybreaks
\begin{align*}
\text{Fig.(\ref{2PN_Term_108}$\vert$g)} &= \intdt \frac{3 \Gamma
\left[\frac{d-1}{2}\right]^3}{1024 (d-2)^3 \pi^{\frac{3}{2} (d-1)}}
\nonumber \\
&\qquad \times \frac{M_a^2 M_b M_c}{ \mpl^{3 (d-2)} R_{ac}^{2(d-3)}
R_{bc}^{d-3} } \nonumber \\
\text{Fig.(\ref{2PN_Term_110}$\vert$h)} &= \intdt \frac{\Gamma
\left[\frac{d-1}{2}\right]^3}{512 (d-2)^3 \pi^{ \frac{3}{2}(d-1)
}} \nonumber \\
&\qquad \times \frac{M_a^2 M_b M_c}{\mpl^{3 (d-2)} R_{ab}^{d-3} R_{ac}^{d-3}
R_{bc}^{d-3}} \nonumber \\
\text{Fig.(\ref{2PN_Term_79}$\vert$i)} &= \intdt \frac{\Gamma
\left[\frac{d-1}{2}\right]^3}{512 (d-2)^3 \pi^{ \frac{3}{2}(d-1) }} \nonumber \\
&\qquad \times \frac{M_a^2 M_b M_c}{\mpl^{3 (d-2)} R_{ab}^{d-3}
R_{ac}^{2(d-3)}}
\end{align*}}

\begin{figure}
\begin{center}
\includegraphics[width=1.5in]{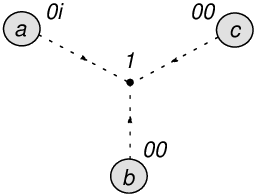} \\
\text{(a) 3 permutations: $a$, $b$, or $c$ carry the $0i$ indices.} \\
\includegraphics[width=1.5in]{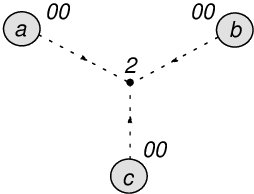} \\
\text{(b) No permutations necessary.} \\
\includegraphics[width=1.5in]{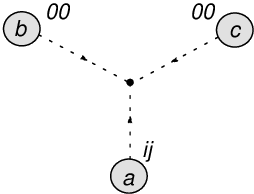} \\
\text{(c) 3 permutations: $a$, $b$, or $c$ carry the $ij$ indices.} \\
\includegraphics[width=1.5in]{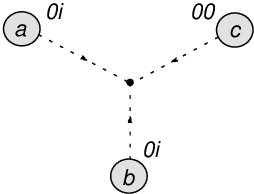} \\
\text{(d) 3 permutations: $a$, $b$, or $c$ carry the $00$ indices.} \\
\includegraphics[width=1.5in]{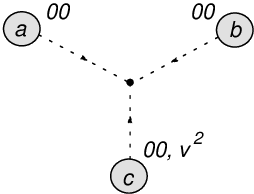} \\
\text{(e) 3 permutations: $a$, $b$, or $c$ carry the $v^2$.} \\
\caption{2 PN three body diagrams with graviton vertices: 1 of 2}
\label{2PN_Term_71} \label{2PN_Term_75} \label{2PN_Term_56} \label{2PN_Term_62}
\label{2PN_Term_67}
\end{center}
\end{figure}

\begin{figure}
\begin{center}
\includegraphics[width=3in]{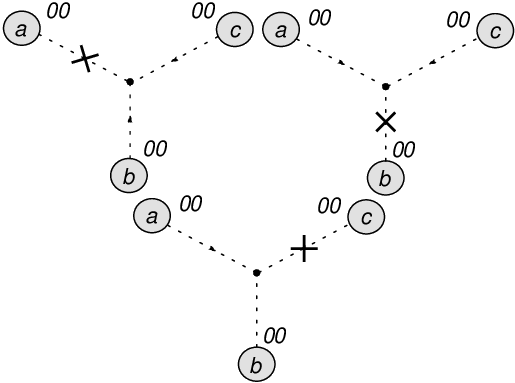} \\
\text{(a) No permutations necessary.} \\
\includegraphics[width=2in]{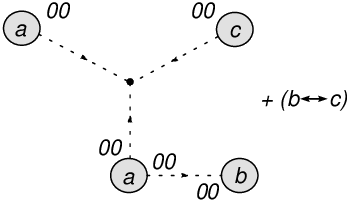} \\
\text{(b) 9 permutations. See text for discussion.} \\
\includegraphics[width=1.2in]{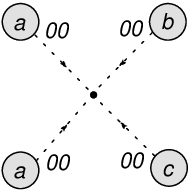} \\
\text{(c) 3 permutations: $a$, $b$, or $c$ for the repeated label.} \\
\includegraphics[width=1.8in]{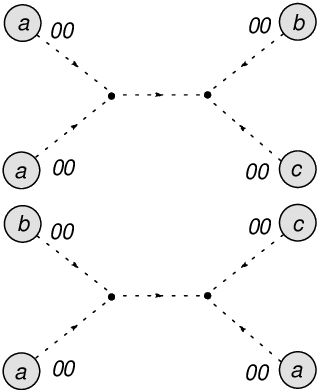} \\
\text{(d) 6 permutations. See text for discussion.} \\
\caption{2 PN three body diagrams with graviton vertices: 2 of 2}
\label{2PN_Term_117} \label{2PN_Term_118} \label{2PN_Term_143}
\label{2PN_Term_149}
\end{center}
\end{figure}


\textsl{Graviton vertices} \quad The rest of the 3 body diagrams contain
graviton vertices.

{\bf Fig.(\ref{2PN_Term_71}$\vert$a)} \quad Fig.(\ref{2PN_Term_71}$\vert$a)
requires $I_{0i0000}$ from (\ref{masterintegralI0i0000}). For the 3 body case,
one sees that $I_{0i0000}$ can be expressed in terms of time- and
space-derivatives on $I_3$. Specifically,

{\allowdisplaybreaks
\begin{align*}
&I_{0i0000}[a,b,c] \nonumber \\
&= \frac{i}{2(d-2)\mpl^{(d-2)/2}} \int\dx{t} \int\dx{t}_a \int\dx{t}_b \int\dx{t}_c \nonumber \\
&\times \delta[t-t_a] \delta[t-t_b] \delta[t-t_c] \nonumber \\
&\times \bigg( \{\left( 2(d-4) v_a^i \partial_i^b + (d-5) v_a^i
\partial_i^c \right) I_3[a,b,c]\} \nonumber \\
&\quad \times \Dd{t_b} \delta[t_b - t] \delta[t_c - t] \nonumber \\
&+ \{\left( 2(d-4) v_a^i \partial_i^c + (d-5) v_a^i \partial_i^b
\right) I_3[a,b,c]\} \nonumber \\
&\quad \times \Dd{t_c} \delta[t_b - t] \delta[t_c - t] \bigg),
\end{align*}}

which means

{\allowdisplaybreaks
\begin{align*}
\text{Fig.(\ref{2PN_Term_71}$\vert$a)} &= \frac{M_a M_b M_c}{4
\mpl^{\frac{3}{2}(d-2)}} I_{0i0000}[a,b,c]
\end{align*}}

{\bf Fig.(\ref{2PN_Term_75}$\vert$b)} \quad As discussed for
Fig.(\ref{2PN_Term_106}$\vert$a), we have a vanishing diagram when $d=4$ for

\begin{align*}
\text{Fig.(\ref{2PN_Term_75}$\vert$b)} &\stackrel{d=4}{=} 0
\end{align*}

{\bf Fig.(\ref{2PN_Term_56}$\vert$c)} \quad The master integral for
Fig.(\ref{2PN_Term_56}$\vert$c) is $I_{ij0000}$ in
(\ref{masterintegralIij0000}), whose 3 body expression in terms of $I_3$ is

{\allowdisplaybreaks
\begin{align*}
&I_{ij0000}[a,b,c] \nonumber \\
&= i\int\dx{t} \bigg\{ \frac{\Gamma\left[ \frac{d-3}{2} \right]^2}{16 \pi^{d-1} (d-2)^2\mpl^{(d-2)/2}} \vec{v}_a^2 \big( R_{ab}^{3-d} R_{ac}^{3-d} \nonumber \\
&\qquad -(d-3) \left( R_{ba}^{3-d}  R_{bc}^{3-d} +  R_{ca}^{3-d}  R_{cb}^{3-d} \right) \big) \nonumber \\
&-\frac{d-3}{2(d-2)\mpl^{(d-2)/2}} v_a^i v_a^j \nonumber \\
&\qquad \times \left( \partial_i^a \partial_j^a + \partial_i^b \partial_j^b +
\partial_i^c \partial_j^c \right) I_3[a,b,c] \bigg\},
\end{align*}}

so that

{\allowdisplaybreaks
\begin{align*}
\text{Fig.(\ref{2PN_Term_56}$\vert$c)} &= \frac{M_a M_b M_c}{8
\mpl^{\frac{3}{2}(d-2)}} I_{ij0000}[a,b,c]
\end{align*}}

{\bf Fig.(\ref{2PN_Term_62}$\vert$d)} \quad The 3 body master integral for
Fig.(\ref{2PN_Term_62}$\vert$d) is $I_{0i0j00}$ from
(\ref{masterintegralI0i0j00}), which reads

{\allowdisplaybreaks
\begin{align*}
&I_{0i0j00}[a,b,c] \nonumber \\
&= i\int\dx{t}\bigg\{ \frac{\Gamma\left[ \frac{d-3}{2} \right]^2 \vec{v}_a \cdot \vec{v}_b}{64 \pi^{d-1} (d-2)\mpl^{(d-2)/2}} \nonumber \\
&\times \big( (d-4) \left( R_{ab}^{3-d} R_{ac}^{3-d} + R_{ba}^{3-d} R_{bc}^{3-d} \right) \nonumber \\
&\qquad + (d-2) R_{ca}^{3-d} R_{cb}^{3-d} \big) \nonumber \\
&- \frac{1}{2(d-2) \mpl^{(d-2)/2}} v_a^i v_b^j \nonumber \\
&\qquad \times \big( (d-4) \partial_i^b
\partial_j^a \nonumber \\
&\qquad + (d-3)(\partial_i^c \partial_j^a + \partial_i^b \partial_j^c) \big)
I_3[a,b,c] \bigg\},
\end{align*}}

leading us to

\begin{align*}
\text{Fig.(\ref{2PN_Term_62}$\vert$d)} &= \frac{M_a M_b
M_c}{2\mpl^{\frac{3}{2}(d-2)}} I_{0i0j00}[a,b,c]
\end{align*}

{\bf Fig.(\ref{2PN_Term_67}$\vert$e)} \quad Fig.(\ref{2PN_Term_67}$\vert$e)
involves products of $\vec{v}^2$ with the 1 PN 3-graviton master integral
$I_{000000}$.

{\allowdisplaybreaks
\begin{align*}
\text{Fig.(\ref{2PN_Term_67}$\vert$e)} &= \int\dx{t} \frac{M_a M_b M_c}{16 \mpl^{\frac{3}{2}(d-2)}} \vec{v}_c^2 I_{000000}[a,b,c] \nonumber \\
&= -\intdt \frac{(d-3)^2 \Gamma \left[\frac{d-3}{2}\right]^2 M_a M_b M_c}{256
\pi^{d-1}
(d-2)^2 \mpl^{2(d-2)}} \vec{v}_c^2 \nonumber \\
&\times \left( R_{ab}^{3-d} R_{ac}^{3-d} + R_{bc}^{3-d} R_{ac}^{3-d} +
R_{ab}^{3-d} R_{bc}^{3-d} \right)
\end{align*}}

{\bf Fig.(\ref{2PN_Term_117}$\vert$a)} \quad The 3 body master integral for
Fig.(\ref{2PN_Term_117}$\vert$a) is $I_{000000\times}$ from
(\ref{masterintegralI000000x}):

{\allowdisplaybreaks
\begin{align*}
&I_{000000\times}[a,b,c] = \widetilde{I}_{\times}[a,b,c] + \widetilde{I}_{\times}[b,a,c] + \widetilde{I}_{\times}[c,b,a] \nonumber \\
&\widetilde{I}_{\times}[a,b,c] \nonumber \\
&\equiv \frac{i(d-3)^2}{(d-2)^2 \mpl^{(d-2)/2}} \int\dx{t} \int\dx{t}_a \int\dx{t}_b \int\dx{t}_c \nonumber \\
&\times \bigg( I_3[a,b,c] \nonumber \\
&\qquad + \frac{\Gamma\left[ \frac{d-3}{2} \right] \Gamma\left[ \frac{d-5}{2} \right]}{64 \pi^{d-1}} \left( R_{ba}^{5-d} R_{bc}^{3-d} +  R_{ca}^{5-d} R_{cb}^{3-d} \right) \bigg) \nonumber \\
&\times \frac{\dd^2}{\dx{t}_a^2} \delta[t-t_a] \delta[t-t_b] \delta[t-t_c]
\end{align*}}

In terms of $I_{000000\times}[a,b,c]$,

{\allowdisplaybreaks
\begin{align*}
\text{Fig.(\ref{2PN_Term_117}$\vert$a)} &= \frac{M_a M_b M_c}{8
\mpl^{\frac{3}{2}(d-2)}} I_{000000\times}[a,b,c]
\end{align*}}

{\bf Fig.(\ref{2PN_Term_118}$\vert$b)} \quad For the class of diagrams
represented by Fig.(\ref{2PN_Term_118}$\vert$b), the 9 permutations include
both types of diagrams where the 2 world line sources attached to the 3
graviton vertex not contracted with an additional world line can either belong
to the same particle (6 of them) or 2 distinct particles (3 of them).
(Fig.(\ref{2PN_Term_118}$\vert$b) itself belongs to the latter.) We observe, as
we did in the 2 body case, that these diagrams are products of the lowest order
$\langle h_{00} h_{00} \rangle$ with the 1 PN 3-graviton master integral
$I_{000000}$ in (\ref{masterintegralI000000}).

{\allowdisplaybreaks
\begin{align}
\label{2PN_Term_118_eqn} &\text{Fig.(\ref{2PN_Term_118}$\vert$b)} \nonumber \\
&= \frac{\Gamma\left[ \frac{d-1}{2} \right] M_a^2 M_b M_c}{64 (d-2) \pi^{\frac{d-1}{2}} \mpl^{\frac{5}{2}(d-2)}} \int\dx{t} \nonumber \\
&\times \left( R_{ac}^{3-d} I_{000000}[a,b,a] + R_{ab}^{3-d} I_{000000}[a,c,a] \right) \nonumber \\
&= -\intdt \frac{(d-3)^2 \Gamma\left[\frac{d-3}{2}\right]^2 \Gamma
\left[\frac{d-1}{2}\right] M_a^2 M_b M_c}{1024 \mpl^{3(d-2)} \pi^{\frac{3}{2}(d-1)} (d-2)^3} \nonumber \\
&\quad \times\left( R_{ac}^{3-d} R_{ab}^{2(3-d)} + R_{ac}^{2(3-d)} R_{ab}^{3-d}
\right),
\end{align}}

For the other 6 permutations where the 2 world line sources attached to the 3
graviton vertex not contracted with an additional world line belong to the same
particle, the analog to the above 2 terms in (\ref{2PN_Term_118_eqn}) have
different numerical factors in front of them.

{\bf Fig.(\ref{2PN_Term_143}$\vert$c)} \quad Fig.(\ref{2PN_Term_143}$\vert$c)
is a straightforward application of $I_{00000000}$ in
(\ref{masterintegralI00000000}) and (\ref{integral_PropI}).

{\allowdisplaybreaks
\begin{align*}
\text{Fig.(\ref{2PN_Term_143}$\vert$c)} &= \int\dx{t} \frac{M_a^2 M_b M_c}{32 \mpl^{2(d-2)}} I_{00000000}[a,a,b,c] \nonumber \\
&= -\intdt \frac{ (d-3) (d(7 d-51)+86) \Gamma
\left[\frac{d-3}{2}\right]^3}{8192
\pi^{\frac{3}{2}(d-1)} (d-2)^3} \nonumber \\
&\times \frac{M_a^2 M_b M_c}{\mpl^{3(d-2)}} \left( R_{bc}^{3-d} R_{ab}^{2(3-d)}
+ R_{ac}^{2(3-d)} R_{bc}^{3-d} \right)
\end{align*}}

{\bf Fig.(\ref{2PN_Term_149}$\vert$d)} \quad Referring to the discussion under
the 2 body counterpart of Fig.(\ref{2PN_Term_149}$\vert$d), if $a$ is the
repeated label for a given 3 body diagram, then the 6 permutations of particle
labels are: $2 \times (aa|bc)$, $2 \times (ab|ac)$, and $2 \times (ac|ab)$.

The master integrals for Fig.(\ref{2PN_Term_149}$\vert$d) can be found in
(\ref{masterintegralI0000-0000},\ref{masterintegralI0000-0000_I},\ref{masterintegralI0000-0000_II_orig},\ref{masterintegralI0000-0000_III}).
$I_{0000-0000}^{\text{I}}$ is the linear combination of products of
(\ref{integral_PropI}). As we did in the 2 body case,
$I_{0000-0000}^{\text{II}}$ can be done using (\ref{integral_PropI}), its
derivatives, and derivatives on $I_3$. When $a$ is the repeated particle label,
$I_{0000-0000}^{\text{II}}$ is

{\allowdisplaybreaks
\begin{align}
\label{masterintegralI0000-0000_II_3Body} &I_{0000-0000}^{\text{II}}[a,a,b,c] \nonumber \\
&\stackrel{d=4}{=} \frac{i}{256 \mpl^2 \pi^3} \int\dx{t} \bigg( \frac{R_{ab}^2-R_{ac}^2}{2 R_{ac}^2 R_{bc}^3} - \frac{R_{ab}^2 + R_{ac}^2 - R_{bc}^2}{2 R_{ab}^3 R_{ac}^2} \nonumber \\
&\quad \qquad - \frac{R_{ab}^2+R_{ac}^2-R_{bc}^2}{2 R_{ab}^2 R_{ac}^3} + \frac{R_{ab}-R_{bc}}{R_{ac}^3 R_{bc}} \nonumber \\
&\quad \qquad + \frac{R_{ac}-R_{bc}}{R_{ab}^3 R_{bc}} +
\frac{R_{ac}^2-R_{ab}^2}{2 R_{ab}^2 R_{bc}^3} \bigg)
\end{align}}

We will leave $I_{0000-0000}^{\text{III}}[q,q,r,s]$ for possible future work.

In terms of the $I_{0000-0000}$ in
(\ref{masterintegralI0000-0000},\ref{masterintegralI0000-0000_I},\ref{masterintegralI0000-0000_II_orig},\ref{masterintegralI0000-0000_III}),
we have

\begin{align*}
\text{Fig.(\ref{2PN_Term_149}$\vert$d)} &= \frac{M_a^2 M_b M_c}{64
\mpl^{2(d-2)}} I_{0000-0000}[a,a,b,c]
\end{align*}

\subsubsection{4 body diagrams}

\begin{figure}
\begin{center}
\includegraphics[width=1.5in]{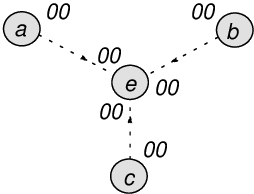} \\
\text{(a) 4 permutations: $a$, $b$, $c$, or $e$ for the center label.} \\
\includegraphics[width=3in]{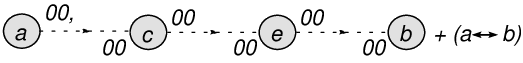} \\
\text{(b) 6 permutations. See text for discussion.} \\
\caption{2 PN four body diagrams with no graviton vertices} \label{2PN_Term_76}
\label{2PN_Term_81}
\end{center}
\end{figure}


\textsl{No graviton vertices} \quad The vertex-less diagrams are:

{\allowdisplaybreaks
\begin{align*}
\text{Fig.(\ref{2PN_Term_76}$\vert$a)} &= \intdt \frac{3 \Gamma
\left[\frac{d-1}{2}\right]^3}{512 (d-2)^3 \pi^{ \frac{3}{2}(d-1) }}
\nonumber \\
&\quad \times \frac{M_a M_b M_c M_e}{\mpl^{3 (d-2)} R_{ae}^{d-3} R_{be}^{d-3}
R_{ce}^{d-3}}
\end{align*}}

For the class of diagrams in Fig.(\ref{2PN_Term_81}$\vert$b), there are 12 ways
to choose 2 out of the $\{a,b,c,e\}$ for the middle two labels, but there is a
reflection symmetry, and hence there are 6 distinct permutations of the
particle labels.

{\allowdisplaybreaks
\begin{align*}
\text{Fig.(\ref{2PN_Term_81}$\vert$b)} &= \intdt
\frac{\Gamma\left[\frac{d-1}{2}\right]^3}{512 (d-2)^3 \pi^{\frac{3}{2}(d-1)}}
\frac{M_a M_b M_c M_e}{\mpl^{3 (d-2)}} \nonumber \\
&\times \left( R_{ae}^{3-d} R_{bc}^{3-d} R_{ce}^{3-d} + R_{ac}^{3-d}
R_{be}^{3-d} R_{ce}^{3-d} \right)
\end{align*}}

\begin{figure}
\begin{center}
\includegraphics[width=2in]{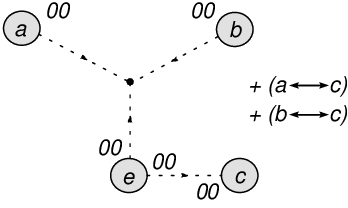} \\
\text{(a) 4 permutations. $a,b,c$ or $e$ has 2 graviton fields.} \\
\includegraphics[width=1.2in]{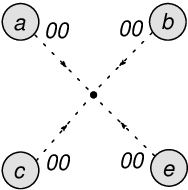} \\
\text{(b) No permutations necessary.} \\
\includegraphics[width=2.0in]{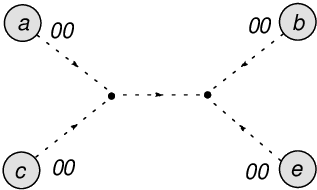} \\
\text{(c) 6 permutations. See text for discussion.} \\
\caption{2 PN four body diagrams with graviton vertices} \label{2PN_Term_121}
\label{2PN_Term_125} \label{2PN_Term_152}
\end{center}
\end{figure}


\textsl{Graviton vertices} \quad The rest of the 4 body diagrams have graviton
vertices.

{\bf Fig.(\ref{2PN_Term_121}$\vert$a)} \quad The class of diagrams in
Fig.(\ref{2PN_Term_121}$\vert$a), like those of its 2- and 3-body counterparts,
involve products of the lowest order $\langle h_{00} h_{00} \rangle$ with the 1
PN 3-graviton vertex integral $I_{000000}$. Also, there are 4 distinct
permutations, with the world line operator with 2 graviton fields associated
with either $a$, $b$, $c$, or $e$.

{\allowdisplaybreaks
\begin{align*}
&\text{Fig.(\ref{2PN_Term_121}$\vert$a)} \nonumber \\
&= \int\dx{t} \frac{\Gamma \left[\frac{d-1}{2}\right] M_a M_b M_c M_e}{64 (d-2) \mpl^{\frac{5 (d-2)}{2}} \pi^{\frac{d-1}{2}}} \nonumber \\
&\times\big( R_{ae}^{3-d} I_{000000}[b,c,e] + R_{ce}^{3-d} I_{000000}[a,b,e] \nonumber \\
&\qquad + R_{be}^{3-d} I_{000000}[a,c,e]\big) \nonumber \\
&= -\intdt \frac{(d-3)^2 \Gamma \left[\frac{d-1}{2}\right] \Gamma
\left[\frac{d-3}{2}\right]^2 M_a M_b M_c M_e}{1024 \pi^{\frac{3}{2}(d-1)} \mpl^{3(d-2)} (d-2)^3} \nonumber \\
&\times \bigg( \left( R_{bc}^{3-d} R_{be}^{3-d} + R_{ce}^{3-d} R_{be}^{3-d} +
R_{bc}^{3-d} R_{ce}^{3-d} \right) R_{ae}^{3-d} \nonumber \\
&\qquad + \left( R_{ab}^{3-d} R_{ae}^{3-d} + R_{be}^{3-d} R_{ae}^{3-d} +
R_{ab}^{3-d} R_{be}^{3-d} \right) R_{ce}^{3-d} \nonumber \\
&\qquad + \left( R_{ac}^{3-d} R_{ae}^{3-d} + R_{ce}^{3-d} R_{ae}^{3-d} +
R_{ac}^{3-d} R_{ce}^{3-d} \right) R_{be}^{3-d} \bigg)
\end{align*}}

{\bf Fig.(\ref{2PN_Term_125}$\vert$b)} \quad Another straightforward
application of $I_{00000000}$ in (\ref{masterintegralI00000000}) and
(\ref{integral_PropI}) provides us with

{\allowdisplaybreaks
\begin{align*}
&\text{Fig.(\ref{2PN_Term_125}$\vert$b)} \nonumber \\
&= \int\dx{t} \frac{M_a M_b M_c M_e}{16 \mpl^{2(d-2)}} I_{00000000}[a,b,c,e] \nonumber \\
&= -\intdt \frac{(d-3) (d (7 d-51)+86) \Gamma \left[\frac{d-3}{2}\right]^3}{4096 \pi^{\frac{3}{2}(d-1)} (d-2)^3} \nonumber \\
&\quad \times \frac{M_a M_b M_c M_e}{ \mpl^{3(d-2)}} \bigg( R_{ab}^{3-d}
R_{ac}^{3-d} R_{ae}^{3-d} + R_{be}^{3-d}
R_{ce}^{3-d} R_{ae}^{3-d} \nonumber \\
&\qquad + R_{ab}^{3-d} R_{bc}^{3-d} R_{be}^{3-d} + R_{ac}^{3-d} R_{bc}^{3-d}
R_{ce}^{3-d} \bigg)
\end{align*}}

{\bf Fig.(\ref{2PN_Term_152}$\vert$c)} \quad Referring to the discussion under
the 2 body counterpart of Fig.(\ref{2PN_Term_152}$\vert$c), the 6 permutations
of particle labels are: $2 \times (ab|ce)$, $2 \times (ac|be)$, and $2 \times
(ae|bc)$.

The master integrals for Fig.(\ref{2PN_Term_152}$\vert$c) can be found in
(\ref{masterintegralI0000-0000},\ref{masterintegralI0000-0000_I},\ref{masterintegralI0000-0000_II_orig},\ref{masterintegralI0000-0000_III}).
$I_{0000-0000}^{\text{I}}$ is the linear combination of products of
(\ref{integral_PropI}). As we did in the 2 body case,
$I_{0000-0000}^{\text{II}}$ can be done using (\ref{integral_PropI}), its
derivatives, and derivatives on $I_3$. We have given the explicit expression
for $I_{0000-0000}^{\text{II}}$ in the 3 body case
(\ref{masterintegralI0000-0000_II_3Body}), but the 4 body one is too lengthy to
display.

$I_{0000-0000}^{\text{III}}[q,r,s,u]$ is left for possible future work.

In terms of the $I_{0000-0000}$ in
(\ref{masterintegralI0000-0000},\ref{masterintegralI0000-0000_I},\ref{masterintegralI0000-0000_II_orig},\ref{masterintegralI0000-0000_III}),
we have

\begin{align*}
\text{Fig.(\ref{2PN_Term_152}$\vert$c)} &= \frac{M_a M_b M_c M_e}{32
\mpl^{2(d-2)}} I_{0000-0000}[a,b,c,e]
\end{align*}

\subsubsection{$\mathcal{O}[v^4]$ Effective Lagrangian}

Adding all the relevant diagrams, their permutations and the second order
relativistic correction to kinetic energy from the $\eta_{\mu\nu} v^\mu v^\nu$
in the infinitesimal proper time $\dx{s}$ now gives us the effective lagrangian
describing the gravitational dynamics of $n$ point masses at $\mathcal{O}[v^4]$
relative to Newtonian gravity in 3+1 dimensions:

{\allowdisplaybreaks
\begin{widetext}
\begin{align}
\label{2PNnBody} L^{\text{(2 PN)}}_{\text{eff}} &\stackrel{d=4}{=} L_4^{\text{2 Body}} + L_4^{\text{3 Body}} + L_4^{\text{4 Body}} \\
\label{2PN2Body} L_4^{\text{2 Body}} &\equiv \frac{1}{2} \sum_{\substack{1 \leq
a, b
\leq n \\ a \neq b}} \bigg\{ \frac{M_a}{16} \vec{v}_a^6 + \frac{M_b}{16} \vec{v}_b^6 \nonumber \\
&+ \frac{G_{\text{N}} M_a M_b}{R_{ab}} \bigg( \vec{R}_{ab} \cdot \vec{v}_a \left( \frac{7}{4} \vec{v}_a \cdot \dot{\vec{v}}_b - \frac{3}{2} \vec{v}_b \cdot \dot{\vec{v}}_b \right) + \vec{R}_{ba} \cdot \vec{v}_b \left( \frac{7}{4} \vec{v}_b \cdot \dot{\vec{v}}_a - \frac{3}{2} \vec{v}_a \cdot \dot{\vec{v}}_a \right) \nonumber \\
&\qquad - \frac{1}{8} \left( \frac{\vec{R}_{ab} \cdot \vec{v}_a}{R_{ab}} \right)^2 \left( \vec{R}_{ba} \cdot \dot{\vec{v}}_b + \vec{v}_b^2 \right) - \frac{1}{8} \left( \frac{\vec{R}_{ba} \cdot \vec{v}_b}{R_{ab}} \right)^2 \left( \vec{R}_{ab} \cdot \dot{\vec{v}}_a + \vec{v}_a^2 \right) \nonumber \\
&\qquad + \frac{3}{4} \left( \vec{v}_a^2 + \vec{v}_b^2 - 2 \ \vec{v}_a \cdot \vec{v}_b \right) \frac{\vec{R}_{ab} \cdot \vec{v}_a}{R_{ab}} \frac{\vec{R}_{ba} \cdot \vec{v}_b}{R_{ab}} + \frac{3}{8} \frac{(\vec{R}_{ab} \cdot \vec{v}_a)^2 (\vec{R}_{ba} \cdot \vec{v}_b)^2}{R_{ab}^4} \nonumber \\
&\qquad + \frac{1}{8} \left( \vec{R}_{ba} \cdot \dot{\vec{v}}_b \ \vec{v}_a^2 +
\vec{R}_{ab} \cdot \dot{\vec{v}}_a \ \vec{v}_b^2 \right) + \frac{1}{8}
\vec{R}_{ab} \cdot \dot{\vec{v}}_a \ \vec{R}_{ba} \cdot \dot{\vec{v}}_b \nonumber \\
&\qquad + \frac{15}{8} \dot{\vec{v}}_a \cdot \dot{\vec{v}}_b \ R_{ab}^2 + \frac{7}{8} (\vec{v}_a^4 + \vec{v}_b^4) + \frac{1}{4} (\vec{v}_a \cdot \vec{v}_b)^2 + \frac{3}{8} \vec{v}_a^2 \vec{v}_b^2 - \frac{5}{4} (\vec{v}_a^2 + \vec{v}_b^2) \vec{v}_a \cdot \vec{v}_b \bigg) \nonumber \\
&+ \frac{G_{\text{N}}^2 M_a M_b}{R_{ab}^2} \bigg( - \frac{3}{2} \frac{M_a (\vec{R}_{ba} \cdot \vec{v}_b)^2 + M_b (\vec{R}_{ab} \cdot \vec{v}_a)^2}{R_{ab}^2} - 2 (M_a + M_b) \frac{\vec{R}_{ab} \cdot \vec{v}_a \ \vec{R}_{ba} \cdot \vec{v}_b}{R_{ab}^2} \nonumber \\
&\qquad - (2 M_a + M_b) \vec{R}_{ab} \cdot \dot{\vec{v}}_a - (2 M_b + M_a) \vec{R}_{ba} \cdot \dot{\vec{v}}_b \nonumber \\
&\qquad + \vec{v}_a^2 \left( 2 M_a + \frac{11}{4} M_b \right) + \vec{v}_b^2 \left( 2 M_b + \frac{11}{4} M_a \right) - \frac{9}{2} \vec{v}_a \cdot \vec{v}_b \left( M_a + M_b \right) \bigg) \nonumber \\
&- \frac{G_{\text{N}}^3 M_a M_b}{R_{ab}^3} \left( M_a M_b + \frac{3}{2} (M_a^2 + M_b^2) \right) \bigg\} \\
\label{2PN3Body} L_4^{\text{3 Body}} &\equiv \frac{1}{3!} \sum_{\substack{1 \leq a,b,c \leq n \\ a,b,c \text{ distinct}}} \bigg\{ G_{\text{N}}^2 M_a M_b M_c \nonumber \\
&\qquad \times \bigg( \frac{1}{R_{ab} R_{ac}} \left( \frac{9}{2} \vec{v}_a^2 + 8 \vec{v}_b \cdot \vec{v}_c \right) + \frac{1}{R_{ab} R_{bc}} \left( \frac{9}{2} \vec{v}_b^2 + 8 \vec{v}_a \cdot \vec{v}_c \right) + \frac{1}{R_{ac} R_{bc}} \left( \frac{9}{2} \vec{v}_c^2 + 8 \vec{v}_a \cdot \vec{v}_b \right) \nonumber \\
&\qquad - \frac{8}{(R_{ab} + R_{ac} + R_{bc})^2} \left( \frac{\vec{R}_{ba} \cdot \vec{v}_b \vec{R}_{ca} \cdot \vec{v}_c}{R_{ab} R_{ac}} + \frac{\vec{R}_{ab} \cdot \vec{v}_a \vec{R}_{cb} \cdot \vec{v}_c}{R_{ab} R_{bc}} + \frac{\vec{R}_{ac} \cdot \vec{v}_a \vec{R}_{bc} \cdot \vec{v}_b}{R_{ac} R_{bc}} \right) \nonumber \\
&\qquad + \frac{4}{R_{ab} + R_{ac} + R_{bc}} \left( \frac{\vec{v}_a^2}{R_{bc}} + \frac{\vec{v}_b^2}{R_{ac}} + \frac{\vec{v}_c^2}{R_{ab}} \right) \nonumber \\
&\qquad + \bigg[ \frac{1}{2 R_{ab} R_{ac}^3} \left( \vec{R}_{ac} \cdot \vec{v}_a \vec{R}_{ba} \cdot \vec{v}_b + \vec{R}_{ac} \cdot \vec{v}_a \vec{R}_{ca} \cdot \vec{v}_c + \vec{R}_{ba} \cdot \vec{v}_b \vec{R}_{ca} \cdot \vec{v}_c + 2 (\vec{R}_{ca} \cdot \vec{v}_c)^2 \right) \nonumber \\
&\qquad - \frac{1}{R_{ab} R_{ac}} \left( \vec{R}_{ba}\cdot\dot{\vec{v}}_b + \frac{7}{2} \vec{v}_a \cdot \vec{v}_b + \frac{5}{2} \vec{v}_b^2 \right) \nonumber \\
&\qquad + \frac{1}{(R_{ab}+R_{ac}+R_{bc})^2} \bigg( \frac{1}{R_{ab}^2} \left( 4 \vec{R}_{ab} \cdot \vec{v}_c \vec{R}_{ba} \cdot \vec{v}_b + 8 \vec{R}_{ab} \cdot \vec{v}_a \vec{R}_{ab} \cdot \vec{v}_c - 2 (\vec{R}_{ab} \cdot \vec{v}_a)^2 - 2 (\vec{R}_{ab} \cdot \vec{v}_c)^2 \right) \nonumber \\
&\qquad \qquad + \frac{1}{R_{ab} R_{ac}} \left( 4 \vec{R}_{ac} \cdot \vec{v}_b \vec{R}_{ba} \cdot \vec{v}_b - 8 \vec{R}_{ac} \cdot \vec{v}_b \vec{R}_{ba} \cdot \vec{v}_c + 12 \vec{R}_{ac} \cdot \vec{v}_a \vec{R}_{ba} \cdot \vec{v}_c - 12 \vec{R}_{ac} \cdot \vec{v}_a \vec{R}_{ba} \cdot \vec{v}_b \right) \bigg) \nonumber \\
&\qquad + \frac{1}{R_{ab}+R_{ac}+R_{bc}} \bigg( \frac{1}{R_{ab}^3} \left( 8 \vec{R}_{ab} \cdot \vec{v}_a \vec{R}_{ab} \cdot \vec{v}_c + 4 \vec{R}_{ab} \cdot \vec{v}_c \vec{R}_{ba} \cdot \vec{v}_b - 2 (\vec{R}_{ab} \cdot \vec{v}_c)^2 - 2 (\vec{R}_{ab} \cdot \vec{v}_a)^2 \right) \nonumber \\
&\qquad \qquad + \frac{1}{R_{ab}} \left( 2 \vec{v}_a^2 - 4 \vec{v}_a \cdot \vec{v}_c - 2 \vec{R}_{ab} \cdot \dot{\vec{v}}_a \right) \bigg) + \text{$5$ other permutations of $\{a,b,c\}$} \bigg] \bigg) \nonumber \\
&+ G_{\text{N}}^3 M_a M_b M_c \bigg( \bigg[ \frac{(M_a + M_c) R_{ab}^2}{R_{ac}^2 R_{bc}^3} + \frac{2 M_b R_{ab}}{R_{ac} R_{bc}^3} - \frac{3 M_a}{R_{ab}^3} \nonumber \\
&\qquad - \frac{1}{R_{ab} R_{ac}^2} \left( M_a + \frac{3}{2} M_c \right) + \text{$5$ other permutations of $\{a,b,c\}$} \bigg] \nonumber \\
&\qquad + \frac{1}{16 \pi^2} \left( M_a I_{22}[a,a,b,c] + M_b I_{22}[b,b,a,c] + M_c I_{22}[c,c,a,b] \right) - 2 \left( \frac{M_a}{R_{bc}^3} + \frac{M_b}{R_{ac}^3} + \frac{M_c}{R_{ab}^3} \right) \bigg) \bigg\} \\
\label{2PN4Body} L_4^{\text{4 Body}} &\equiv \frac{1}{4!} \sum_{\substack{1 \leq a,b,c,e \leq n \\ a,b,c,e \text{ distinct}}} G_{\text{N}}^3 M_a M_b M_c M_e \nonumber \\
&\qquad \times \bigg\{ \frac{I_{22}[a,b,c,e]}{8 \pi^2} - 3 \left( \frac{1}{R_{ab} R_{ac} R_{ae}} + \frac{1}{R_{ba} R_{bc} R_{be}} + \frac{1}{R_{ca} R_{cb} R_{ce}} + \frac{1}{R_{ea} R_{eb} R_{ec}} \right) \nonumber \\
&\qquad \qquad + \bigg[ \frac{1}{R_{ab} + R_{ac} + R_{bc}} \left( \frac{R_{bc}}{R_{ab} R_{ac} R_{ae}} + \frac{2 R_{ae}^2}{R_{ab} R_{be}^3} - \frac{2 R_{ab}}{R_{ae}^3} \right) \nonumber \\
&\qquad \qquad \qquad + \text{$23$ other permutations of $\{a,b,c,e\}$} \bigg] \bigg\} \\
\label{I22Def} I_{22}[q,r,s,u] &\equiv \lim_{\substack{\varepsilon \to 0^+ \\ \text{after integration}}} \int \dd^{3-2\varepsilon}y \int \dd^{3-2\varepsilon}z \ \delta^{ij} \delta^{mn} \bigg( (\partial_i^q R_{qy}^{-1+2\varepsilon})(\partial_m^r R_{ry}^{-1+2\varepsilon})(R_{yz}^{-1+2\varepsilon})(\partial_j^s R_{sz}^{-1+2\varepsilon})(\partial_n^u R_{uz}^{-1+2\varepsilon}) \nonumber \\
&\qquad \qquad \qquad \qquad + (\partial_i^q R_{qy}^{-1+2\varepsilon})(\partial_m^r R_{ry}^{-1+2\varepsilon})(R_{yz}^{-1+2\varepsilon})(\partial_n^s R_{sz}^{-1+2\varepsilon})(\partial_j^u R_{uz}^{-1+2\varepsilon}) \bigg) \nonumber \\
&\qquad \qquad \qquad \qquad + \text{5 permutations of }
\{\vec{x}_q,\vec{x}_r,\vec{x}_s,\vec{x}_u \}, \qquad R_{qz} \equiv |\vec{x}_q -
\vec{z}|
\end{align}
\end{widetext}}

The permutations in the definitions of $L_4^{\text{3 Body}}$ and $L_4^{\text{4
Body}}$ means one would have to take the terms in the given square brackets
$[\dots]$, consider the resulting expressions obtained from permuting the
particle labels as stated, and sum them all up at the end. For
$I_{22}[q,r,s,u]$, if $(qr|su)$ represents the term with $\vec{x}_q$ and
$\vec{x}_r$ occurring in the $\vec{y}$ integration and the $\vec{x}_s$ and
$\vec{x}_u$ in the $\vec{z}$ integration, then the 6 permutations in the
definition of $I_{22}$ are: $2 \times (qr|su)$, $2 \times (qs|ru)$ and $2
\times (qu|rs)$.

\textsl{Relation to L$_{\text{ADM}}$} \quad As a (partial) check of these
results, we shall construct here a coordinate transformation that would bring
the 2 body portion of $L^{\text{(2 PN)}}_{\text{eff}}$ into the 2 body,
acceleration-independent, lagrangian $L_{\text{ADM}}$ in the literature; for
example, eq. (178) of Blanchet \cite{Blanchet2BodyLADM}. (This construction can
be found in Damour and Sch\"{a}fer
\cite{DamourSchaeferNBody,DamourSchaeferRedefinition}.) First, we note that
defining

\begin{align*}
x^i_a[t] &\equiv z^i_a[t] + \delta z^i_a[t],
\end{align*}

where $\delta z_a^i$ is assumed to be small relative to $z_a^i$ (an assumption
to be justified shortly), would modify the form of the lagrangian $L[\{x_a,
\vec{v}_a,\dots\}]$ up to first order in $\delta z$ in the following manner:

\begin{align*}
L[\{\vec{x}_a,&\vec{v}_a,\dot{\vec{v}}_a,\dots\}] \nonumber \\
&= L[\{\vec{z}_a,\dot{\vec{z}}_a,\ddot{\vec{z}}_a,\dots\}] + \frac{\delta L}{\delta z} \delta z + \text{total derivative} \\
\frac{\delta L}{\delta z} \delta z &\equiv \sum_{1 \leq a \leq n} \sum_{s =
0}^\infty \left( \frac{\dd^s}{\dx{t}^s} \frac{\partial L}{\partial (\dd^s
\vec{z}_a/\dx{t}^s)} \right) \cdot \delta \vec{z}_a
\end{align*}

In particular, varying the Newtonian lagrangian $L_{\text{0 PN}}$ gives us

\begin{align}
\label{newtonianlagrangianvaried} &\sum_a \frac{\delta L_{\text{0 PN}}}{\delta \vec{x}_a} \cdot \delta \vec{x}_a \nonumber \\
&= -\sum_{a \neq b} M_a \left( \ddot{\vec{x}}_a + \frac{G_{\text{N}}
M_b}{R_{ab}^3} \vec{R}_{ab} \right) \cdot \delta \vec{x}_a
\end{align}

Before proceeding with any coordinate transformation, however, one needs to
first re-write the terms quadratic in accelerations, $\sum_{i,j;a,b}
\ddot{x}_a^i \ddot{x}_b^j \tilde{L}^{ij} \subset L^{\text{(2
PN)}}_{\text{eff}}$ as

\begin{align}
&\sum_{1 \leq i,j \leq 3} \sum_{1 \leq a,b \leq n} \ddot{x}_a^i \ddot{x}_b^j \tilde{L}^{ij} \nonumber \\
\label{linearizeacc} &=\sum_{i,j;a,b} \bigg\{ \left( \ddot{x}_a^i + \frac{G_{\text{N}} M_b}{R_{ab}^3} R_{ab}^i \right) \left( \ddot{x}_b^j + \frac{G_{\text{N}} M_a}{R_{ba}^3} R_{ba}^j \right) \tilde{L}^{ij} \nonumber \\
&\quad - \ddot{x}_a^i \frac{G_{\text{N}} M_a}{R_{ba}^3} R_{ba}^j \tilde{L}^{ij} - \frac{G_{\text{N}} M_b}{R_{ba}^3} R_{ab}^i \ddot{x}_b^j \tilde{L}^{ij} \nonumber \\
&\quad - \frac{G_{\text{N}} M_b}{R_{ba}^3} R_{ab}^i \frac{G_{\text{N}}
M_a}{R_{ba}^3} R_{ba}^j \tilde{L}^{ij} \bigg\}
\end{align}

Because the term in the first line on the right hand side of
(\ref{linearizeacc}) contains the ``square" of the Newtonian equations of
motion, namely $(1/M_a) (\delta L_{\text{0 PN}}/\delta z_a^i) (1/M_b) (\delta
L_{\text{0 PN}}/\delta z_b^j)$, and because $(1/M)(\delta L_{\text{0
PN}}/\delta z + \delta L_{\text{1 PN}}/\delta z + \dots) = 0$, we see that this
first term on the right hand side of (\ref{linearizeacc}) scales as $[(1/M)
\delta L_{\text{0 PN}}/\delta z]^2 \sim [(1/M) \delta L_{\text{1 PN}}/\delta
z]^2 \sim [v^4/r]^2$ and therefore can be discarded at the 2 PN order.

{\bf Removal of accelerations} \quad The reason for linearizing the
acceleration dependent terms in $L_{\text{2 PN}}$, keeping only the second and
third lines on the right hand side of (\ref{linearizeacc}) is this. Denoting
this linearized form of $L_{\text{2 PN}}$ as $L^\ell_{\text{2 PN}}$ and
referring to the $-M_a \ddot{\vec{x}}_a \cdot \delta \vec{x}_a$ piece in
(\ref{newtonianlagrangianvaried}), we see that the remaining acceleration
dependent terms in $L^\ell_{\text{2 PN}}$ can now be removed by defining

\begin{align}
\label{accelerationremoval} \delta \vec{z}_a^{\text{(I)}} \equiv \left.
\frac{1}{M_a} \frac{\partial L^\ell_{\text{2 PN}}}{\partial
\ddot{\vec{x}}_a}\right\vert_{\vec{x}_a = \vec{z}_a}
\end{align}

{\bf Further transformations} \quad One can perform further coordinate
transformations without re-introducing acceleration dependent terms. The key is
to make the $-M_a \ddot{\vec{x}}_a \cdot \delta \vec{x}_a$ piece in
(\ref{newtonianlagrangianvaried}) part of a total time derivative. Observe
that, by having some arbitrary functional $F$ depend only on positions and
velocities,\footnote{We also exclude the possibility that $F$ depend on time
explicitly, since our post-Newtonian lagrangian does not.} we have the identity

\begin{align*}
&\sum_a \ddot{\vec{x}}_a \cdot \frac{\partial F}{\partial \dot{\vec{x}}_a}[\{\vec{x}_a, \dot{\vec{x}}_a\}] \\
&= \frac{\dd F}{\dx{t}}[\{\vec{x}_a, \dot{\vec{x}}_a\}] - \sum_a
\dot{\vec{x}}_a \cdot \frac{\partial F}{\partial \vec{x}_a}[\{\vec{x}_a,
\dot{\vec{x}}_a\}]
\end{align*}

Therefore, by putting

\begin{align*}
\delta \vec{z}_a^{\text{(II)}} &\equiv \frac{1}{M_a} \frac{\partial F}{\partial
\dot{\vec{z}}_a}[\{\vec{z}_b, \dot{\vec{z}}_b\}],
\end{align*}

we can replace $-\sum_a M_a \ddot{\vec{z}}_a \cdot \delta
\vec{z}_a^{\text{(II)}} = -\sum_a \ddot{\vec{z}}_a \cdot \partial F/\partial
\dot{\vec{z}}_a$ with $\sum_a \dot{\vec{z}}_a \cdot \partial F/\partial
\vec{z}_a$.

At this point, let us note that the alterations to the form of the lagrangian
due to $\delta z$ (\ref{accelerationremoval}) occurs solely at the 2 PN order:
$\delta \vec{z} \sim r v^4$. Hence $\delta z$ is indeed small relative to
$\vec{z}_a$, and it is only necessary to consider $\delta L_{\text{0
PN}}/\delta z$ and not $\delta L_{\text{1 PN}}/\delta z$, $\delta L_{\text{2
PN}}/\delta z$, nor any corrections that are quadratic or higher polynomials of
$\delta z$.

Altogether, the 2 body lagrangian after linearizing the accelerations and after
the transformation $\vec{x}_a \equiv \vec{z} + \delta \vec{z}_a^{\text{(I)}} +
\delta \vec{z}_a^{\text{(II)}}$, less total derivative terms, now reads

\begin{align}
&L[\{\vec{x}_a,\vec{v}_a,\dot{\vec{v}}_a,\dots\}] \nonumber \\
&= L_{\text{0 PN}}[\{\vec{z}_a,\dot{\vec{z}}_a\}] + L_{\text{1 PN}}[\{\vec{z}_a,\dot{\vec{z}}_a\}] \nonumber \\
&\qquad + L^\ell_{\text{2 PN}}[\{\vec{z}_a,\dot{\vec{z}}_a,\ddot{\vec{z}}_a\}] \nonumber \\
&\label{deltaL_accremoval} \quad - \left. \sum_{a \neq b} \left(
\ddot{\vec{x}}_a + \frac{G_{\text{N}} M_b}{R_{ab}^3}
\vec{R}_{ab} \right) \cdot \frac{\partial L^\ell_{\text{2 PN}}}{\partial \ddot{\vec{x}}_a} \right\vert_{\vec{x}_a = \vec{z}_a} \\
&\label{deltaL_furthertrans} \quad + \left. \sum_{a \neq b} \left(
\dot{\vec{x}}_a \cdot \frac{\partial F}{\partial \vec{x}_a} -
\frac{G_{\text{N}} M_b}{R_{ab}^3} \vec{R}_{ab} \cdot \frac{\partial F}{\partial
\dot{\vec{x}}_a} \right) \right\vert_{\vec{x}_a = \vec{z}_a}
\end{align}

It remains to construct $F[\{\vec{x}_a, \dot{\vec{x}}_a\}]$. From
$(1/M)(\partial F/\partial \dot{x}) = \delta z \sim r v^4$, we must have $F
\sim M r v^5$. From (\ref{deltaL_furthertrans}), we also must have $F \propto
M_a M_b / \mpl^2 \sim G_{\text{N}} M_a M_b \sim M r v^2$, since all terms in
$L^{\text{(2 PN)}}_{\text{eff}}$ need to contain at least one power of the mass
of each of the 2 point particles and at least one power of $G_{\text{N}}$ is
required because all the terms (less the $M \vec{v}^6/16$) in $L^{\text{(2
PN)}}_{\text{eff}}$ are at least linear in $G_{\text{N}}$. To supply the
additional $v^3$ needed, we have to consider all possible products of the
dimensionless scalars built out of terms occurring at 2 PN order, namely,
$\{\vec{v}_a^2, \vec{v}_b^2, \vec{v}_a \cdot \vec{v}_b, G_{\text{N}}
M_a/R_{ab}, G_{\text{N}} M_b/R_{ab}\}$ for the $v^2$ terms and $\{\vec{v}_a
\cdot \vec{R}_{ab}/R_{ab}, \vec{v}_b \cdot \vec{R}_{ba}/R_{ab}\}$ for the $v^1$
terms. The most general $F$ is thus

{\allowdisplaybreaks
\begin{align}
\label{arbitraryF} &F[\{\vec{x}_s, \dot{\vec{x}}_s | s=1,2,\dots,n\}] \nonumber \\
&= \frac{1}{2} \sum_{a \neq b} \bigg\{ \frac{G_{\text{N}}^2 M_a M_b (c_1 M_a + c_2 M_b)}{R_{ab}^2} \vec{v}_a \cdot \vec{R}_{ab} \nonumber \\
&\quad + \frac{G_{\text{N}} M_a M_b}{R_{ab}} \vec{v}_a \cdot \vec{R}_{ab} \bigg( c_3 \vec{v}_a^2 + c_4 \vec{v}_b^2 + c_5 \vec{v}_a \cdot \vec{v}_b \nonumber \\
&\qquad + c_6 \left( \frac{\vec{v}_a \cdot \vec{R}_{ab}}{R_{ab}} \right)^2 + c_7 \left( \frac{\vec{v}_b \cdot \vec{R}_{ba}}{R_{ab}} \right)^2 \nonumber \\
&\qquad + c_8 \left( \frac{\vec{v}_a \cdot \vec{R}_{ab}}{R_{ab}} \right) \left(
\frac{\vec{v}_b \cdot \vec{R}_{ba}}{R_{ab}} \right) \bigg) + (a \leftrightarrow
b) \bigg\},
\end{align}}

where the $\{c_i|i=1,2,\dots,8\}$ are arbitrary real numbers, and $(a
\leftrightarrow b)$ means one would have to take the terms occurring before it
and swap all the particle labels $a \leftrightarrow b$.

Computing (\ref{deltaL_accremoval}) and (\ref{deltaL_furthertrans}) with such a
$F$ reveals that one would recover the $L_{\text{ADM}}$ in Blanchet
\cite{Blanchet2BodyLADM} from (\ref{2PN2Body}) for

\begin{align*}
c_1 = 0, \quad c_2 = -\frac{3}{4}, \quad c_3 = 0, \quad c_4 = \frac{1}{2} & \nonumber \\
c_5 = 0, \quad c_6 = 0,            \quad c_7 = - c_8 &,
\end{align*}

where $c_8$ can be an arbitrary real number.

\section{Summary and Discussion}

In this paper we have, following \cite{nrgr}, used a point mass approximation
for the $n$-body in general relativity, allowing us to obtain a lagrangian
description at the cost of introducing an infinite number of terms in the
action. Because we are seeking the 2 PN effective lagrangian, however, only the
minimal terms $\{-M_a \int\dx{s}_a\}$ are necessary. By examining the physical
scales in the problem, we have described how to organize our action
(\ref{fullaction}) and outlined an algorithm that would allow us in principle
to generate all the necessary Feynman diagrams up to an arbitrary PN order for
a given set of point particle actions (minimal or not); as well as automate the
computation process so that as few of the Feynman diagrams as possible are left
for human evaluation. This way, the post-Newtonian program can be pursued in an
efficient and systematic manner within the framework of perturbative field
theory, and the necessary software may be developed to tackle the effective 2
body lagrangian calculation at 4 PN and beyond. In the bulk of this work, we
obtained in closed form the conservative portion of the effective lagrangian
$L_{\text{eff}}[\{\vec{x}_a,\vec{v}_a,\dot{\vec{v}}_a\}]$ up to 1 PN for the
general case of $n$ point masses in $d \geq 4$ spacetime dimensions and up to 2
PN for 2 point masses in (3+1)-dimensions.

It is apparent that the primary bottleneck of higher post-Newtonian
calculations is one of calculus. For the $n$-body problem, it is the analytic
evaluation of integrals such as $I_{22}$ (\ref{I22Def}) and the $I_{1\dots N}$
(\ref{npoint}) for $N \geq 4$. At the same time, it is possible that choosing a
different gauge from the one used in (\ref{gaugefixing}) and/or a different
parametrization of $h_{\mu\nu}$ may help reduce the number of diagrams and the
amount of work needed in manipulating the momentum dot products from the tensor
contraction of graviton vertices in fourier space. The reduction of diagrams at
2 PN was recently demonstrated in the 2 body case computed by Gilmore and Ross
\cite{GilmoreRoss}. They used the full de Donder gauge\footnote{This will
modify the $N$-graviton interaction in the Einstein-Hilbert action to all
orders in $h_{\mu\nu}$.} $S_{\text{gf}} = \intx{d}{x} \sqrt{g} g^{\alpha\beta}
g^{\mu\nu} g^{\sigma\rho} \Gamma_{\sigma\mu\nu} \Gamma_{\rho\alpha\beta}$ and
the Kaluza-Klein parametrization for $h_{\mu\nu}$, first introduced to the PN
problem by Kol and Smolkin \cite{KolSmolkin}.\footnote{Comparing the calculus
involved, however, when using the Kol-Smolkin parametrization for $h_{\mu\nu}$,
one notes that Gilmore and Ross \cite{GilmoreRoss}, for the 2 body problem,
encountered the same master integrals as the ones used in this paper,
(\ref{integral_PropI}) and (\ref{integral_OneLoopRd}). Moreover, for the
$n$-body case, the most difficult integrals at 2 PN, arising from the
contraction of two 3-graviton vertices, namely
$I_{0000-0000}^{\text{III}}[q,r,s,u]$ in (\ref{masterintegralI0000-0000_III}),
are identical in form to the ones in their equation (52), keeping all particles
distinct.} Some other possible choices include the ADM variables normally
associated with the (3+1)-decomposition of the spacetime metric. Yet another
possibility is to employ the gravitational lagrangian constructed by Bern and
Grant \cite{BernGrant_EH=YMYM} using quantum chromodynamics gluon amplitudes,
up to the 5-graviton interaction; this is sufficient, however, only to 3 PN.

We end with a cautionary remark against taking superposition too literally
within the post-Newtonian framework. By using the 1 PN lagrangian
(\ref{1PNLeff}) in 3 spatial dimensions, and taking the continuum limit, the
force $\vec{F}$ experienced by a stationary point mass $M_x$ located at
$\vec{r}$ away from the center of a static, spherical, hollow shell of surface
mass density $\sigma$ and coordinate radius $R$ can be shown to
be\footnote{This computation arose out of discussions on Birkhoff's theorem in
GR with Dai De-Chang and Glenn Starkman. The following formula can also be
found in their recent paper with Matsuo \cite{FDivergesAtShell}. Its derivation
actually only requires the 2 body portion of the $L_{\text{eff}}$, because the
3-body portion integrates to a constant.}

\begin{align*}
\vec{F} &\equiv M_x \frac{\dd^2 \vec{r}}{\dx{t}^2} \nonumber \\
&= -M_x^2 \frac{\pi R G_{\text{N}}^2 \sigma}{1 + 12 \pi R G_{\text{N}} \sigma}
\frac{\partial}{\partial \vec{r}} \left( \frac{1}{|\vec{r}|} \ln\left\vert
\frac{|\vec{r}|+R}{|\vec{r}|-R} \right\vert \right),
\end{align*}

which evidently diverges as the point mass approaches the surface of the shell.
Because the force would vanish if gravity were purely Newtonian, such a result
for a first order calculation most likely indicates the breakdown of
perturbation theory in this regime, since the post-Newtonian lagrangian was
derived with an implicit assumption that the point masses involved were well
separated, i.e. $r_{\text{s}} \ll r$.

\section{Acknowledgements}

I thank Tanmay Vachaspati for encouraging me to complete this work and for
comments on the draft. I would like to thank Luc Blanchet, Alessandra Buonanno
and Harsh Mathur for their help and discussions. I also wish to thank Umberto
Cannella for bringing to my attention several typographical errors.

Most of the calculations were done using \mma \ \cite{mathematica}, often in
conjunction with the package {\sf FeynCalc} \cite{feyncalc}. All Feynman
diagrams were drawn with {\sf JaxoDraw} \cite{JaxoDraw}. This work was
supported in part by the U.S. Department of Energy and FQXi at Case Western
Reserve University.

\appendix

\section{The $N$--graviton Feynman Rule}

Here we outline an algorithm that can be implemented on symbolic and tensor
manipulation software such as {\sf Mathematica} \cite{mathematica} and the
package {\sf FeynCalc} \cite{feyncalc}, to generate the Feynman rule for the $N
\geq 2$ graviton vertex in Minkowski space.

Given a product of two function(al)s $f[h] g[h]$ the term that contains exactly
$n$ powers of $h$ is a discrete convolution

\begin{align*}
(f \cdot g | n)[h] = \sum_{m=0}^{n} (f|m)[h] (g|n-m)[h]
\end{align*}

where $(A|m)[h]$ denotes the term in $A$ that contains exactly $m$ powers of
$h$. Here we also assume that both $f$ and $g$ can be developed as power series
expansions starting from the zeroth power in $h$.

With this observation, the term in the Einstein-Hilbert lagrangian containing
exactly $n$ powers of the graviton field is given by

\begin{align}
\label{ngravitonEH} - 2 \mpl^{d-2} \sum_{m=0}^{n} \intx{d}{x} \left( \left.
\sqrt{|g|} \right\vert m \right) \left( R \vert n - m \right)
\end{align}

where\footnote{We are absorbing the $\mpl^{1-(d/2)}$ into the $h_{\mu\nu}$ to
save clutter. The full $N$ graviton rule would therefore be multiplied by a
factor of $\mpl^{N(1-(d/2))}$; for instance, the 2-graviton ``vertex" would
contain no $\mpl$.}

{\allowdisplaybreaks \label{ngravitonformulas}
\begin{align}
g_{\mu \nu} &\equiv \eta_{\mu \nu} + h_{\mu \nu} \\
\left( \left. \sqrt{|g|} \right\vert n \right) &= \frac{1}{n!} \left. \frac{\text{d}^n}{\dx{\epsilon}^n} \right\vert_{\epsilon = 0} \nonumber \\
& \quad \text{exp} \left[ \frac{1}{2} \sum_{s=1}^{n} \left\{ \frac{(-)^{s+1} \epsilon^s }{s} \text{Tr} \left( \eta^{-1} h \right)^s \right\} \right] \nonumber \\
\text{Tr} \left( \eta^{-1} h \right)^s &= h^{\nu_s}_{\phantom{\nu_s} \nu_1} h^{\nu_1}_{\phantom{\nu_1} \nu_2} \dots h^{\nu_{s-1}}_{\phantom{\nu_{s-1}} \nu_s} \nonumber \\
(\mathcal{R} | n) &= \sum_{m=0}^{n} (g|m)^{\beta \nu} (R|n-m)^{\alpha}_{\phantom{\alpha}\beta \alpha \nu} \nonumber \\
(g | n)^{\alpha \beta} &= \left\{
\begin{array}{ll} \eta^{\alpha \beta} & \text{if $n = 0$} \\
(-)^n h^{\alpha}_{\phantom{\alpha} \mu_1} h^{\mu_1}_{\phantom{\alpha} \mu_2}
\dots h^{\mu_{n-1} \beta} & \text{if $n > 0$} \nonumber
\end{array} \right. \nonumber \\
(\Gamma | n )^{\alpha}_{\phantom{\alpha} \mu \nu} &= \frac{1}{2} (g|n-1)^{\alpha \lambda} \left( \partial_{\mu} h_{\nu \lambda} + \partial_{\nu} h_{\mu \lambda} - \partial_{\lambda} h_{\mu \nu} \right) \nonumber \\
(R|n)^{\alpha}_{\phantom{\alpha}\beta \mu \nu} &= \partial_{\mu} (\Gamma | n )^{\alpha}_{\phantom{\alpha} \beta \nu} - ( \mu \leftrightarrow \nu ) \nonumber \\
&\quad + \sum_{m = 0}^n (\Gamma | m )^{\alpha}_{\phantom{\alpha} \mu \lambda}
(\Gamma | n - m )^{\lambda}_{\phantom{\lambda} \beta \nu} - ( \mu
\leftrightarrow \nu ) \nonumber
\end{align}}

Note that the action for GR contains a total $d$-derivative term $-2 \mpl^{d-2}
\intx{d}{x} \ \eta^{\mu\nu} \left( \partial_{\alpha} \Gamma^{\alpha}_{\mu \nu}
- \partial_{\mu} \Gamma^{\alpha}_{\nu \alpha} \right)$, which needs to be
discarded when deriving the Feynman rules. Furthermore, these Feynman rules
will also be modified accordingly when a gauge fixing term is added.

Because the graviton field is symmetric in its indices, to obtain the $N$
graviton vertex with external indices $\{ \alpha_1, \beta_1 \}, \dots, \{
\alpha_N, \beta_N \}$ given the action (\ref{ngravitonEH}) containing exactly
$N$ powers of the graviton field, we first choose one particular set of
contractions between the graviton fields in (\ref{ngravitonEH}) with the $N$
external ones. We then replace each field $h_{\mu \nu}$ in (\ref{ngravitonEH})
with the identity tensor $\mathbb{I}^{\alpha_\ell
\beta_\ell}_{\phantom{\alpha_\ell \beta_\ell}\mu\nu}$ carrying the indices
$\{\alpha_\ell,\beta_\ell\}$ that correspond to those on the $\ell$th external
field $h^{\text{ext}}_{\alpha_\ell \beta_\ell}$ it is being contracted with.
The identity tensor reads

\begin{align*}
\mathbb{I}^{\alpha_\ell \beta_\ell}_{\phantom{\alpha_\ell \beta_\ell}\mu\nu}
&\equiv \frac{1}{2} \left( \delta^{\alpha_\ell}_{\phantom{\alpha_\ell}\mu}
\delta^{\beta_\ell}_{\phantom{\beta_\ell}\nu} +
\delta^{\alpha_\ell}_{\phantom{\alpha_\ell}\nu}
\delta^{\beta_\ell}_{\phantom{\beta_\ell}\mu} \right)
\end{align*}

In momentum $(k-)$space, if we define the direction of momentum to be always
flowing into the $N$ graviton vertex, we would also replace partial derivatives
occurring in (\ref{ngravitonEH}) using the prescription

\begin{align*}
\partial_{\lambda} h_{\mu\nu} \to i k_{\lambda_\ell} \mathbb{I}^{\alpha_\ell
\beta_\ell}_{\phantom{\alpha_\ell \beta_\ell}\mu\nu},
\end{align*}

where $\lambda_\ell$ is the $\lambda$th component of the $d$-vector $k$ of the
$\ell$th external graviton $h_{\alpha_\ell \beta_\ell}$ that is contracted with
$h_{\mu\nu}$.\footnote{The sign in front of the momentum vector, i.e. $-i
k_{\lambda_\ell} \dots$ vs. $+ i k_{\lambda_\ell} \dots$, is actually
immaterial because every term in the Einstein-Hilbert action contains two
derivatives; what is important is to maintain a consistent sign convention for
the arguments of the exponentials, either $\exp[i p^0 x^0 - i \vec{p} \cdot
\vec{x}]$ or $\exp[-i p^0 x^0 + i \vec{p} \cdot \vec{x}]$, in the fourier
transforms.}

The complete Feynman rule for the $N$ graviton vertex would be found by summing
up the results from the above procedure for all the $N!$ possible permutations
of the external indices. An example featuring the 3-graviton Feynman rule can
be found in appendix B of \cite{nrgr}.

\section{Integrals}
\label{integralssection}

In this section we review the techniques involved in performing the Feynman
integrals encountered in the $n$ body problem at 2 PN.\footnote{A comprehensive
textbook on evaluating Feynman integrals is Smirnov \cite{SmirnovBook}.}

The starting point is the observation (usually attributed to Schwinger) that
one may employ the integral representation of the Gamma function,
$\Gamma[z]/b^z = \int_0^\infty t^{z-1} e^{-bt} \dx{t}$, for Re$[b]>0$, to first
give us the formula for combining multiple denominators,

\begin{align}
\label{feynmanparameters} \frac{1}{A_1^{\sigma_1} \dots A_N^{\sigma_N}} &=
\left( \prod_{s=1}^N
\frac{1}{\Gamma[\sigma_s]} \int_0^1 \text{d}\alpha_s \ \alpha_s^{\sigma_s-1} \right) \nonumber \\
&\times \frac{\Gamma\left[ \sum_{r=1}^N \sigma_r \right] \delta\left[ 1 -
\sum_{r=1}^N \alpha_r \right]}{\left( \sum_{r=1}^N \alpha_r A_r
\right)^{\sigma_1+\dots+\sigma_N}},
\end{align}

and second, together with the gaussian integral $\int \exp[-x^2] \dx{x} =
\pi^{1/2}$, to further yield

\begin{align}
\label{integral_Delta} \int \frac{\text{d}^\lambda z}{(\vec{z}^2 +
\Delta)^\sigma} &= \frac{\pi^{\lambda/2} \Gamma\left[ \sigma -
\frac{\lambda}{2} \right]}{\Gamma[\sigma] \Delta^{\sigma - \frac{\lambda}{2}}},
\end{align}

and

\begin{align}
\label{integral_PropI} \int \frac{\text{d}^\lambda p}{(2 \pi)^\lambda}
\frac{e^{i \vec{p} \cdot \vec{x}}}{[\vec{p}^2]^\sigma} &= \frac{\Gamma\left[
\frac{\lambda}{2} - \sigma\right]}{4^\sigma \pi^{\lambda/2} \Gamma[\sigma]
|\vec{x}|^{\lambda-2\sigma}}, & \text{for $\vec{x} \neq \vec{0}$} \nonumber \\
&= 0, & \text{for $\vec{x} = \vec{0}$},
\end{align}

where the second equality is to be understood within the framework of
dimensional regularization.

An important corollary of (\ref{feynmanparameters}) and (\ref{integral_Delta})
is

\begin{align}
\label{integral_OneLoopRd} &\int \frac{\text{d}^\lambda z}{
[\vec{z}^2]^{\sigma_1} [(\vec{z}-\vec{x})^2]^{\sigma_2} } \nonumber \\
&= \frac{ \pi^{\lambda/2} \Gamma[\frac{\lambda}{2}-\sigma_1]
\Gamma[\frac{\lambda}{2}-\sigma_2] \Gamma[\sigma_1+\sigma_2-\frac{\lambda}{2}]
}{ \Gamma[\sigma_1]\Gamma[\sigma_2] \Gamma[\lambda-\sigma_1-\sigma_2]
[\vec{x}^2]^{\sigma_1+\sigma_2-\frac{\lambda}{2}}}.
\end{align}

By considering single and double spatial derivatives of
(\ref{integral_OneLoopRd}), we may also obtain the formulas:

{\allowdisplaybreaks
\begin{align}
\label{integral_OneLoopRvmu} &\intx{\lambda}{z}
\frac{(x-z)^i}{|\vec{z}|^{\rho_1}|\vec{x}-\vec{z}|^{\rho_2}} \\
&= \frac{ \pi^{\lambda/2} \Gamma\left[ \frac{\lambda-\rho_1}{2} \right]
\Gamma\left[ \frac{\lambda-\rho_2+2}{2} \right] \Gamma\left[
\frac{\rho_1+\rho_2-\lambda}{2} \right] }{ \Gamma\left[ \frac{\rho_1}{2}
\right] \Gamma\left[ \frac{\rho_2}{2} \right] \Gamma \left[ \lambda-\frac{
\rho_1+\rho_2-2}{2} \right] }
\frac{x^i}{|\vec{x}|^{\rho_1+\rho_2-\lambda}} \nonumber \\
\label{integral_OneLoopRvmunu} &\intx{\lambda}{z} \frac{(x - z)^i (x -
z)^j}{|\vec{z}|^{\tau_1} |\vec{x} - \vec{z}|^{\tau_2}} \\
&= \frac{\pi^{\lambda/2} \Gamma\left[ \frac{\lambda-\tau_1}{2} \right]
\Gamma\left[ \frac{\tau_1+\tau_2-2-\lambda}{2} \right] \Gamma\left[
\frac{\lambda - \tau_2 + 2}{2} \right]}{2 \Gamma\left[ \frac{\tau_1}{2} \right]
\Gamma\left[ \frac{\tau_2}{2} \right] \Gamma \left[ \lambda-\frac{
\tau_1+\tau_2-2}{2} \right] (2(\lambda+1) -(\tau_1+\tau_2))} \nonumber \\
&\times \bigg( (\lambda - \tau_2 + 2)(\tau_1+\tau_2-2-\lambda) \frac{x^i
x^j}{|\vec{x}|^{\tau_1+\tau_2-\lambda}} \nonumber \\
&\qquad + \delta^{ij}
\frac{\lambda-\tau_1}{|\vec{x}|^{\tau_1+\tau_2-2-\lambda}} \bigg) \nonumber
\\
\label{integral_OneLoopRvmunuAB} &\intx{\lambda}{z} \frac{(x_a - z)^i (x_b -
z)^j}{|\vec{z} - \vec{x}_a|^{\rho_a} |\vec{z} - \vec{x}_b|^{\rho_b}} \\
&= \frac{ \pi^{\lambda/2} \Gamma\left[ \frac{\lambda- \rho_a + 2}{2} \right]
\Gamma\left[ \frac{\lambda-\rho_b+2}{2} \right] \Gamma\left[ \frac{\rho_a -
2+\rho_b-\lambda}{2} \right] }{2 \Gamma\left[ \frac{\rho_a}{2} \right]
\Gamma\left[ \frac{\rho_b}{2} \right] \Gamma \left[ \lambda-\frac{ \rho_a +
\rho_b - 4}{2} \right] } \nonumber \\
&\times \bigg( \frac{\delta^{ij}}{|\vec{x}_b - \vec{x}_a|^{\rho_a + \rho_b - 2
- \lambda}} \nonumber \\
&\qquad - \left( \rho_a + \rho_b - \lambda - 2 \right) \frac{(x_a - x_b)^i (x_a
- x_b)^j}{|\vec{x}_b - \vec{x}_a|^{\rho_a + \rho_b - \lambda}} \bigg) \nonumber
\end{align}}

The last formula (\ref{integral_OneLoopRvmunuAB}) has been derived from
(\ref{integral_OneLoopRd}) by re-defining $\vec{x} \equiv \vec{x}_a -
\vec{x}_b$ and shifting integration variables before performing the appropriate
derivatives.

{\bf N point integrals} \quad We next review the evaluation of the ``$N$-point"
integrals first carried out by Boos and Davydychev
\cite{BoosDavydychev3NointMasslessVertexType,
DavydychevNPointMasslessVertexType}.

\begin{align}
\label{npoint} I_{1 \dots N} \equiv \int_{\mathbb{R}^\lambda}
\frac{\text{d}^\lambda z}{\prod_{s=1}^N |(\vec{x}_s-\vec{z})^2|^{\rho_s}}.
\end{align}

These integrals can be viewed as the higher $N$ generalizations of the $N=2$
case in (\ref{integral_OneLoopRd}).

Applying (\ref{feynmanparameters}) transforms it into

\begin{align*}
I_{1 \dots N} &= \int_{\mathbb{R}^\lambda} \text{d}^\lambda z \left(
\prod_{s=1}^N
\frac{1}{\Gamma[\rho_s]} \int_0^1 \text{d}\alpha_s \ \alpha_s^{\rho_s-1} \right) \\
&\times \frac{\Gamma[\rho_1+\dots+\rho_N] \delta\left[1 - \sum_{r=1}^N \alpha_r
\right]}{\left( \vec{z}^2 + \sum_{r,s=1}^N \left( \alpha_r \alpha_s \vec{x}_s^2
- \alpha_r \alpha_s \vec{x}_r \cdot \vec{x}_s \right)
\right)^{\rho_1+\dots+\rho_N}},
\end{align*}

where the constraint $\sum_r \alpha_r = 1$ as well as a shift in the variable
$\vec{z}$ have been employed. Writing

\begin{align*}
\sum_{r,s=1}^N \left( \alpha_r \alpha_s \vec{x}_s^2 - \alpha_r \alpha_s
\vec{x}_r \cdot \vec{x}_s \right) &= \frac{1}{2} \sum_{r,s=1}^N \alpha_r
R_{rs}^2 \alpha_s, \\
R_{rs} &\equiv |\vec{x}_r - \vec{x}_s|
\end{align*}

and recalling (\ref{integral_Delta}) then allow us to deduce

\begin{align}
\label{NPointMasslessVertexPartI} I_{1 \dots N} &= \pi^{\lambda/2}
\frac{\Gamma\left[\sum_r \rho_r-\frac{\lambda}{2}\right]}{\prod_r
\Gamma[\rho_r]} \left( \prod_{s=1}^N
\int_0^\infty \text{d}\alpha_s \alpha_s^{\rho_s-1} \right) \nonumber \\
& \quad \times \frac{\delta[1 - \sum_r \alpha_r]}{[\frac{1}{2}\sum_p \sum_q
\alpha_p \alpha_q R_{pq}^2]^{\sum_r \rho_r - \frac{\lambda}{2}}}
\end{align}

By viewing $\alpha_p R_{pq} \alpha_q$ as components of a $N \times N$ symmetric
matrix with zeros on the diagonal, one notes that there are $N(N-1)/2$ distinct
terms in the sum in the denominator of (\ref{NPointMasslessVertexPartI}). To
make further progress, one needs to iterate $(L-2)$ times the Mellin-Barnes
(MB) integral representation

\begin{align*}
\frac{1}{(X+Y)^\tau} &= \frac{1}{\Gamma[\tau]} \frac{1}{2\pi i}
\int_{-i\infty}^{i\infty} \dx{s} \ \Gamma[s+\tau] \Gamma[-s]
\frac{Y^s}{X^{s+\tau}},
\end{align*}

to obtain the $(L-1)$-fold MB representation for the sum of $L$ terms in a
denominator raised to some power,

\begin{align}
\label{LfoldMB} \frac{1}{[ \sum_{a=1}^L u_a ]^\tau} &= \frac{1}{\Gamma[\tau]
u_L^\tau}
\frac{1}{(2\pi i)^{L-1}} \nonumber \\
&\times \left( \prod_{a=1}^{L-1} \int_{-i\infty}^{i\infty} \dx{s}_a
\Gamma[-s_a]
\left( \frac{u_a}{u_L} \right)^{s_a} \right) \nonumber \\
&\times \Gamma\left[s_1 + s_2 + \dots + s_{L-1} +\tau\right]
\end{align}

In these integrals, the contour for the $i$th variable $s_i$ is chosen such
that the poles of the Gamma functions of the form $\Gamma[\dots - s_i]$ lie to
the right and those of the Gamma functions of the form $\Gamma[\dots + s_i]$
lie to the left.

Using (\ref{LfoldMB}) on (\ref{NPointMasslessVertexPartI}), followed by
(\ref{feynmanparameters}) with $A_1 = A_2 = \dots = A_{N(N-1)/2} = 1$ and some
careful algebraic reasoning, one arrives at the final form of the MB
representation for the $N$-point integral,

{\allowdisplaybreaks
\begin{widetext}
\begin{align}
\label{NPointMasslessVertexIntegral} \int_{\mathbb{R}^\lambda}
\frac{\text{d}^\lambda z}{\prod_{s=1}^N |(\vec{z}-\vec{x}_s)^2|^{\rho_s}} &=
\frac{\pi^{\lambda/2} [R^2_{i'j'}]^{\frac{\lambda}{2}-\sum_r \rho_r}
}{\Gamma[\lambda - \sum_r \rho_r] \prod_{r=1}^N \Gamma[\rho_r]} \frac{1}{(2\pi
i)^{(N/2)(N-1)-1}} \prod_{\substack{\{i,j\} \in \text{upper $\Delta$} \nonumber \\
\text{of $N \times N$ matrix} \\ \{i,j\} \neq \{i',j'\}}}^{(N/2)(N-1)-1} \left(
\int_{-i\infty}^{i\infty} \dx{s}_{ij} \Gamma[-s_{ij}]  \left(
\frac{R^2_{ij}}{R^2_{i'j'}} \right)^{s_{ij}} \right) \nonumber \\
&\times \Gamma\left[\sum_{\substack{\{k,l\} \in \text{upper $\Delta$}
\\ \{k,l\} \neq \{i',j'\}}}^{(N/2)(N-1)-1} s_{kl} + \sum_{r=1}^N \rho_r - \frac{\lambda}{2} \right] \prod_{\substack{1 \leq s \leq N \\ s \neq i'; s \neq
j'}} \Gamma\left[ \rho_s + \sum_{\substack{\{j,k\} \in s^{\rm th} \ \text{row and } s^{\rm th} \ \rm{column} \\ \text{upper }\Delta}} s_{jk} \right] \nonumber \\
&\times \Gamma\left[ \frac{\lambda}{2} - \sum_{r=1}^N \rho_r + \rho_{i'} -
\sum_{\substack{\{k,l\}\in\text{upper }\Delta
\\ k \neq i'; l \neq i'}} s_{kl} \right] \Gamma\left[ \frac{\lambda}{2} - \sum_{r=1}^N \rho_r + \rho_{j'} -
\sum_{\substack{\{k,l\}\in\text{upper }\Delta
\\ k \neq j'; l \neq j'}} s_{kl} \right]
\end{align}
\end{widetext}}

Here, $\{i',j'\}$ is some fixed pair of numbers chosen from the ``upper
triangular" portion of the $N \times N$ matrix of number pairs corresponding to
their coordinates on the matrix; namely, the first row on the ``upper $\Delta$"
reads from left to right, $\{1,2\}, \{1,3\},\dots,\{1,N\}$, the second reads
$\{2,3\}, \{2,4\},\dots,\{2,N\}$, and so on until the $(N-1)$th row, which has
only one element, $\{N-1,N\}$.

\textsl{N = 3} \quad The MB integrals for the $N=3$ case have been explicitly
evaluated by Boos and Davydychev \cite{BoosDavydychev3NointMasslessVertexType}.
One has from (\ref{NPointMasslessVertexIntegral}),

\begin{widetext}
\begin{align}
\label{3PointMasslessVertexIntegralMB} I_{123} &= \frac{\pi^{\lambda/2}
[R_{23}^2]^{\frac{\lambda}{2}-\sum_r\rho_r}}{\Gamma[\lambda - \sum_r \rho_r]
\prod_r \Gamma[\rho_r]} \frac{1}{(2\pi i)^2} \int_{-i\infty}^{i\infty} \dx{u}
\int_{-i\infty}^{i\infty} \dx{v} \Gamma[-u] \Gamma[-v] \left( \frac{R^2_{12}}{R_{23}^2} \right)^u \left( \frac{R^2_{13}}{R_{23}^2} \right)^v \nonumber \\
&\times \Gamma\left[u+v + \sum_r \rho_r - \frac{\lambda}{2} \right]
\Gamma\left[ \rho_1 + u + v \right] \Gamma\left[ \frac{\lambda}{2} - \rho_1 -
\rho_3 - v \right] \Gamma\left[ \frac{\lambda}{2} - \rho_1 - \rho_2 - u \right]
\end{align}
\end{widetext}

Assuming there exists a series representation of the integral $I_{123}$ in
powers of $R_{12}$ and $R_{13}$, we close the $u$- and $v$-contours on the
right, turning each integral into an infinite sum over its residues by noting
that $\Gamma[-z]$ has singularities on the complex $z$ plane only in the form
of simple poles at zero and the positive integers. (Whether one should close
the contour to the left or to the right really depends on the numerical range
of $R_{12}$ and $R_{13}$ considered. For instance, the MB representation
$(1+z)^{-\lambda} = (2\pi i\Gamma[\lambda])^{-1}
\int_{-i\infty}^{+i\infty}\dx{u} \Gamma[u+\lambda]\Gamma[-u]z^u$ can be
converted into a power series in $1/z$ or $z$, for $\lambda
> 0$, by closing the contour to the left or right, depending on whether $|z|>1$ or $|z|<1$. Here we
will simply assume, for each choice (left or right), there is some region of
$R_{12}, R_{13} \in \mathbb{R}$ in which it is valid.) The residues of
$\Gamma[-z]$ at these locations are

\begin{align*}
\left. {\rm Res} \ \Gamma[-z] \right\vert_{z = m} = \frac{(-)^m}{m!}, \quad
m=0,+1,+2,\dots
\end{align*}

Because there are 2 Gamma functions of the form $\Gamma[\dots-u]$ and 2 of the
form $\Gamma[\dots-v]$, this converts the 2-fold MB integrals into a 2-fold
infinite sum of 4 terms. One can proceed to change summation variables and
manipulate the Gamma functions in the summands using the definitions for the
Gauss hypergeometric function and the Pochhammer symbol

\begin{align*}
_2F_1[a,b;c;z] &= \sum_{u = 0}^\infty \frac{(a)_u (b)_u}{u! (c)_u} z^u \\
(a)_u &\equiv a(a+1) \dots (a+(u-1)) = \frac{\Gamma[a+u]}{\Gamma[a]}
\end{align*}

and the relation

\begin{align*}
\Gamma[\tau-m] &= \frac{(-)^m \Gamma[\tau]}{(1-\tau)_m}, \ m \in \mathbb{Z},
\tau \in \mathbb{C}
\end{align*}

to further reduce the 2-fold sum into a single sum:

{\allowdisplaybreaks
\begin{widetext}
\begin{align}
\label{3PointMasslessVertexIntegralAsSum} I_{abc}[\rho_a, \rho_b, \rho_c]
&\equiv \int \frac{\text{d}^\lambda z}{[(\vec{z}-\vec{x}_a)^2]^{\rho_a}
[(\vec{z}-\vec{x}_b)^2]^{\rho_b}
[(\vec{z}-\vec{x}_c)^2]^{\rho_c}} \nonumber \\
&= \frac{\pi^{\lambda/2}}{\Gamma[\lambda - \sum_r \rho_r] \prod_r \Gamma[\rho_r]} \nonumber \\
&\times \sum_{\ell = 0}^\infty \bigg\{ [R_{bc}^2]^{\frac{\lambda}{2}-\sum_r
\rho_r} \frac{1}{\ell!} \left( \frac{R_{ab}^2}{R_{bc}^2} \right)^\ell
\Gamma\left[\frac{\lambda}{2} - \rho_a - \rho_b \right]
\Gamma\left[\frac{\lambda}{2} - \rho_a - \rho_c \right] \Gamma\left[ \sum_r \rho_r - \frac{\lambda}{2} \right] \Gamma\left[ \rho_a \right] \nonumber \\
&\qquad \times \frac{\left(\sum_r \rho_r - \frac{\lambda}{2}\right)_\ell \left(
\rho_a \right)_\ell}{\left(1-\frac{\lambda}{2} + \rho_a + \rho_b \right)_\ell}
\ _2F_1\left[ -\ell, \frac{\lambda}{2} - \rho_a - \rho_b - \ell;
1- \frac{\lambda}{2} + \rho_a + \rho_c; \frac{R_{ac}^2}{R_{ab}^2} \right] \nonumber \\
&+ [R_{bc}^2]^{- \rho_c} \left[ R_{ab}^2
\right]^{\frac{\lambda}{2}-\rho_a-\rho_b} \frac{1}{\ell !} \left(
\frac{R_{ab}^2}{R_{bc}^2} \right)^\ell \Gamma\left[\rho_a + \rho_b -
\frac{\lambda}{2}\right] \Gamma\left[ \frac{\lambda}{2} - \rho_a - \rho_c
\right] \Gamma\left[\rho_c\right] \Gamma\left[
\frac{\lambda}{2} - \rho_b \right] \nonumber \\
&\qquad \times \frac{\left(\rho_c\right)_\ell \left( \frac{\lambda}{2} - \rho_b
\right)_\ell}{(1 + \frac{\lambda}{2} - \rho_a - \rho_b)_\ell} \ _2F_1\left[
-\ell, \rho_a + \rho_b -\frac{\lambda}{2} - \ell; 1
- \frac{\lambda}{2} + \rho_a + \rho_c; \frac{R_{ac}^2}{R_{ab}^2} \right] \nonumber \\
&+ [R_{bc}^2]^{-\rho_b} \left[ R_{ac}^2 \right]^{\frac{\lambda}{2} - \rho_a -
\rho_c} \frac{1}{\ell !} \left( \frac{R_{ab}^2}{R_{bc}^2} \right)^\ell
\Gamma\left[\frac{\lambda}{2} - \rho_a - \rho_b \right] \Gamma\left[
-\frac{\lambda}{2} + \rho_a + \rho_c \right] \Gamma\left[ \rho_b \right]
\Gamma\left[
\frac{\lambda}{2} - \rho_c \right] \nonumber \\
&\qquad \times \frac{\left( \rho_b \right)_\ell \left( \frac{\lambda}{2} -
\rho_c \right)_\ell}{(1-\frac{\lambda}{2} + \rho_a + \rho_b )_\ell} \
_2F_1\left[ -\ell, \frac{\lambda}{2} - \rho_a - \rho_b - \ell;
1 + \frac{\lambda}{2} - \rho_a - \rho_c; \frac{R_{ac}^2}{R_{ab}^2} \right] \nonumber \\
&+ [R_{bc}^2]^{\rho_a - \frac{\lambda}{2}} \left[ R_{ab}^2
\right]^{\frac{\lambda}{2} - \rho_a - \rho_b} \left[ R_{ac}^2
\right]^{\frac{\lambda}{2} - \rho_a - \rho_c} \frac{1}{\ell !} \left(
\frac{R_{ab}^2}{R_{bc}^2} \right)^\ell \nonumber \\
&\qquad \times \Gamma\left[ -\frac{\lambda}{2} + \rho_a + \rho_b \right]
\Gamma\left[ -\frac{\lambda}{2} + \rho_a + \rho_c \right] \Gamma\left[
\frac{\lambda}{2} - \rho_a \right]
\Gamma\left[ \lambda - \sum_r \rho_r \right] \nonumber \\
&\qquad \times \frac{\left( \frac{\lambda}{2} - \rho_a \right)_\ell \left(
\lambda - \sum_r \rho_r \right)_\ell}{(1 + \frac{\lambda}{2} - \rho_a - \rho_b
)_\ell} \ _2F_1\left[ -\ell, -\frac{\lambda}{2} + \rho_a + \rho_b - \ell;
1+\frac{\lambda}{2} - \rho_a - \rho_c; \frac{R_{ac}^2}{R_{ab}^2} \right]
\bigg\}
\end{align}
\end{widetext}}

As described in Boos and Davydychev
\cite{BoosDavydychev3NointMasslessVertexType}, this sum has a closed form
expression in terms of the Appell hypergeometric function $F_4$ of two
variables, which has a perturbative definition of the form

\begin{align*}
F_4\left[ \alpha, \beta; \gamma, \delta; x, y \right] &= \sum_{m=0}^\infty
\sum_{n=0}^\infty \frac{(\alpha)_{m+n} (\beta)_{m+n}}{(\gamma)_m (\delta)_n}
\frac{x^m}{m!} \frac{y^n}{n!}.
\end{align*}

Through the relation

\begin{align*}
\sum_{j=0}^\infty \frac{x^j}{j!} \frac{(\alpha)_j (\beta)_j}{(\gamma)_j} & \
_2F_1\left[ -j, 1-\gamma-j; \delta; y \right] \\
&= F_4\left[ \alpha, \beta; \gamma, \delta; x, xy \right],
\end{align*}

we now have

{\allowdisplaybreaks
\begin{widetext}
\begin{align*}
I_{abc}[\rho_a, \rho_b, \rho_c] &\equiv \int \frac{\text{d}^\lambda
z}{[(\vec{z}-\vec{x}_a)^2]^{\rho_a} [(\vec{z}-\vec{x}_b)^2]^{\rho_b}
[(\vec{z}-\vec{x}_c)^2]^{\rho_c}} \\
&= \frac{\pi^{\lambda/2}}{\Gamma[\lambda - \sum_r \rho_r] \prod_r \Gamma[\rho_r]} \\
&\times \bigg\{ [R_{bc}^2]^{\frac{\lambda}{2}-\sum_r \rho_r}
\Gamma\left[\frac{\lambda}{2} - \rho_a - \rho_b \right]
\Gamma\left[\frac{\lambda}{2} - \rho_a - \rho_c \right] \Gamma\left[ \sum_r \rho_r - \frac{\lambda}{2} \right] \Gamma\left[ \rho_a \right] \\
&\qquad \times F_4\left[\sum_r \rho_r - \frac{\lambda}{2}, \rho_a;
1-\frac{\lambda}{2} + \rho_a +
\rho_b, 1- \frac{\lambda}{2} + \rho_a + \rho_c; \frac{R_{ab}^2}{R_{bc}^2}, \frac{R_{ac}^2}{R_{bc}^2} \right] \\
&+ [R_{bc}^2]^{- \rho_c} \left[ R_{ab}^2
\right]^{\frac{\lambda}{2}-\rho_a-\rho_b} \Gamma\left[\rho_a + \rho_b -
\frac{\lambda}{2}\right] \Gamma\left[ \frac{\lambda}{2} - \rho_a - \rho_c
\right] \Gamma\left[\rho_c\right] \Gamma\left[
\frac{\lambda}{2} - \rho_b \right] \\
&\qquad \times F_4\left[ \rho_c, \frac{\lambda}{2} - \rho_b; 1 +
\frac{\lambda}{2} - \rho_a -
\rho_b, 1 - \frac{\lambda}{2} + \rho_a + \rho_c; \frac{R_{ab}^2}{R_{bc}^2}, \frac{R_{ac}^2}{R_{bc}^2} \right] \\
&+ [R_{bc}^2]^{-\rho_b} \left[ R_{ac}^2 \right]^{\frac{\lambda}{2} - \rho_a -
\rho_c} \Gamma\left[\frac{\lambda}{2} - \rho_a - \rho_b \right] \Gamma\left[
-\frac{\lambda}{2} + \rho_a + \rho_c \right] \Gamma\left[ \rho_b \right]
\Gamma\left[
\frac{\lambda}{2} - \rho_c \right] \\
&\qquad \times F_4\left[ \rho_b, \frac{\lambda}{2} - \rho_c;
1-\frac{\lambda}{2} + \rho_a +
\rho_b, 1 + \frac{\lambda}{2} - \rho_a - \rho_c; \frac{R_{ab}^2}{R_{bc}^2}, \frac{R_{ac}^2}{R_{bc}^2} \right] \\
&+ [R_{bc}^2]^{\rho_a - \frac{\lambda}{2}} \left[ R_{ab}^2
\right]^{\frac{\lambda}{2} - \rho_a - \rho_b} \left[ R_{ac}^2
\right]^{\frac{\lambda}{2} - \rho_a - \rho_c} \Gamma\left[ -\frac{\lambda}{2} +
\rho_a + \rho_b \right] \Gamma\left[ -\frac{\lambda}{2} + \rho_a + \rho_c
\right] \Gamma\left[ \frac{\lambda}{2} - \rho_a \right]
\Gamma\left[ \lambda - \sum_r \rho_r \right] \\
&\qquad \times F_4\left[ \frac{\lambda}{2} - \rho_a, \lambda - \sum_r \rho_r; 1
+ \frac{\lambda}{2} - \rho_a - \rho_b, 1+\frac{\lambda}{2} - \rho_a - \rho_c;
\frac{R_{ab}^2}{R_{bc}^2}, \frac{R_{ac}^2}{R_{bc}^2} \right] \bigg\}
\end{align*}
\end{widetext}}

However, $F_4$ is not defined in {\sf Mathematica} \cite{mathematica}, whereas
the sum in (\ref{3PointMasslessVertexIntegralAsSum}) can easily be entered. In
particular, at 2 PN, the $n$ body problem requires the knowledge of

\begin{align*}
I_3[a,b,c] = \left( \frac{\Gamma\left[ \frac{d-3}{2} \right]}{4
\pi^{\frac{d-1}{2}}} \right)^3 I_{123}\left[ \rho_1 = \rho_2 = \rho_3 =
\frac{d-3}{2} \right].
\end{align*}

Applying, in $3 - 2\varepsilon$ spatial dimensions, the Laurent expansion for
the Gamma function about negative integers or zero,

\begin{align*}
\Gamma[-m + \varepsilon] &= \frac{(-)^m}{m!} \left( \frac{1}{\varepsilon} -
\gamma_{\rm E} + \sum_{r=1}^m \frac{1}{r} + \mathcal{O}[\varepsilon] \right), \\
m &= 0, 1, 2, \dots,
\end{align*}

to the summand in (\ref{3PointMasslessVertexIntegralAsSum}), before employing
the \mma \ command {\sf FullSimplify} on the summation
(\ref{3PointMasslessVertexIntegralAsSum}), yields the final form for
$I_3[a,b,c]$,

\begin{align}
&\label{3PointMasslessVertexIntegralLog} I_3[a,b,c] \nonumber \\
&= \frac{1}{64\pi^2} \bigg( - \frac{1}{\varepsilon} + 2 - 2 \gamma_{\rm E} - 2 \ln[ \pi ] \nonumber \\
&\qquad - 4 \ln \left| R_{ab} + R_{ac} + R_{bc} \right| +
\mathcal{O}[\varepsilon] \bigg)
\end{align}

where $\gamma_{\text{E}} = 0.57721\dots$ is the Euler-Mascheroni constant and
the hyperbolic function identity $\tanh^{-1}[z] = (1/2)(\ln|1+z|-\ln|1-z|)$ was
used. An alternate derivation of this result can be found in Blanchet et al.
\cite{BlanchetEtalIntegral}.

A direct computation would show that this result is consistent with the Poisson
equation obeyed by the $N=3$ integral in 3 spatial dimensions,

\begin{align*}
\delta^{ij} \partial_i^a \partial_j^a I_{123}\left[\rho_a = \rho_b = \rho_c =
1/2 \right] = -4 \pi (R_{ab} R_{ac})^{-1}.
\end{align*}

\section{3 PN Diagrams}
\label{3PNDiagramsSection}

In this section, we collect the fully distinct Feynman diagrams necessary for
the computation of the effective lagrangian for $n$ non-rotating,
structure-less point masses as described by the minimal action in
(\ref{fullaction}) at the 3 PN order. Fully distinct here means that, to obtain
the full 3 PN lagrangian one would have to, whenever applicable:

\begin{itemize}

\item Consider all possible permutations of the particle labels of the diagrams
displayed.

\item For the $n=3,4$ and $5$ diagrams, consider all possible ways of setting some
of the particle labels equal to each other, so that from the $n=3$ diagrams one
would obtain their $n = 2$ counterparts; from the $n=4$ their $n=2$ and $3$
counterparts; and from the $n=5$ their $n=2,3$ and $4$ counterparts.

\end{itemize}

The 2 body diagrams are in Fig. (\ref{3PN2BodyDiagrams}). The 3 body diagrams
with graviton vertices are Fig. (\ref{3PN3BodyDiagramsVtx3IofII},
\ref{3PN3BodyDiagramsVtx3IIofII}); and those with no graviton vertices are Fig.
(\ref{3PN3BodyDiagramsNoVtx3}). The 4 body diagrams with graviton vertices are
Fig. (\ref{3PN4BodyDiagramsIofIII}, \ref{3PN4BodyDiagramsIIofIII},
\ref{3PN4BodyDiagramsIIIofIII}); and those without graviton vertices are Fig.
(\ref{3PN4BodyDiagramsNoVtxIofII}, \ref{3PN4BodyDiagramsNoVtxIIofII}). Finally
the 5 body diagrams with graviton vertices can be found in Fig.
(\ref{3PN5BodyDiagramsVtx}), whereas those with none can be found in Fig.
(\ref{3PN5BodyDiagramsNoVtx}).


\begin{figure*}
\begin{center}
\includegraphics[width=1.5in]{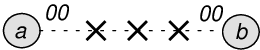} \
\includegraphics[width=1.5in]{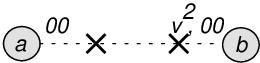} \
\includegraphics[width=1.5in]{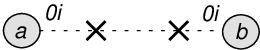} \
\includegraphics[width=1.5in]{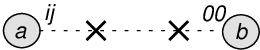} \\
\includegraphics[width=1.5in]{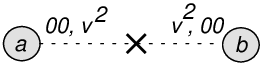} \
\includegraphics[width=1.5in]{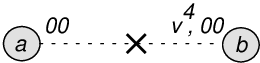} \
\includegraphics[width=1.5in]{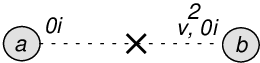} \
\includegraphics[width=1.5in]{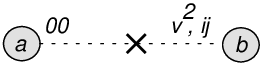} \\
\includegraphics[width=1.5in]{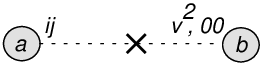} \
\includegraphics[width=1.5in]{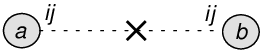} \
\includegraphics[width=1.5in]{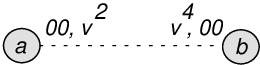} \
\includegraphics[width=1.5in]{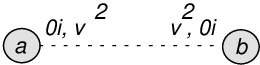} \\
\includegraphics[width=1.5in]{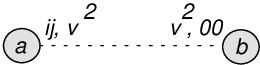} \
\includegraphics[width=1.5in]{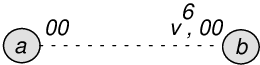} \
\includegraphics[width=1.5in]{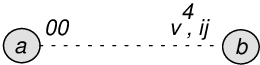} \
\includegraphics[width=1.5in]{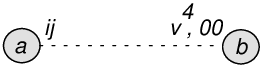} \\
\includegraphics[width=1.5in]{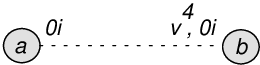} \
\includegraphics[width=1.5in]{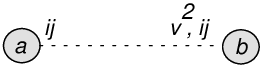}
\end{center}
\caption{3 PN 2 body diagrams.} \label{3PN2BodyDiagrams}
\end{figure*}


\begin{figure*}
\begin{center}
\includegraphics[width=3in]{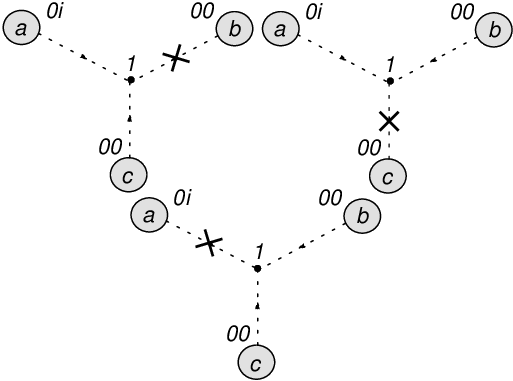}     \
\includegraphics[width=3in]{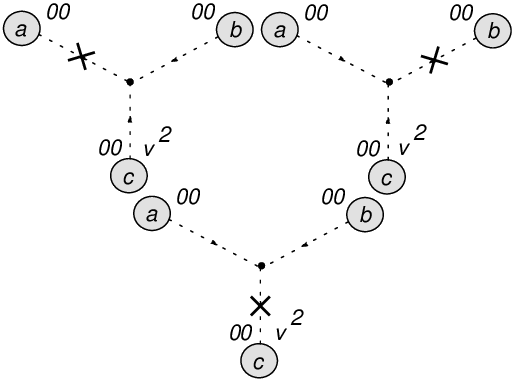}
\end{center}
\caption{3 PN 3 body diagrams containing 3 graviton vertices, 1 of 2.}
\label{3PN3BodyDiagramsVtx3IofII}
\end{figure*}

\begin{figure*}
\begin{center}
{\allowdisplaybreaks
\includegraphics[width=3in]{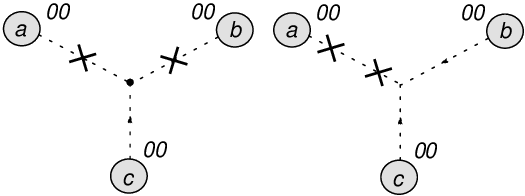}     \
\includegraphics[width=1.5in]{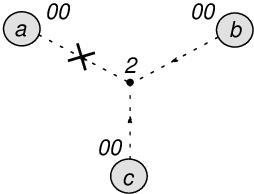}   \
\includegraphics[width=1.5in]{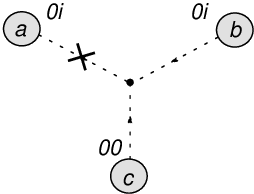}   \\
\includegraphics[width=1.5in]{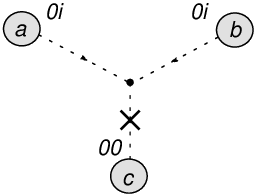}  \
\includegraphics[width=1.5in]{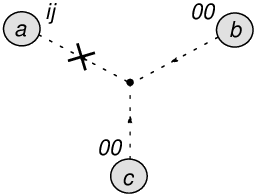}   \
\includegraphics[width=1.5in]{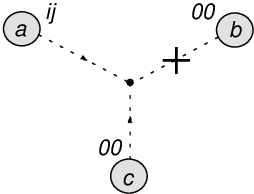}  \
\includegraphics[width=1.5in]{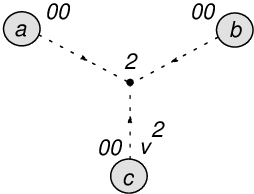}   \\
\includegraphics[width=1.5in]{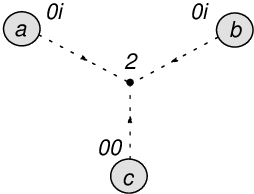}   \
\includegraphics[width=1.5in]{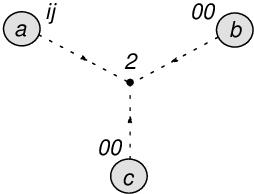}   \
\includegraphics[width=1.5in]{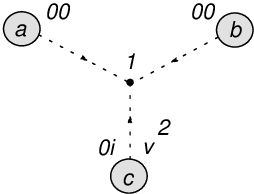}   \
\includegraphics[width=1.5in]{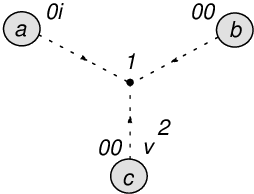}   \\
\includegraphics[width=1.5in]{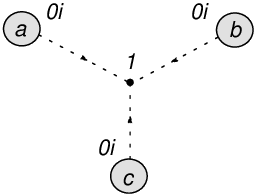}   \
\includegraphics[width=1.5in]{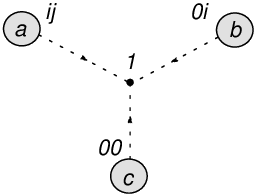}   \
\includegraphics[width=1.5in]{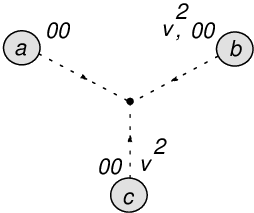}   \
\includegraphics[width=1.5in]{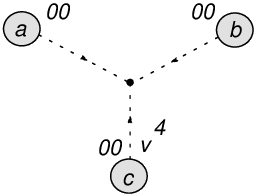}   \\
\includegraphics[width=1.5in]{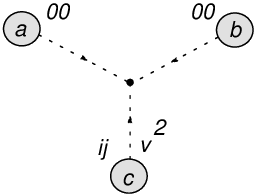}   \
\includegraphics[width=1.5in]{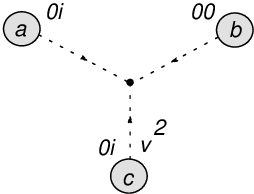}   \
\includegraphics[width=1.5in]{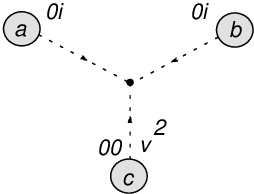}   \
\includegraphics[width=1.5in]{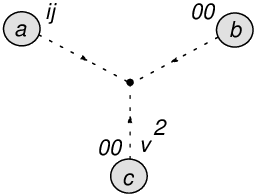}   \\
\includegraphics[width=1.5in]{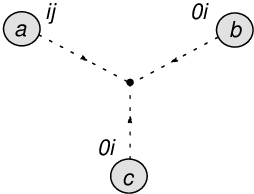}   \
\includegraphics[width=1.5in]{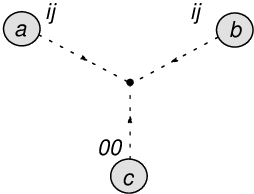}
}
\end{center}
\caption{3 PN 3 body diagrams containing 3 graviton vertices, 2 of 2.}
\label{3PN3BodyDiagramsVtx3IIofII}
\end{figure*}

\begin{figure*}
\begin{center}
\includegraphics[width=2.5in]{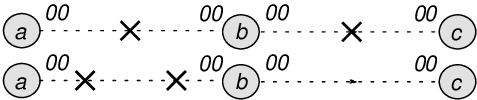} \\
\includegraphics[width=2.5in]{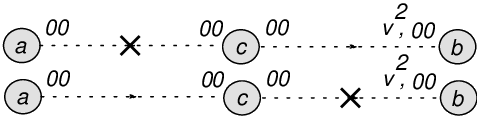} \
\includegraphics[width=2.5in]{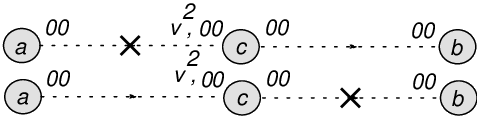} \\
\includegraphics[width=2.5in]{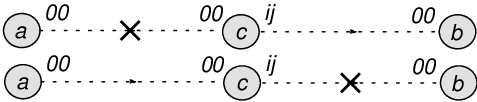} \
\includegraphics[width=2.5in]{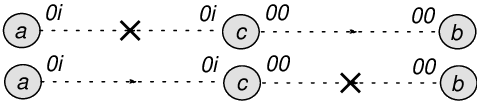} \\
\includegraphics[width=2.5in]{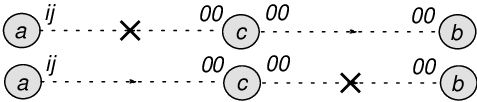} \
\includegraphics[width=2.5in]{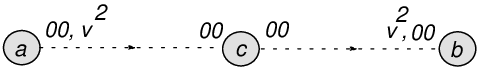} \\
\includegraphics[width=2.5in]{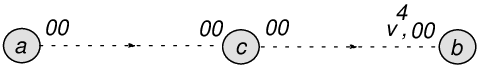} \
\includegraphics[width=2.5in]{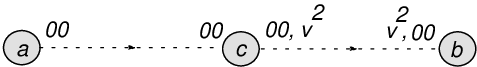} \\
\includegraphics[width=2.5in]{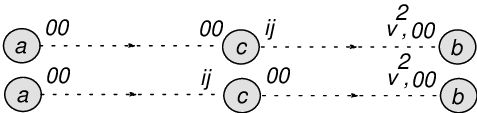} \
\includegraphics[width=2.5in]{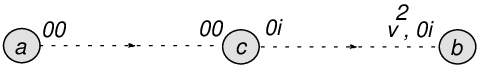} \\
\includegraphics[width=2.5in]{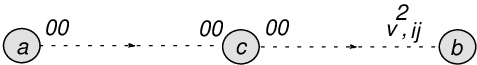} \
\includegraphics[width=2.5in]{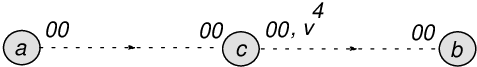} \\
\includegraphics[width=2.5in]{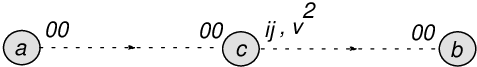} \
\includegraphics[width=2.5in]{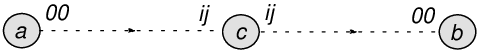} \\
\includegraphics[width=2.5in]{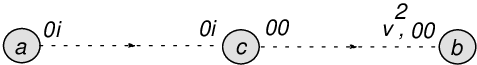} \
\includegraphics[width=2.5in]{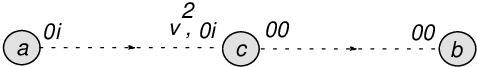} \\
\includegraphics[width=2.5in]{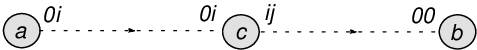} \
\includegraphics[width=2.5in]{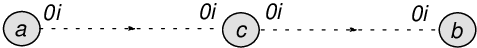} \\
\includegraphics[width=2.5in]{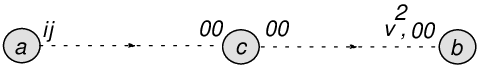} \
\includegraphics[width=2.5in]{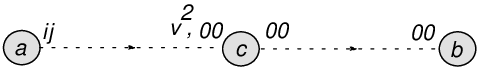} \\
\includegraphics[width=2.5in]{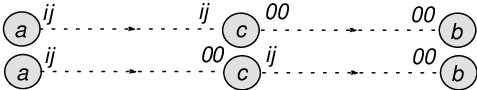} \
\includegraphics[width=2.5in]{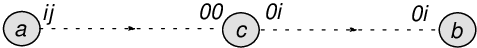} \\
\includegraphics[width=2.5in]{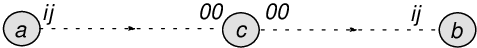}
\end{center}
\caption{3 PN 3 body diagrams with no graviton vertices.}
\label{3PN3BodyDiagramsNoVtx3}
\end{figure*}


\begin{figure*}
\begin{center}
\includegraphics[width=1.8in]{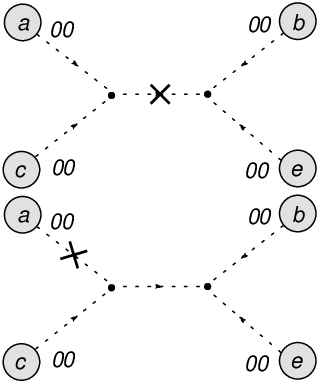} \
\includegraphics[width=1.8in]{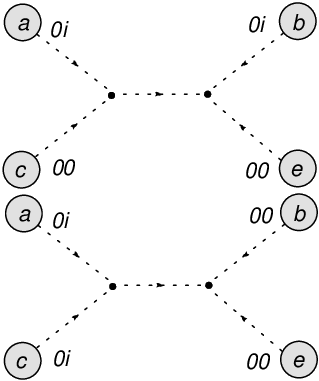} \
\includegraphics[width=1.8in]{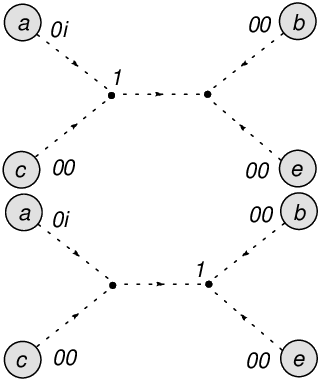} \\
\includegraphics[width=1.8in]{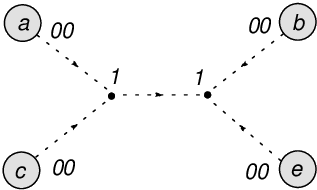} \
\includegraphics[width=1.8in]{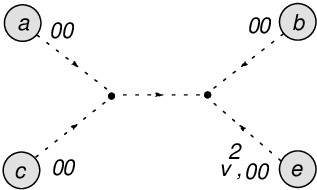} \
\includegraphics[width=1.8in]{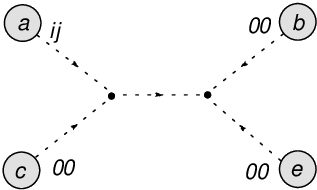} \\
\includegraphics[width=1.8in]{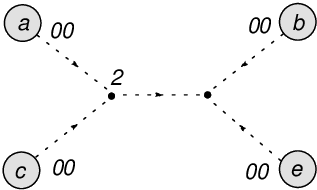} \
\end{center}
\caption{3 PN 4 body diagrams with graviton vertices, 1 of 3.}
\label{3PN4BodyDiagramsIofIII}
\end{figure*}

\begin{figure*}
\begin{center}
\includegraphics[width=1.2in]{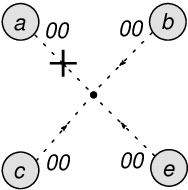} \
\includegraphics[width=1.2in]{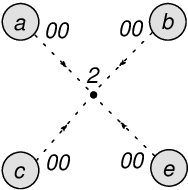} \
\includegraphics[width=1.2in]{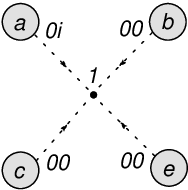} \\
\includegraphics[width=1.2in]{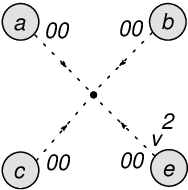} \
\includegraphics[width=1.2in]{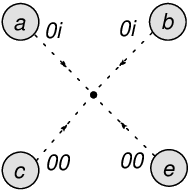} \
\includegraphics[width=1.2in]{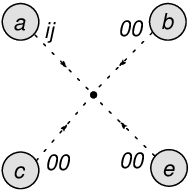}
\end{center}
\caption{3 PN 4 body diagrams with graviton vertices, 2 of 3.}
\label{3PN4BodyDiagramsIIofIII}
\end{figure*}

\begin{figure*}
\begin{center}
\includegraphics[width=3in]{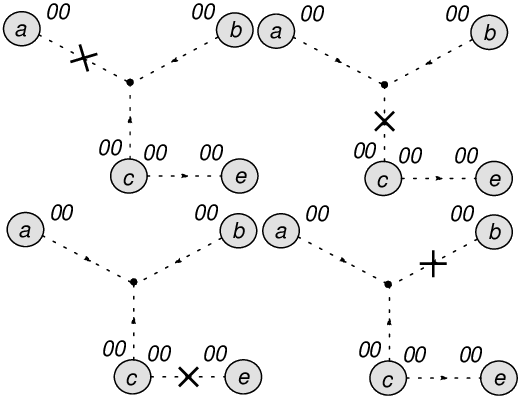} \
\includegraphics[width=3in]{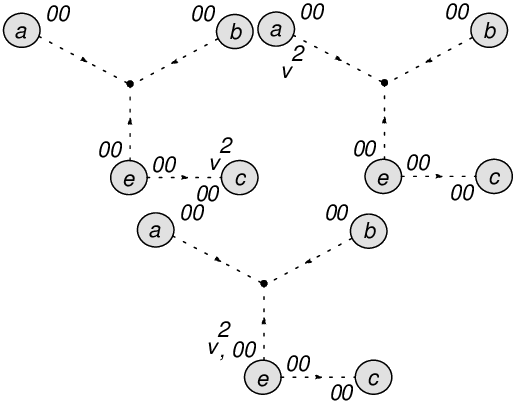} \\
\includegraphics[width=1.5in]{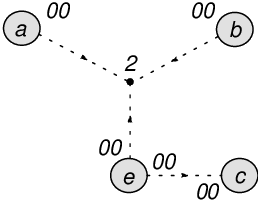} \
\includegraphics[width=1.5in]{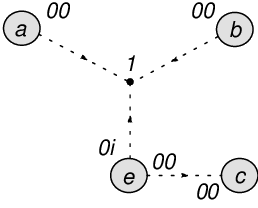} \
\includegraphics[width=1.5in]{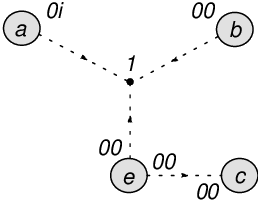} \
\includegraphics[width=1.5in]{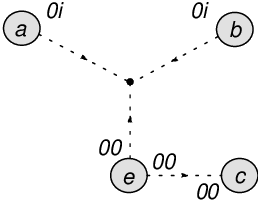} \\
\includegraphics[width=1.5in]{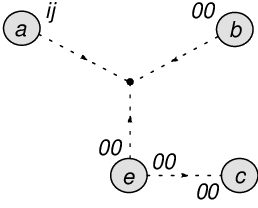}  \
\includegraphics[width=1.5in]{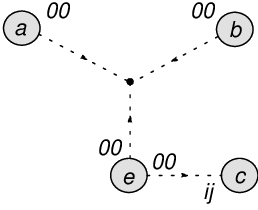} \
\includegraphics[width=3in]{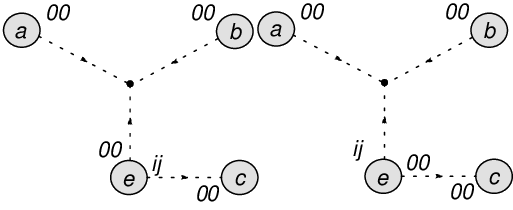}    \
\includegraphics[width=3in]{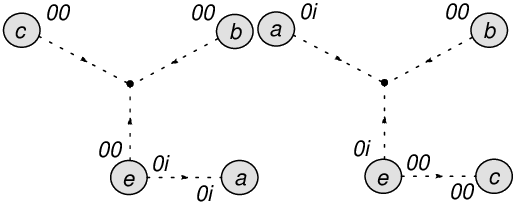}
\end{center}
\caption{3 PN 4 body diagrams with graviton vertices, 3 of 3.}
\label{3PN4BodyDiagramsIIIofIII}
\end{figure*}

\begin{figure*}
\begin{center}
\includegraphics[width=1.5in]{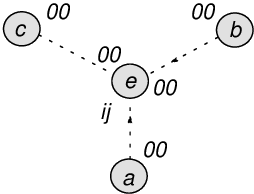} \
\includegraphics[width=1.5in]{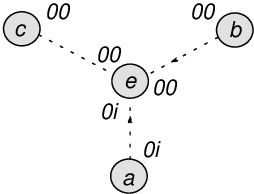} \
\includegraphics[width=1.5in]{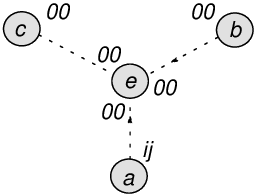} \\
\includegraphics[width=1.5in]{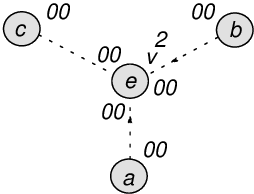} \
\includegraphics[width=1.5in]{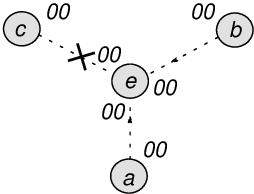} \
\includegraphics[width=1.5in]{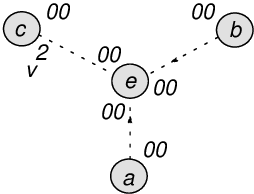}
\end{center}
\caption{3 PN 4 body diagrams with no graviton vertices, 1 of 2.}
\label{3PN4BodyDiagramsNoVtxIofII}
\end{figure*}

\begin{figure*}
\begin{center}
\includegraphics[width=2.5in]{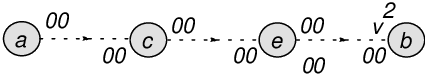} \
\includegraphics[width=2.5in]{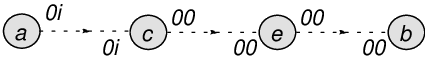} \\
\includegraphics[width=2.5in]{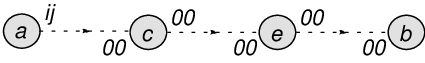} \
\includegraphics[width=2.5in]{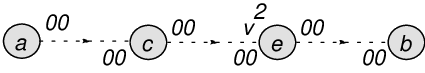} \\
\includegraphics[width=2.5in]{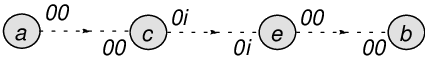} \
\includegraphics[width=2.5in]{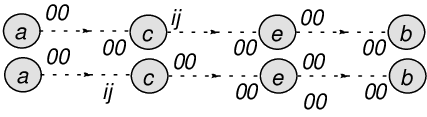} \\
\includegraphics[width=2.5in]{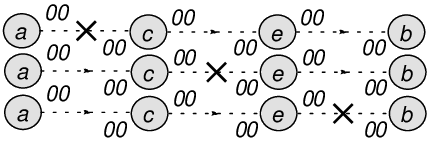}
\end{center}
\caption{3 PN 4 body diagrams with no graviton vertices, 2 of 2.}
\label{3PN4BodyDiagramsNoVtxIIofII}
\end{figure*}


\begin{figure*}
\begin{center}
\includegraphics[width=3.2in]{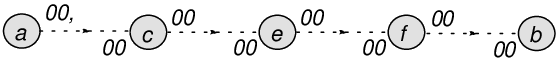}    \\
\includegraphics[width=1.2in]{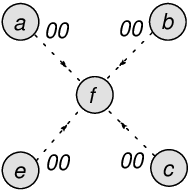}    \
\includegraphics[width=1.7in]{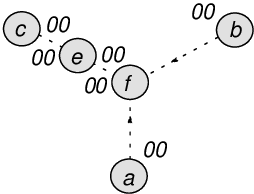}
\end{center}
\caption{3 PN 5 body diagrams with no graviton vertices.}
\label{3PN5BodyDiagramsNoVtx}
\end{figure*}

\begin{figure*}
\begin{center}
\includegraphics[width=2in]{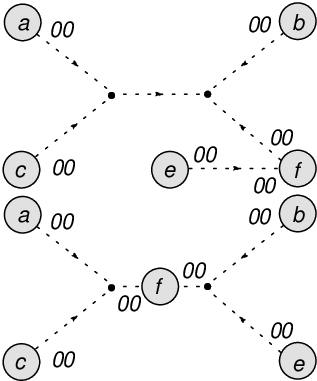}    \
\includegraphics[width=2.4in]{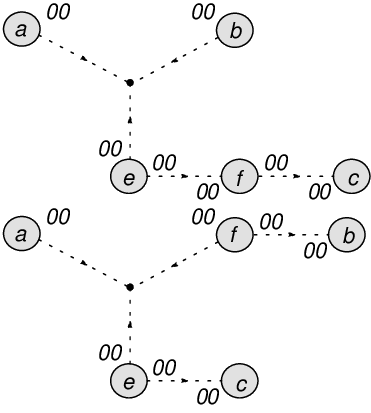}  \\
\includegraphics[width=1.7in]{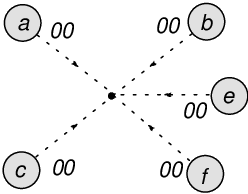}  \
\includegraphics[width=2.4in]{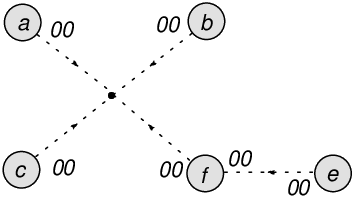}  \\
\includegraphics[width=2.5in]{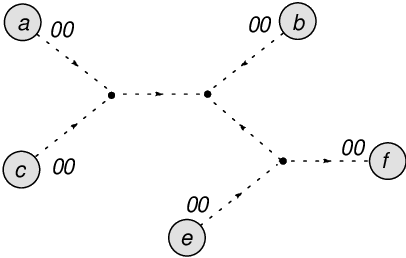}  \\
\includegraphics[width=2in]{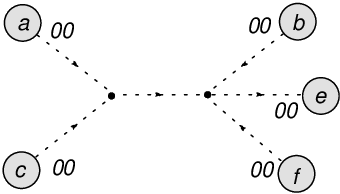}    \
\includegraphics[width=2in]{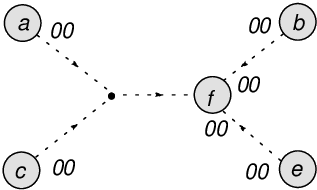}    \
\end{center}
\caption{3 PN 5 body diagrams with graviton vertices.}
\label{3PN5BodyDiagramsVtx}
\end{figure*}

\end{document}